\documentclass[showpacs,prb,superscriptaddress,twocolumn]{revtex4}

	\usepackage{textcomp}
	\usepackage{amsmath}
	\usepackage{amssymb}
	\usepackage{graphicx} 
	\usepackage[lofdepth,lotdepth,caption=false]{subfig}
	\usepackage[breaklinks=true,colorlinks,citecolor=blue,linkcolor=blue,urlcolor=blue]{hyperref}
	\usepackage{mathtools}
	\usepackage{bm}
	\usepackage{color}
	\usepackage{ulem}
	\usepackage{textcomp}
	\usepackage[percent]{overpic}

	\newcommand\at[2]{\left.#1\right|_{#2}}

	\newcommand{\sign}{\text{sign}}

\begin{document}

	\title{Thermoelectric properties of an interacting quantum dot-based heat engine}

	\author{Paolo Andrea Erdman}
	\affiliation{NEST, Scuola Normale Superiore and Istituto Nanoscienze-CNR, I-56127 Pisa, Italy}
	\email{paolo.erdman@sns.it}

	\author{Francesco Mazza}
	\affiliation{NEST, Scuola Normale Superiore and Istituto Nanoscienze-CNR, I-56127 Pisa, Italy}

	\author{Riccardo Bosisio}
	\affiliation{NEST, Scuola Normale Superiore and Istituto Nanoscienze-CNR, I-56127 Pisa, Italy}

	\author{Giuliano Benenti}
\affiliation{Center for Nonlinear and Complex Systems,
Dipartimento di Scienza e Alta Tecnologia,
Universit\`a degli Studi dell'Insubria, via Valleggio 11, 22100 Como, Italy}
\affiliation{Istituto Nazionale di Fisica Nucleare, Sezione di Milano,
via Celoria 16, 20133 Milano, Italy}
\affiliation{NEST, Istituto Nanoscienze-CNR, I-56126 Pisa, Italy}

	\author{Rosario Fazio}
	\affiliation{ICTP, Strada Costiera 11, I-34151 Trieste, Italy}
	\affiliation{NEST, Scuola Normale Superiore and Istituto Nanoscienze-CNR, I-56127 Pisa, Italy}

	\author{Fabio Taddei}
	\affiliation{NEST, Istituto Nanoscienze-CNR and Scuola Normale Superiore, I-56126 Pisa, Italy}

	\begin{abstract}
We study the thermoelectric properties and heat-to-work conversion performance of an interacting, multi-level quantum dot (QD) weakly coupled to electronic reservoirs. We focus on the sequential tunneling regime.
The dynamics of the charge in the QD is studied by means of master equations for the probabilities of occupation. From here we compute the charge and heat currents in the linear response regime.
Assuming a generic multi-terminal setup, and for low temperatures (quantum limit), we obtain analytical expressions for the transport coefficients which account for the interplay between interactions (charging energy) and level quantization. In the case of systems with two and three terminals we derive formulas for the power factor $Q$ and the figure of merit $ZT$ for a QD-based heat engine, identifying optimal working conditions which maximize output power and efficiency of heat-to-work conversion. Beyond the linear response we concentrate on the two-terminal setup. 
We first study the thermoelectric non-linear coefficients assessing the consequences of large temperature and voltage biases, focusing on the breakdown of the Onsager reciprocal relation between thermopower and Peltier coefficient. We then investigate the conditions which optimize the performance of a heat engine, finding that in the quantum limit output power and efficiency at maximum power can almost be simultaneously maximized by choosing appropriate values of electrochemical potential and bias voltage. At last we study how energy level degeneracy can increase the output power.
	\end{abstract}



	\maketitle

	\section{Introduction}

The study of thermoelectric effects in nanostructures~\cite{bib:dresselhaus,bib:vineis,bib:snyder,PRreview} is attracting increasing interest.
Heat-to-work conversion based on thermoelectricity promises an enhanced efficiency as a consequence of the reduction of the phonon contribution 
to thermal conductance in disordered nanostructures~\cite{Hochbaum} and 
of the ``energy filtering'' effect that can result from confinement and 
quantum effects \cite{bib:hicks,bib:mahan}.
In particular, an increase of the electron contribution to the figure of merit $ZT$ (which controls the maximum efficiency and the efficiency at maximum power) is possible if one can ``filter'' the electrons participating in the transport to a narrow energy range~\cite{bib:mahan}.

A heat engine composed of a quantum dot (QD) is a paradigmatic example, since it is characterized by a spectrum of discrete levels which maximizes energy filtering.
The thermoelectric properties of QD systems~\cite{bib:zianni_2007,bib:zianni_2008,bib:beenakker1,bib:beenakker2,rejec2012,muralidharan2013,jacquet2009,wohlman2012,bjorn2013,kennes2013,billings2010,kubala2006,koch2004,andreev2001,dutt2013,costi2010,turek2002,fazio2001,jukka2008,mani2011,sanchezPRB2013,Sanchez2014,zimbovskaya2016} and the performance of QD-based heat engines~\cite{weymann2013,donsa2014,sothmann2012,liu2011,kuo2013,kuo2010,wierzbicki2010,rossello2017,perroni2016,szukiewicz2016,Mazza2014,kuo2016,karol2006,murphy2008,kuo2009,dubi2009, imry2010,karlstrom2011,trocha2012,tseng2015,taylor2015,esposito2009,swirkowicz2009,leijnse2010,nakpathomkun2010,buttiker2011,muralindharan2012,buttiker2013,de2016,agarwal2014,zianni2010,liu2010} has been studied theoretically by a number of authors (see Ref.~\onlinecite{sothmann2015} for a review).
The vast majority of the papers dealing with QD-based heat engines consider a single degenerate energy level or two non-degenerate levels~\cite{weymann2013,donsa2014,sothmann2012,liu2011,kuo2013,wierzbicki2010,rossello2017,perroni2016,szukiewicz2016,Mazza2014,kuo2016,karol2006,murphy2008,esposito2009,kuo2009,swirkowicz2009,dubi2009, imry2010,leijnse2010, nakpathomkun2010,buttiker2011,karlstrom2011,trocha2012, muralindharan2012,buttiker2013,tseng2015,taylor2015,de2016}.
The case of QDs with many levels has been addressed only in few papers~\cite{agarwal2014,zianni2010,liu2010,kuo2010}.
Moreover, the performance of QD-based heat engine has been mostly studied within the linear response regime~\cite{weymann2013,donsa2014,liu2011,rossello2017,kuo2013,wierzbicki2010,Mazza2014,kuo2016,karol2006,murphy2008,kuo2009,dubi2009,imry2010,karlstrom2011,trocha2012,tseng2015,taylor2015,zianni2010,liu2010}, where the thermoelectric performance of the system is entirely characterized by $ZT$.
The case of an interacting multi-level QD beyond the linear response has not been addressed so far.
On the one hand, the presence of many levels is expected to yield important consequences.
Indeed, already in the linear-response regime they have an impact on the thermopower, the thermal conductance and $ZT$~\cite{bib:beenakker2,liu2010,bib:zianni_2007,bib:zianni_2008}.
On the other, non-linear effects, relevant when larger temperature and voltage biases are applied, are important as far as power and efficiency are concerned.
We emphasize that a number of experiments assessing the thermoelectric properties of QDs have been reported in Refs.~\onlinecite{llaguno2003,thierschmann2013,scheibner2007,godijn1999,reddy2007,dzurak1998,svilansPhysE2016,svensson2012,svilans2016,scheibner2005,dzurak1993,harman2002,molenkamp1992, staring1993, pogosov2006, scheibner2008, svensson2013, matthews2014,glattli2015,hartmann2015,molenkamp2015}.

In this paper we fill this gap by studying the thermoelectric properties and heat-to-work conversion performance of a multi-level QD in a multi-terminal configuration within the Coulomb blockade regime.
We consider the limit of small tunnelling rates (sequential tunnelling regime) and we study both the linear and non-linear response regimes.
Coulomb interaction among electrons is accounted for by a finite and small capacitance $C$ whose associated energy scale is its charging energy $(Ne)^2/2C$, where $N$ is the number of electrons in the QD and $e$ is the electron charge.
Moreover, we consider a generic multi-terminal structure, whereby the QD is connected to many (two or more) reservoirs.
We will concentrate only 
on the optimization of the thermoelectric properties of the electronic system, 
neglecting the parasitic phononic contribution to heat transport.
Our results therefore set an upper bound
to the thermoelectric efficiency of the QD, approachable 
only in the limit in which suitable strategies to strongly
reduce phonon transport are implemented. 

Generalizing Refs.~\onlinecite{bib:beenakker1} and \onlinecite{bib:beenakker2} to the multi-terminal case, by solving a set of kinetic equations one can determine the probability of occupation of the energy levels of the QD in a multi-terminal setup, thus allowing to calculate the charge and heat currents for given values of the electrochemical potentials and temperatures of the reservoirs.

In the linear response regime, where voltage and temperature biases are small, we derive closed-form expressions for the charge and heat currents and specify their limits of validity.
We define local and non-local transport coefficients and express them in terms of a generating function under the assumption that the tunneling rates are energy-independent. We then derive, in the low temperature limit, analytical expressions for all transport coefficients as a function of the electrochemical potential $\mu$.
Along with the main features of the transport coefficients (located around values of $\mu$ equal to the dominant transition energies required to add or remove an electron from the QD), such expressions also describe a fine structure arising from the interplay between interaction and level quantization (controlled by the two energy scales: charging energy and level spacing).
Furthermore, for the calculation of the thermal conductance we find that it is crucial to consider the presence of many levels.
Within the linear response, we consider both the two-terminal and the three-terminal system aiming at addressing the performance of heat-to-work conversion.
In the former case we obtain analytical expressions for the power factor $Q$ and the figure of merit $ZT$ finding, remarkably, that those quantities are simultaneously maximized for values of the electrochemical potential which differ by about $2.40k_BT$ with respect to the dominant transition energies.
In addition, $ZT$ shows a fine structure of secondary peaks whose height is independent of the system's parameters and can take values as large as $ZT=9$.
We compare the figure of merit with a non-interacting system, finding that Coulomb interactions dramatically increase $ZT$ by strongly suppressing the thermal conductance. For the case of three terminals with energy-independent tunneling rates, we derive analytic expressions for the maximum power and corresponding efficiency, finding that the addition of a third terminal at an intermediate temperature decreases the efficiency at maximum power, but can increase the power. We also find particular intermediate temperatures where the third terminal increases the maximum power and achieves the same efficiency of a two terminal system.

We analyze the regime beyond the linear response by numerically solving the kinetic equations, focusing on the two-terminal setup.
Going beyond the linear response, i.e. considering large temperature and voltage biases, $\Delta T$ and $V$, is interesting for various reasons. On one hand, it allows to increase the Carnot efficiency $\eta_C$ and the power generated by a heat engine (in our case the peak power scales approximately as $(\Delta T)^2$
also beyond the linear response regime).
On the other hand, the efficiency at maximum power is not bounded by $\eta_C/2$, as in the linear response, and can even go above the Curzon-Alhborn 
efficiency~\cite{bib:yvon,bib:chambadal,bib:novikov,bib:curzon,bib:broeck}.
In literature, the scattering theory of 
nonlinear thermoelectric transport in quantum conductors
has been developed only recently
\cite{meair2013,whitney2013,Sanchez2013}.
The regime beyond linear response 
for QD-based heat engines has been theoretically addressed 
in Refs.~\onlinecite{perroni2016,de2016,agarwal2014,buttiker2013,sothmann2012,buttiker2011,kuo2010,nakpathomkun2010,swirkowicz2009,esposito2009, leijnse2010, muralindharan2012, szukiewicz2016}, but limited to single or double level quantum dots.

In discussing the results, we first focus on the behavior of the non-linear Seebeck and Peltier coefficients aiming at assessing the interplay between charging energy and level spacing on these two quantities and how the Onsager reciprocity relation that connects them is violated beyond linear response.
Second, we study the efficiency and output power of a heat engine.
In particular, we calculate the maximum efficiency and maximum power by maximizing such quantities with respect to the applied bias voltage, for fixed values of temperature bias and electrochemical potential.
The maximum efficiency shows only quantitative changes, with respect to the linear response, by increasing the temperature bias. The efficiency at maximum power instead develops peaks which go beyond the $\eta_C/2$ linear-response limit and approach $\eta_C$ for large temperature differences. Remarkably, the maximum power, normalized to its peak value, only slightly depends on the temperature bias and can be well approximated by the analytic expression obtained for the linear response regime.
Moreover, we find that efficiency at maximum power and maximum power take approximately their peak values simultaneously, under the same conditions found for the linear response. 
Finally, we assess the impact of interactions by comparing the efficiency at maximum power in two situations, namely the case of a doubly degenerate level with interaction and the case of two non-degenerate levels without interaction.
We find that, especially when charging energy and level spacing are of the order of the thermal energy, the efficiency at maximum power is much higher in the interacting case and goes above the Curzon-Alhborn efficiency.
We also find that in the doubly degenerate interacting case the maximum power is enhanced by almost a factor 2 with respect to the non-degenerate case.

The paper is organized as follows: in Sec.~\ref{theory} we describe the system under investigation and we detail the theoretical model.
In Sec.~\ref{chap:linear} the model is specified to the linear response regime.
Here we report analytic expressions obtained in the quantum limit, Sec.~\ref{qr}, and calculate power and efficiency in the cases of a two-terminal system, Sec.~\ref{2terms}
(with the efficiencies of interacting and non-interacting QDs compared in
Sec.~\ref{sec:comparisonintnonint}), 
and of a three-terminal system, Sec.~\ref{3term}.
Sec.~\ref{chap:beylinear} is devoted to the discussion of the regime beyond the linear response in a two-terminal system: in Sec.~\ref{sec-SP} we study the non-linear Seebeck and Peltier coefficients and in Sec.~\ref{blEP} we discuss efficiency and output power of a heat engine.
Finally, in Sec.~\ref{conc} we draw our conclusions and discuss future developments.
In addition, the paper includes four appendices where the details of some calculations and the analytic non linear study of a single energy level QD system are reported.

	\section{Multi-level interacting QD}
	\label{theory}
\begin{figure}[!tb]
		\centering
		\includegraphics[width=1\columnwidth]{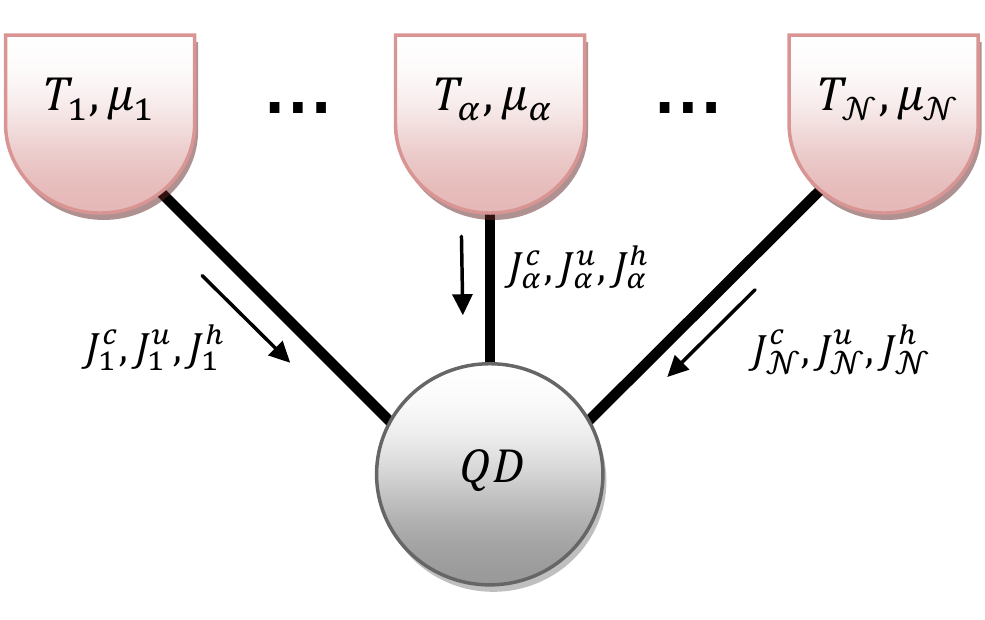}
		\includegraphics[width=1\columnwidth]{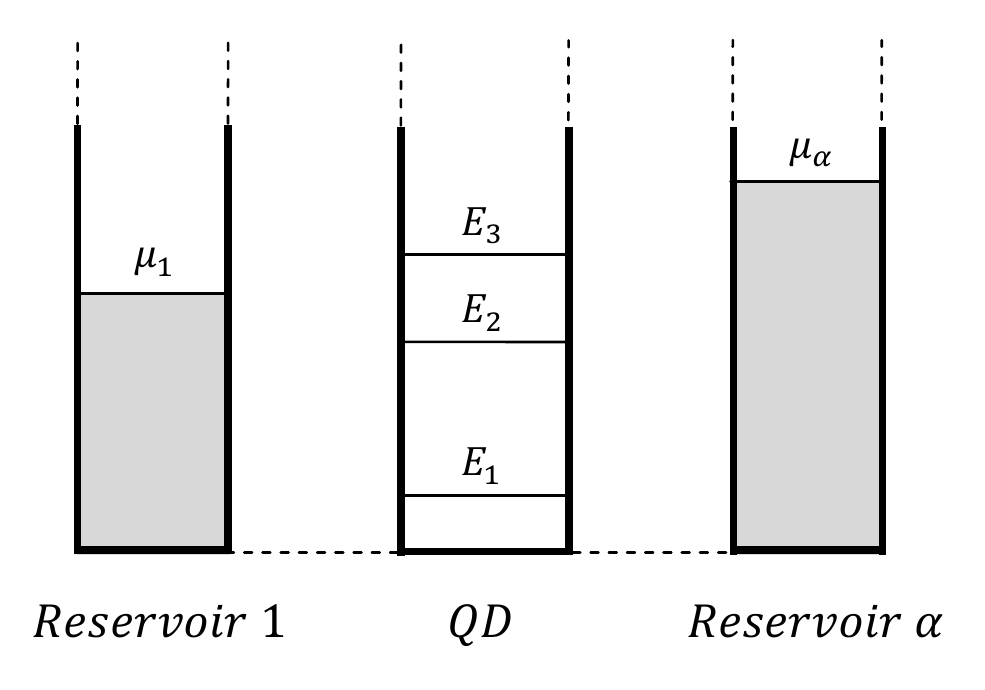}
		\caption{(Color online) Top: A quantum dot (QD) is {\it tunnel-coupled to ${\cal N}$ reservoirs}, each kept at a temperature $T_\alpha$ and at an electrochemical potential $\mu_\alpha$, with $\alpha=1,...,{\cal N}$. Arrows represent charge, energy and heat currents ($J^c_\alpha$, $J^u_\alpha$, and $J^h_\alpha$, respectively) flowing from the reservoirs $\alpha$ to the QD.
	Bottom: Schematic energy representation of a multilevel QD. $E_1$, $E_2$, etc. are the single-electron energy levels of the QD, while $\mu_1$ and $\mu_\alpha$ are the electrochemical potentials relative to reservoir 1 and $\alpha$, respectively.}
		\label{fig:qd_scheme}
	\end{figure}

	In this Section we briefly outline the formalism used to describe the thermoelectric properties of a {\it multilevel interacting QD}.
	We will only consider electron transport, neglecting any contribution due to phonons.
	As shown in Fig.~\ref{fig:qd_scheme} (top), the QD is tunnel-coupled to ${\cal N}$ electron reservoirs, each characterized by a given temperature $T_\alpha$ and electrochemical potential $\mu_\alpha$, so that the occupation of the electrons within reservoir $\alpha$ follows the Fermi distribution
	\begin{equation}
	\label{eq:res_fermi_distr_def}
	  \begin{aligned}
	  &f_\alpha(E) = \left[1+\exp\left(\frac{E-\mu_\alpha}{k_BT_\alpha}\right)\right]^{-1},
	  \end{aligned}
	\end{equation}
	where $k_B$ is Boltzmann's constant.
	In Fig.~\ref{fig:qd_scheme} (bottom), $E_p$ (with $p=1,2,\dots$ labeled in ascending order) are the QD single-electron energy levels.
	These levels can be shifted by means of an applied gate voltage.

	The QD is weakly coupled to the reservoirs through large tunneling barriers.
	More precisely, we assume that thermal energy $k_BT$, level spacing and charging energy are much larger than the coupling energy between reservoirs and QD [$\hbar\sum_\alpha\Gamma_\alpha(p)$, where $\Gamma_\alpha(p)$ is the tunneling rate from level $p$ to reservoir $\alpha$, which we assume independent of the number $N$ of electrons inside the dot].
	As a consequence, the charge on the QD is quantized, i.e. each energy level $E_p$ can have either zero or one electron, $n_p=0$ or $n_p=1$ (any degeneracy, like electron spin, can be taken into account counting each level multiple times), and transport occurs due to single-electron tunneling processes ({\it sequential tunneling regime}).
	The electrostatic energy associated with the electrons within the QD is given by $U(N) = E_C N^2$, where $E_C = e^2/2C$, $N = \sum_i n_i$ is the total number of electrons within the QD, and $C$ is the capacitance of the QD.
	The QD is described by states characterized by a set of occupation numbers $\{n_i\}$ relative to the energy levels.
	The QD changes state whenever a single-electron tunneling process takes place.
	The non-equilibrium probability for a given state $\{n_i\}$ to occur, $P(\{n_i\})$, can be computed \cite{bib:beenakker1,bib:nazarov} by writing a straightforward set of balance equations for $P\left(\{n_i\}\right)$.
	Our aim is to compute in stationary conditions the charge, energy and heat currents out of the electron reservoirs 
	[denoted in Fig.~\ref{fig:qd_scheme} (top) by $J_\alpha^c$, $J_\alpha^u$ and $J_\alpha^h$, 
	respectively],  
	induced by the temperature and electrochemical potential differences.

\subsection{Kinetic equations}
	In what follows we describe a generalization of the method put forward by Beenakker in Refs.~\onlinecite{bib:beenakker1} and \onlinecite{bib:beenakker2}.
	The single-electron tunneling processes that contribute to changing over time the probability $P(\{n_i\})$ are due to electrons that tunnel from the QD to the reservoirs and vice-versa.
	For an electron exiting the QD, initially with $N$ electrons, from energy level $E_p$ and going into reservoir $\alpha$ at energy $E^{\text{fin}}$, energy conservation 
imposes that
	\begin{equation}
	\label{eq:energy_conservation_fin}
		E_p  +  U(N) = E^{\text{fin}}(N)+ U(N-1) .
	\end{equation}
	On the contrary, for an electron that tunnels from an initial state in reservoir $\alpha$ at energy $E^{\text{in}}$ to the level $E_p$ in the QD that initially had $N$ electrons, energy conservation imposes that
	\begin{equation}
	\label{eq:energy_conservation_in}
		E^{\text{in}}(N) + U(N) = E_p + U(N+1) .
	\end{equation}
	$P(\{n_i\})$ can then be determined by the following set of kinetic equations, one for each configuration $\{n_i\}$:
	\begin{widetext}
	\begin{multline}
	\label{eq:kinetic_def}
			\frac{\partial}{\partial t} P\left(\{n_i\}\right) = 
			-\sum\limits_{p\alpha} \delta_{n_p,0} P\left(\{n_i\}\right) \Gamma_\alpha(p) f_\alpha\left(E^{\text{in}}(N)\right) 
			-\sum\limits_{p\alpha} \delta_{n_p,1} P\left(\{n_i\}\right) \Gamma_\alpha(p) \left[ 1-f_\alpha\left(E^{\text{fin}}(N)\right)\right]  \\
			+\sum\limits_{p\alpha} \delta_{n_p,0} P\left(\{n_i\},n_p=1\right) \Gamma_\alpha(p) \left[ 1-f_\alpha\left(E^{\text{fin}}(N+1) \right)\right] 
			+\sum\limits_{p\alpha} \delta_{n_p,1} P\left(\{n_i\},n_p=0\right) \Gamma_\alpha(p) f_\alpha\left(E^{\text{in}}(N-1) \right)  ,
	\end{multline}
	\end{widetext}
	where we have introduced the notation
	\begin{equation}
		P\left(\{n_i\},n_p=1\right) = P\left(\{n_1,\dots,n_{p-1},1,n_{p+1},\dots\}\right) 
	\end{equation}
	and
	\begin{equation}
		P\left(\{n_i\},n_p=0\right) = P\left(\{n_1,\dots,n_{p-1},0,n_{p+1},\dots\}\right)
	\end{equation}
	for the QD states. The first term in Eq.~(\ref{eq:kinetic_def}) accounts for the decrease of the probability $P(\{n_i\})$, with the QD initially in the state $\{n_i\}$, due to an electron coming from a reservoir and occupying an empty level in the QD. The rate of electrons coming from reservoir $\alpha$ will be given by a sum over all empty levels $p$ (such that $n_p = 0$) of the tunnel rate $\Gamma_\alpha (p)$, multiplied by the probability of finding the QD in this state, $P\left(\{n_i\}\right)$, and multiplied by the reservoir's occupation $f_\alpha\left(E^{\text{in}} (N) \right)$ at the correct energy $E^{\text{in}} (N)$ to tunnel to level $p$.
The second term accounts for the decrease of the probability $P(\{n_i\})$, with the QD initially in the state $\{n_i\}$, due an electron leaving the QD from an occupied level to tunnel into a reservoir.
The third term accounts for the increase of the probability $P(\{n_i\})$ if the QD is in a state with an extra electron in level $p$ with respect to $\{n_i\}$, and if this electron leaves the QD, tunneling to the reservoirs.
The forth term accounts for the increase of the probability $P(\{n_i\})$ if the QD is in a state with a missing electron in level $p$ with respect to $\{n_i\}$, and if this electron enters the QD in level $p$, tunneling from the reservoirs.
	The {\it stationary solution} of the kinetic equations, obtained imposing $\partial P/\partial t = 0$, together with the normalization request
	\begin{equation}
			\sum\nolimits_{\{n_i\}} P\left(\{n_i\}\right) = 1
			\label{eq:normalize_probabilities}
	\end{equation}
	provides a complete set of equations that uniquely defines $P\left(\{n_i\}\right)$.
	The sum over $\{n_i\}$ means the sum over $n_i = 0,1$, with $i = 1,2,...$.

\subsection{Charge, energy, and heat currents}

	Charge $J^c_\alpha$ and energy $J^u_\alpha$ currents flowing from reservoir $\alpha$ to the QD can be calculated as the sum of all possible tunneling processes,
	since the QD can be in any state $\{n_i\}$ with probability $P(\{n_i\})$ and an electron can tunnel into or out of any energy level $E_p$.
	More precisely, for the charge current we have
	\begin{multline}
			J^c_\alpha = e \sum\limits_{p=1}^\infty \sum\limits_{\{n_i\}} P(\{n_i\})\Gamma_\alpha(p)\Big\{\delta_{n_p,0}f_\alpha(E^{\text{in}}(N)) \\
			-\delta_{n_p,1}[1-f_\alpha(E^{\text{fin}}(N))] \Big\},
	\label{eq:jc_general}
	\end{multline}
	$e$ being the electronic charge, while for the energy current we have
	\begin{widetext}
	\begin{equation}
				J^u_\alpha = \sum\limits_{p=1}^\infty \sum\limits_{\{n_i\}} P(\{n_i\})\Gamma_\alpha(p)\Big\{\delta_{n_p,0}f_\alpha(E^{\text{in}}(N))
				E^{\text{in}}(N)  
				- \delta_{n_p,1}[1-f_\alpha (E^{\text{fin}}(N))]E^{\text{fin}}(N)  \Big\},
	\label{eq:ju_general}
	\end{equation}
	$E^{\text{in}}(N) $ [$E^{\text{fin}}(N) $] being the energy carried by an electron entering (exiting) the QD.
	The heat currents exiting the reservoirs can be calculated as $J^h_{\alpha} =  J^u_{\alpha} -\frac{\mu_\alpha}{e} J^c_{\alpha}$.
	Using Eqs.~(\ref{eq:jc_general}) and (\ref{eq:ju_general}), we find that
	\begin{equation}
				J^h_\alpha = \sum\limits_{p=1}^\infty \sum\limits_{\{n_i\}} P(\{n_i\})\Gamma_\alpha(p)\Big\{\delta_{n_p,0}f_\alpha(E^{\text{in}}(N))
				\left[E^{\text{in}}(N) - \mu_\alpha \right]
				- \delta_{n_p,1}[1-f_\alpha (E^{\text{fin}}(N))]\left[E^{\text{fin}}(N) - \mu_\alpha \right] \Big\}.
	\label{eq:jh_general}
	\end{equation}
	\end{widetext}

	In order to numerically determine the stationary probability distribution 
	$P(\{n_i\})$ from the kinetic equations, 
	we will consider a finite number $L$ of energy levels. 
	\footnote{The results can be interpreted as the exact solution of a system with $L$ energy levels or as an approximate solution of a system with infinite levels.
	In the latter case, one needs to check that the result are stable with increasing number of energy levels.}.
	By organizing the values of $P(\{n_i\})$ into a $2^L$-components vector $\vec{P}$ (two choices $n_i=0,1$ for each level), the kinetic equations (\ref{eq:kinetic_def}) for the 
	stationary probability distribution,  
	$\frac{\partial \vec{P}}{\partial t} =\vec{0}$,
	can be represented as the homogeneous linear system 
	$M \vec{P}=\vec{0}$, where $M$ is a $2^L \times 2^L$ matrix.
	$M$ must have a null space of at least dimension 1, 
	otherwise the only possible solution would be the trivial one ($\vec{P} \equiv 0$).
	This is demonstrated in App.~\ref{app:KE} by showing that 
	summing together all the stationary kinetic equations yields zero. 
	We can thus find the probabilities by including the normalization condition, Eq.~(\ref{eq:normalize_probabilities}).

	By defining $\widetilde{N} = \sum\limits_{i \neq p } n_i$ it is possible to show that the kinetic equations for the stationary probability distribution can be written as
	\begin{multline}
	\label{eq:kinetic_a_b_def}
		\sum\limits_p (\delta_{n_p,1} - \delta_{n_p,0})\left[ P\left( \{n_i\}, n_p=0 \right)A_{\widetilde{N},p}  \right. \\
		\left. - P\left( \{n_i\}, n_p=1 \right)B_{\widetilde{N},p}  \right] = 0,
	\end{multline}
	where
	\begin{equation}
	A_{\widetilde{N},p} =\sum_\alpha \Gamma_\alpha(p) f_\alpha\left(E^{\text{in}} (\widetilde{N}) \right) 
	\label{eq:a_def}
	\end{equation}
	and
	\begin{multline}
	B_{\widetilde{N},p} =\sum_\alpha \Gamma_\alpha(p) \left[ 1- f_\alpha\left(E^{\text{in}} (\widetilde{N}) \right) \right] \\
	=\sum_\alpha \Gamma_\alpha(p)-A_{\widetilde{N},p} .
	\label{eq:b_def}
	\end{multline}
	To derive Eq.~(\ref{eq:kinetic_a_b_def}) we have used the fact that $\widetilde{N}=N$, for the terms in the kinetic equations proportional to $\delta_{n_p,0}$, and $\widetilde{N}=N-1$, for the terms in the kinetic equations proportional to $\delta_{n_p,1}$, and the 
	identity
	\begin{equation}
	E^{\text{fin}} (N+1)= E^{\text{in}} (N) ,
	\end{equation}
	stemming from Eqs.~(\ref{eq:energy_conservation_fin}) and (\ref{eq:energy_conservation_in}).
	It is worth mentioning that using the kinetic equations in the form of Eq.~(\ref{eq:kinetic_a_b_def}), it is possible to prove that the kinetic 
equations always allows a non-trivial solution (see App.~\ref{app:KE}).

\subsection{Detailed balance equations}

	It is clear that the kinetic equations (\ref{eq:kinetic_a_b_def}) are automatically satisfied when the following set of equations
	\begin{equation}
		P\left( \{n_i\}, n_p=0 \right)A_{\widetilde{N},p} - P\left( \{n_i\}, n_p=1 \right)B_{\widetilde{N},p} = 0
		\label{eq:det_bal}
	\end{equation}
	is fulfilled for all values of $p$ and for all sets of occupation numbers $\{n_i\}$. Following Ref.~\onlinecite{bib:beenakker1}, Eqs.~(\ref{eq:det_bal}) are hereafter referred to as detailed balance equations (DBEs).
	Eqs.~(\ref{eq:det_bal}) represent a set of $L\cdot 2^{L-1}$ equations, since $p$ can take $L$ values and, at a given $p$, all other occupation numbers ($n_1,\dots,n_{p-1},n_{p+1},\dots,n_L$) can be chosen in $2^{L-1}$ different ways.
	Of course, if a solution to the DBEs exists, than it is also a solution of the kinetic equations.
	We can show, however, that 
	the DBEs are not in general consistent, i.e. no set of $P(\{n_i\})$ exists 
	that can simultaneously satisfy all the DBEs 
	(see App.~\ref{app:DBE1}).
	This is also true in the linear response regime.
	In this case, however, we could prove (see App.~\ref{app:DBE2}) that the DBEs are consistent
	if $E_C = 0$, or if $\Delta T_\alpha = 0$ for all $\alpha$, or
	when the tunneling rates are proportional to each other, 
	namely when $\Gamma_\alpha(p) = k_{\alpha} \Gamma_1(p)$, for $\alpha>1$, $k_{\alpha}$ being 
	constants.
	Note that this condition is trivially satisfied when the rates $\Gamma_\alpha$ do not depend on $p$.
	As a result, the DBEs do not allow in general a solution, but when they do they are useful in computing analytically the energy and heat currents in the linear response regime (see Sec.~\ref{chap:linear}).

\subsection{Level balance equations}

	We will now derive a set of equations that is always consistent and that can be used in the general case to obtain a closed-form expression of 
	the charge current in the linear response regime (see Sec.~\ref{chap:linear}).
	 We impose that, in stationary conditions, the rate of electrons entering any given QD energy level $p$ must equal the rate of electrons leaving that energy level.
	For electrons tunneling into the QD, initially with $N$ electrons, from any reservoir, one has to require that level $p$ is empty and must consider all possible states $(\{n_i\}, n_p=0)$, where $\widetilde{N}=N$.
	The total rate of electrons entering energy level $E_p$ is given by
	\begin{equation}
		\sum\limits_{\alpha,\{n_i\}_{i\neq p}} P(\{n_i\}, n_p=0) \Gamma_\alpha(p) f_\alpha\left(E^{\text{in}}(N) \right) .
	\end{equation}
	For electrons tunneling out of the QD, initially with $N$ electrons, to any reservoir one has to require that level $p$ is occupied and must consider states with $(\{n_i\}, n_p=1)$, where $\widetilde{N}=N-1$.
	The total rate of electrons leaving level $E_p$ is given by
	\begin{equation}
		\sum\limits_{\alpha, \{n_i\}_{i\neq p}}  P(\{n_i\}, n_p=1)  \Gamma_\alpha(p) \left[1-f_\alpha\left(E^{\text{fin}}(N) \right)\right]  .
	\end{equation}
	If we equate the rates of electron entering and leaving level $p$, and use the notation introduced in Eqs.~(\ref{eq:a_def}) and (\ref{eq:b_def}), we obtain
	\begin{multline}
		\label{eq:lev_bal}
		\sum\limits_{ \{n_i\}_{i\neq p} }\left[ P(\{n_i\}, n_p=0)A_{\widetilde{N},p}  \right. \\
		\left. - P(\{n_i\}, n_p=1)B_{\widetilde{N},p}  \right] = 0.
	\end{multline}
	We will refer to this set of $L$ equations (one for each energy level $p$) as the {\it level balance equations} (LBEs)
	\footnote{Note that the DBEs (\ref{eq:det_bal}) could be obtained using the same arguments, but assuming that the QD is in a specific configuration $(\{n_i\})_{i\neq p}$, instead of summing over all possible such configurations. To obtain the LBEs, the probability that level $p$ is occupied is given by the marginal probability $\sum\nolimits_{ \{n_i\}_{i \neq p} }  P\left( \{n_i\}, n_p=1 \right)$.}.
	Note that, using an argument similar to that put forward in App.~\ref{app:KE}, it is possible to prove that Eqs.~(\ref{eq:lev_bal}) can be obtained from the kinetic equations, thus the LBEs are always consistent with the kinetic equations.
	However, the number of LBEs (equal to $L$) is smaller than the number of kinetic equations (equal to $2^L$) and they might not be sufficient to determine the probabilities $P\left(\{n_i\}\right)$.
	Note that one can prove that Eqs.~(\ref{eq:lev_bal}) yield charge current conservation: $\sum_\alpha J^c_\alpha=0$.

\subsection{Output power and efficiency}
\label{chap:eff_pow_def}
Under steady-state conditions, the output power $P$ of a multi-terminal system is given by the sum of all the heat currents
\begin{equation}
	P = \sum\limits_{\alpha=1}^{\cal N} J^h_\alpha.
	\label{eq:p_def}
\end{equation}
If $P>0$, the system behaves as a heat engine, i.e. converting heat into work.
In this situation the efficiency $\eta$ is defined as the ratio between the output power and the total heat current absorbed by the system
\begin{equation}
	\eta = \dfrac{P}{ \sum\limits_{\alpha'} J^h_{\alpha'} },
	\label{eq:eta_def}
\end{equation}
where the sum over $\alpha'$ runs over all positive heat currents.
For a two-terminal system the efficiency cannot exceed the Carnot efficiency defined as $\eta_C = 1 - T_1/T_2$, with $T_2>T_1$.
In addition, for a multi-terminal system $\eta$ cannot go beyond the two-terminal Carnot efficiency~\cite{Mazza2014} calculated using the hottest and coldest temperatures among $T_1,T_2, ...,T_{\cal N}$. 

We define the temperature and electrochemical potential differences as $T_\alpha = T + \Delta T_\alpha$ and $\mu_\alpha = \mu + \Delta \mu_\alpha$, with $\alpha=1,...,{\cal N}$, and choosing reservoir $1$ as the reference value, i.e. $\Delta T_1 = \Delta \mu_1 = 0$. In what follows we fix the values of $\Delta T_\alpha$ and calculate the maximum output power, $P_{\text{max}}$, and maximum efficiency, $\eta_{\text{max}}$, by varying $\Delta \mu_\alpha$. We also consider the efficiency at maximum power, $\eta(P_{\text{max}})$, which is the efficiency when the values of $\Delta \mu_\alpha$ are chosen to maximize the power.

For a two-terminal system within the linear response regime, i.e. when the charge and heat currents depend linearly on the temperature and electrochemical potential differences, both the output power and efficiency can be written in terms of the transport coefficients, namely the electrical conductance $G$, the thermopower $S$ and the thermal conductance $K$, which will be defined in Eqs.~(\ref{eq:g_def}), (\ref{eq:s_def}) and (\ref{eq:k_def}) by setting $\alpha=\beta=2$.
Defining $\Delta T \equiv \Delta T_{2}>0$, we have the following relations \cite{PRreview,bib:ioffe2}
\begin{align}
	&P_{\text{max}} = \frac{1}{4}Q \Delta T^2, \label{eq:p_max_q} \\
	&\eta\left(P_{\text{max}}\right) = \frac{\eta_C}{2} \frac{ZT}{ZT+2}, \label{eq:eta_p_max_zt} \\
	&\eta_{\text{max}} = \eta_C\frac{\sqrt{1+ZT}-1}{\sqrt{1+ZT}+1}, \label{eq:eta_max_zt}
\end{align}
where $Q=GS^2$ is the \textit{power factor} and $ZT = GS^2T/K$ is the (dimensionless) \textit{figure of merit}. 
  As we can see in Eqs. (\ref{eq:eta_p_max_zt}) and (\ref{eq:eta_max_zt}), both $\eta(P_{\text{max}})$ and $\eta_{\rm max}$ 
	are monotonous growing functions of $ZT$; the only restriction imposed by thermodynamics is $ZT\ge 0$.
	When $ZT=0$ both $\eta_{\rm max}$ and $\eta(P_{\text{max}})$ vanish, while for $ZT\to\infty$, $\eta_{\rm max}\to\eta_C$, and 
	$\eta(P_{\text{max}})\to\eta_{CA}$, where $\eta_{CA}=\eta_C/2$ is the so-called Curzon-Ahlborn efficiency
	\cite{bib:yvon,bib:chambadal,bib:novikov,bib:curzon,bib:broeck} in linear response.

	\section{Linear response regime}
	\label{chap:linear}
	As already mentioned above, in the linear response regime the applied temperature and electrochemical potential biases are small enough so that the currents depend linearly on them.
Assuming that $|\Delta T_\alpha|\ll T$ and $|\Delta \mu_\alpha|\ll k_B T$, we 
	follow Refs.~\onlinecite{bib:beenakker1} and \onlinecite{bib:beenakker2} and suppose the probability $P(\{n_i\})$ to differ from its equilibrium distribution $P_{\text{eq}}(\{n_i\})$ in the following way:
	\begin{equation}
		\label{eq:def_psi}
	 P(\{n_i\}) = P_{\text{eq}}(\{n_i\})\left[1+\psi(\{n_i\})\right],
	\end{equation}
	where $\psi$ is a ``small'' function.
	In Eq.~(\ref{eq:def_psi})	
	\begin{equation}
	\label{eq:probabilities_eq}
		P_{\text{eq}}(\{n_i\}) = \frac{1}{Z} \exp\left[-\frac{1}{k_BT}\left( \sum\limits_{p=1}^\infty E_p n_p + U(N)-\mu N   \right)\right]
	\end{equation}
	is the Gibbs distribution in the grand canonical ensemble, when all reservoirs have the same temperature and electrochemical potential,
	with grand partition function given by
	\begin{equation}
	Z= \sum\limits_{\{n_i\}} \exp\left[-\frac{1}{k_BT}\left( \sum\limits_{p=1}^\infty E_p n_p + U(N)-\mu N   \right)\right].
	\end{equation}
	
	In our expressions we will consider
	terms up to first order in $\psi$, $\Delta T_\alpha/T$, and $\Delta\mu_\alpha/k_BT$.
	By linearizing the LBEs with respect to the above small quantities, Eq.~(\ref{eq:lev_bal}), one finds the relation
	\begin{widetext}
	\begin{multline}
	\sum\limits_{\{n_i\}_{i\neq p}} P_{\text{eq}}(\{n_i\},n_p=0) f\left(E_p+(2\widetilde{N}+1)E_C\right)
	\\ \times
	\sum_{\alpha}\Gamma_\alpha(p)\left\{ \psi(\{n_i\},n_p=0) - \psi(\{n_i\},n_p=1)
	+\frac{1}{k_BT}\left[ \left(E_p+(2\widetilde{N}+1)E_C-\mu \right) \frac{\Delta T_\alpha}{T} +\Delta\mu_\alpha 
	\right] \right\}=0 ,
	\label{LBErel}
	\end{multline}
	where $f(E)$ stands for the Fermi distribution at temperature $T$ and 
	electrochemical potential $\mu$.
	By expressing $P(\{n_i\})$ in terms of $\psi(\{n_i\})$ and linearizing Eq.~(\ref{eq:jc_general}), we can use Eq.~(\ref{LBErel}) to remove $\psi(\{n_i\})$ from the charge current, and we find the following closed-form expression:
	\begin{equation}
	\label{eq:lin_jc}
		J^c_\alpha = \frac{e}{k_BT}\sum\limits_{p=1}^\infty \sum\limits_{N=1}^{\infty} P_{\text{eq}}(N) F_{\text{eq}}(E_p|N)
		\left[ 1-f(\epsilon(N,p))  \right]
		\sum_{\beta}\frac{\Gamma_\alpha(p)\Gamma_\beta(p)}{\Gamma_\text{tot}(p)}
		\left[(\epsilon(N,p)-\mu)\frac{\Delta T_\alpha - \Delta T_\beta}{T} + (\Delta\mu_\alpha-\Delta\mu_\beta) \right] ,
	\end{equation}
	where $\Gamma_\text{tot}(p)=\sum_\alpha \Gamma_\alpha(p)$, and 
	\begin{equation}
	\epsilon(N,p) = 
	E_p+U(N)-U(N-1)=E_p + (2N-1)E_C
	\label{eq:epsilonNp}
	\end{equation}
	is the energy needed to add to the QD, initially occupied by $N-1$ electrons, 
	the $N$-th electron to level $p$ (and equivalently for the inverse process).
	In Eq.~(\ref{eq:lin_jc}) the quantity
	\begin{equation}
		P_{\text{eq}}(N) \equiv \sum\limits_{\{n_i\}} P_{\text{eq}}(\{n_i\}) 
	\delta_{\sum n_i,N} 
		\label{eq:peq_def}
	\end{equation}
	is the equilibrium probability of finding $N$ electrons in the QD, and
	\begin{equation}
		F(E_p|N) \equiv \frac{P_{\text{eq}}(E_p \cap N)}{P_{\text{eq}}(N)} =  \frac{\sum\limits_{\{n_i\}} P_{\text{eq}}(\{n_i\}) \delta_{n_p,1} \delta_{\sum n_i,N}}{P_{\text{eq}}(N)}
		\label{eq:fep_def}
	\end{equation} 
	is the equilibrium conditional probability of finding level $p$ occupied, when $N$ electrons are in the QD.
	Note that expression (\ref{eq:lin_jc}) coincides with the one that can be derived using the DBEs. However, the above derivation which uses the LBEs
shows that Eq.~(\ref{eq:lin_jc}) 
is always valid within the linear response regime. 

	Unfortunately, we were able to derive a closed-form expression for the energy current using Eq.~(\ref{LBErel}) only in the absence of interaction ($E_C=0$).
	For $E_C\ne 0$, the energy current $J_\alpha^u$ can be derived using the relation
	\begin{equation}
	\sum_{\alpha}\Gamma_\alpha(p) 
	\left\{ \psi(\{n_i\},n_p=0) - \psi(\{n_i\},n_p=1) 
	+\frac{1}{k_BT}\left[ \left(E_p+(2\widetilde{N}+1)E_C-\mu \right) \frac{\Delta T_\alpha}{T} +\Delta\mu_\alpha 
	\right]  \right\}=0,
	\label{DBErel}
	\end{equation}
	obtained by linearizing the DBEs, Eq.~(\ref{eq:det_bal})
	(which is equivalent to removing the sum over $\{n_i\}$ from Eq.~(\ref{LBErel})).
	Thus, in the domain of validity of the DBEs, the heat current can be written as
	\begin{multline}
	\label{eq:lin_jh}
		J^h_\alpha = \frac{1}{k_BT}\sum\limits_{p=1}^\infty \sum\limits_{N=1}^{\infty} P_{\text{eq}}(N) F_{\text{eq}}(E_p|N) \left[ 1-f(\epsilon(N,p)) \right] \left[\epsilon(N,p) - \mu \right]  \\ \times
		\sum_{\beta}\frac{\Gamma_\alpha(p)\Gamma_\beta(p)}{\Gamma_\text{tot}(p)}
		\left[(\epsilon(N,p)-\mu)\frac{\Delta T_\alpha - \Delta T_\beta}{T} + (\Delta\mu_\alpha-\Delta\mu_\beta) \right].
	\end{multline}
	\end{widetext}

	We can now define the transport coefficients, namely the electrical conductance $G_{\alpha\beta}$, the thermopower $S_{\alpha\beta}$ and the thermal conductance $K_{\alpha\beta}$ for the multi-terminal case, as \cite{Mazza2014}
	 \begin{equation}
		G_{\alpha\beta} = \Big(\frac{e J_\alpha^c}{\Delta \mu_\beta} \Big)_{\mbox{\tiny{$\begin{array}{l}
		\Delta T_\gamma = 0 \; \forall \gamma, \\
		\Delta \mu_{\gamma} = 0\; \forall \gamma\neq \beta
		\end{array}$}} },
		\label{eq:g_def}
	\end{equation}
	\begin{equation}
		S_{\alpha\beta} = -\Big(\frac{\Delta \mu_\alpha}{e\Delta T_\beta} \Big)_{\mbox{\tiny{$\begin{array}{l}
		J^c_\gamma = 0 \; \forall \gamma, \\
		\Delta T_{\gamma} = 0\; \forall \gamma\neq \beta 
		\end{array}$}} },
		\label{eq:s_def}
	\end{equation}
	and
	\begin{equation}
		K_{\alpha\beta} = \Big(\frac{J^h_\alpha}{\Delta T_\beta} 
	\Big)_{\mbox{\tiny{$\begin{array}{l}
		J^c_\gamma = 0 \; \forall \gamma, \\
		\Delta T_{\gamma} = 0\; \forall \gamma\neq \beta
		\end{array}$}} }.
		\label{eq:k_def}
	\end{equation}
	Note that index $\beta$ takes values in the range $2,...,{\cal N}$, since reservoir 1
	is chosen as the reference. 
	Local and non-local transport coefficients are distinguished depending on whether the two indices are, respectively, equal or different.

	The expressions for the currents [(\ref{eq:lin_jc}) and (\ref{eq:lin_jh})] have an intuitive interpretation.
	Indeed, the currents depend on the probability that a given energy level of the QD is occupied [$P_{\text{eq}}(N) F_{\text{eq}}(E_p|N)$] times the probability that there is an empty state with the correct energy in the reservoir 
	[$1-f(\epsilon(N,p))$]. 
	The sum over all energy levels $p$ and over the total number of electrons $N$ in the QD accounts for all the various tunneling processes that can occur. 
	Moreover, as far as the energy current is concerned, 
	$\epsilon(N,p)$ 
	is the energy
	carried by an electron that leaves the QD from level $p$ when $N$ electrons are present before the tunneling process, or equivalently, 
	$\epsilon(N,p)$ 
	is the energy carried by an electron that enters the QD into level $p$ increasing the number of total electrons to $N$.
	We recall that Eq.~(\ref{eq:lin_jc}) for the charge current is always valid, while Eq.~(\ref{eq:lin_jh}) holds only 
	when the DBEs are valid (see App.~\ref{app:DBE2}).
	The expressions for the charge and heat currents, in the case of two terminals, coincide with the ones obtained in Refs.~\onlinecite{bib:beenakker1,bib:beenakker2,bib:zianni_2007}.

	If we assume that the tunneling rates do not depend on $p$, i.e. $\Gamma_\alpha(p) = \Gamma_\alpha$, we can rewrite the charge and heat currents in Eqs.~(\ref{eq:lin_jc}) and (\ref{eq:lin_jh}) as follows:
	\begin{multline}
			J^c_\alpha = \frac{e}{k_BT} \sum\limits_{\beta} \frac{\Gamma_\alpha\Gamma_\beta}{\Gamma_{\text{tot}}} \\
			\times {\cal P} \left[ (\epsilon - \mu)\frac{\Delta T_\alpha - \Delta T_\beta}{T} + \Delta \mu_\alpha - \Delta \mu_\beta\right] ,
	\label{eq:jc_lin_o}
	\end{multline}
	and
	\begin{multline}
			J^h_\alpha = \frac{1}{k_BT} \sum\limits_{\beta} \frac{\Gamma_\alpha\Gamma_\beta}{\Gamma_{\text{tot}}} \\
			\times {\cal P} \bigg[ (\epsilon-\mu)\left( (\epsilon-\mu)\frac{\Delta T_\alpha - \Delta T_\beta}{T} + \Delta \mu_\alpha - \Delta \mu_\beta\right)\bigg] ,
	\label{eq:ju_lin_o}
	\end{multline}
	where 
	${\cal P}$ is the linear functional
	\begin{equation}
		{\cal P}[x] \equiv  \sum\limits_{p=1}^\infty \sum\limits_{N=1}^{\infty} P_{\text{tot}}(N,p) x(N,p),
		\label{eq:o_def}
	\end{equation}
	with
	\begin{equation}	
		P_{\text{tot}}(N,p) = P_{\text{eq}}(N)F_{\text{eq}}(E_p|N) \left[ 1-f(\epsilon(N,p))  \right].
\label{eq:Ptot}
	\end{equation}
	For a two terminal system it is possible to define analogous equations that do not require the tunneling rates to be energy-independent \cite{paolo2016}, but we will not consider this case.
	We can thus use the definitions of the transport coefficients given in Eqs.~(\ref{eq:g_def}), (\ref{eq:s_def}) and (\ref{eq:k_def}) to write them in terms of the functional ${\cal P}$ as
	\begin{equation}
	\begin{aligned}
		&G_{\alpha\beta} = \frac{e^2}{k_BT}\left( \delta_{\alpha\beta}\Gamma_\alpha - \frac{\Gamma_\alpha\Gamma_\beta}{\Gamma_{\text{tot}}} \right) {\cal P}\left[1\right], \\
		&S_{\alpha\beta} = \frac{1}{eT} \delta_{\alpha\beta} \frac{{\cal P}[\epsilon-\mu]}{{\cal P}\left[1\right]}, \\
		&K_{\alpha\beta} = \frac{1}{k_BT^2}\left(\delta_{\alpha\beta}\Gamma_\alpha  - \frac{\Gamma_\alpha\Gamma_\beta}{\Gamma_{\text{tot}}}\right)\\
		&\times \left( {\cal P}[(\epsilon-\mu)^2] - \frac{{\cal P}^2[\epsilon-\mu]}{{\cal P}\left[1\right]} \right) .
	\end{aligned}
	\label{eq:transport_coeff_o}
	\end{equation}
	These expressions only require the calculation of ${\cal P}[(\epsilon-\mu)^k]$, with 
	$k=0,1,2$, and make manifest various properties of the transport coefficients.
	Namely, i) all three transport coefficients are symmetric matrices (as required by the Onsager relations in the presence of time-reversal symmetry); ii) $G_{\alpha\beta}$ and $K_{\alpha\beta}$ have non-local terms, while the thermopower is only local (non-zero non-local $S_{\alpha\beta}$ occur when relaxing the assumption for which the tunneling rates do not depend on the energy levels \cite{Mazza2014}); iii) $\sum\nolimits_\alpha G_{\alpha\beta} = \sum\nolimits_\alpha K_{\alpha\beta} = 0$, stemming from charge and energy 
	conservation. 

	By defining the generating function
	\begin{equation}
		\Omega[\lambda] \equiv \ln{{\cal P}\left[ e^{\lambda(\epsilon-\mu)} \right]} ,
		\label{eq:omega_def}
	\end{equation}
	we can write the transport coefficients as follows:
	\begin{equation}
	\begin{aligned}
		&G_{\alpha\beta} = \frac{e^2}{k_BT}\left( \delta_{\alpha\beta}\Gamma_\alpha - \frac{\Gamma_\alpha\Gamma_\beta}{\Gamma_{\text{tot}}} \right) e^{\Omega\left[0\right]}, \\
		&S_{\alpha\beta} = \frac{1}{eT} \delta_{\alpha\beta} \at{ \frac{\partial \Omega}{\partial \lambda} }{\lambda=0}, \\
		&K_{\alpha\beta} = \frac{1}{k_BT^2}\left( \delta_{\alpha\beta}\Gamma_\alpha - \frac{\Gamma_\alpha\Gamma_\beta}{\Gamma_{\text{tot}}} \right) e^{\Omega\left[0\right]} \at{\frac{\partial^2\Omega}{\partial\lambda^2}}{\lambda=0}.
	\end{aligned}
	\label{eq:transport_coeff_omega}
	\end{equation}
	In the next Subsection, we will compute an analytic expression for $\Omega[\lambda]$ in the quantum limit.

	\subsection{Quantum limit}
	\label{qr}
\begin{figure}[!htb]
		\centering	
		\includegraphics[width=1\columnwidth]{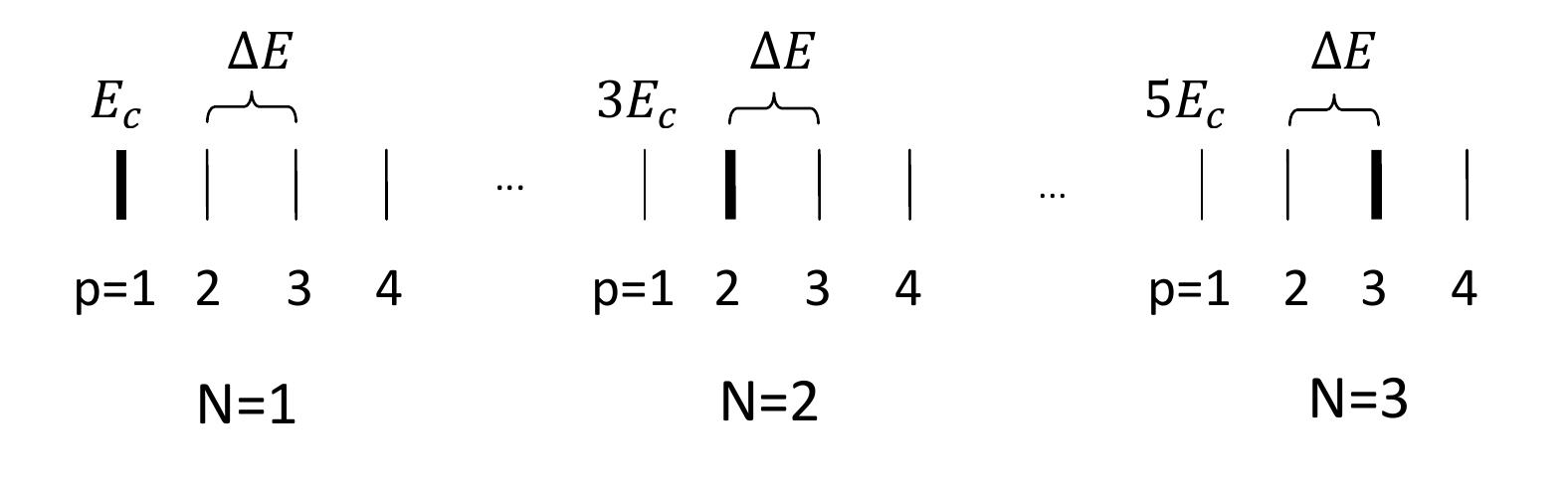}
		\caption{Schematic representation of the transition energies $\epsilon(N,p)$ 
	as $N$ and $p$ vary. In this figure we are assuming $E_C \gg \Delta E$ and 
	equidistant energy levels, $E_p-E_{p-1}=\Delta E$. 
	The bold lines represent the dominant transition energies $\tilde{\epsilon}(N)=
	\epsilon(N,p=N)$.}
		\label{fig:etp_levels}
	\end{figure}

	The quantum limit is characterized by having the energy spacing between levels of the QD and the charging energy much bigger than $k_BT$ [while $k_BT\gg \hbar \Gamma_\alpha (p)$].
	We start by observing that the sum over $p$ and $N$ in Eq.~(\ref{eq:o_def}) accounts for the fact that electrons can enter or leave the QD with energy $\epsilon(N,p)$ through, in principle, any energy level $E_p$ with the QD being occupied by any number of electrons $N$. The transition energies $\epsilon(N,p)$ 
	are schematically shown in Fig.~\ref{fig:etp_levels}.
	At low temperatures we expect the lowest energy levels of the QD to be occupied, so that, if there are initially $N-1$ electrons in the QD, electrons can flow mainly through level $p=N$.
	Such process gives the dominant contribution to transport and is represented by the dominant transition energy $\tilde{\epsilon}(N)\equiv\epsilon(N,p=N)$ (depicted in bold in Fig.~\ref{fig:etp_levels}).
	Therefore, in the quantum limit one expects to get the sum over $N$ and $p$ appearing in ${\cal P}$ [Eq.~(\ref{eq:o_def})] reduced to few dominant terms [the three equilibrium probabilities $P_{\text{eq}}(N)$, $F_{\text{eq}}(E_p|N)$ and $f(\epsilon(N,p))$ becoming very sharp functions].

Following Ref.~\onlinecite{bib:beenakker2},
	one finds (see Ref.~\onlinecite{paolo2016} for details) that the dominant contribution to $P_{\text{tot}}$ in Eq.~(\ref{eq:o_def}) occurs when $N=N_{\text{min}}$ is the integer that minimizes the quantity
	\begin{equation}
		|\tilde{\epsilon}(N) - \mu| ,
	\label{eq:n_min}
	\end{equation}
	and for values of $p$ such that $\epsilon(N_{\text{min}},p)$ is between the electrochemical potential $\mu$ and the dominant transition energy, i.e. 
	such that $\tilde{\epsilon}(N_{\text{min}}) 
	\leq \epsilon(N_{\text{min}},p) < \mu$ or 
	$\mu < \epsilon(N_{\text{min}},p) \leq 
	\tilde{\epsilon}(N_{\text{min}})$.
	In the former case, $p=N_{\text{min}},N_{\text{min}}+1,N_{\text{min}}+2,...,\bar{p}$, 
	where $\bar{p}$ is the largest integer such that $\epsilon(N_\text{min},p)< \mu$. 
	In the latter case, $p=N_{\text{min}},N_{\text{min}}-1,N_{\text{min}}-2,...,\bar{p}$ 
	where $\bar{p}$ is the smallest integer such that $\epsilon(N_\text{min},p)> \mu$.
	We then find that Eq.~(\ref{eq:o_def}) becomes
	\begin{equation}
		{\cal P}[x] = \frac{1}{4\cosh^2{\left(\dfrac{\Delta_{\text{min}}}{2k_BT}\right)}} \sum\limits_{p=N_{\text{min}}}^{\bar{p}} x(N_{\text{min}},p),
		\label{O}
	\end{equation}
	where we have defined $\Delta_{\text{min}}\equiv \tilde{\epsilon}(N_{\text{min}})-\mu$.
	Eq.~(\ref{O}) only keeps the dominant terms in the low-temperature limit.

	This approximation must be improved when $\bar{p}=N_{\text{min}}$, that is,
	the sum in ${\cal P}$ in Eq.~(\ref{O}) reduces to the single term $p=N_{\text{min}}$
	and transport is provided by the dominant transition energy
	$\tilde{\epsilon}(N_{\text{min}})$ only.
	If in this case one imposes that $J^c_\alpha = 0$, then  
	one obtains $J^h_\alpha = 0$, since $J^h_\alpha \propto J^c_\alpha$.
	As a consequence one gets $K_{\alpha\beta}=0$ \footnote{This 
	approximation for two terminals leads to zero thermal conductance 
	and to the Carnot efficiency, as discussed for single-level quantum dots
	\cite{bib:mahan,bib:linke,bib:humphrey}.}.  
	Thus, when $\bar{p}=N_{\text{min}}$, we improve our approximation 
	of ${\cal P}$ by extending the sum over 
	$p$ to the two nearest integers, $p = N_{\text{min}} \pm 1$. We have numerically verified that this approximation is valid when $2E_C > \Delta E$.

	In order to obtain analytical expressions for the transport coefficients, 
	hereafter we focus on the case of equidistant levels, 
	$E_p-E_{p-1}=\Delta E$. After introducing the parameter
	\begin{equation}
		\xi \equiv \frac{4\cosh^{2}\left({\Delta_{\text{min}}}/{2k_B T}\right)}{e^{\Delta E/k_BT}}
	\end{equation}
	and defining the integer $N_J\equiv \bar{p}-N_{\text{min}}$, we obtain  
	\begin{widetext}
	\begin{equation}
	\begin{aligned}
	\Omega[ \lambda ] &
	= -\ln\left[4\cosh^2\left(
	\frac{\Delta_{\text{min}}}{2k_B T}\right)\right] 
	+ \lambda\Delta_{\text{min}}
	\\
	& 
	+\ln\left[\frac{e^{\lambda\Delta E\left( |N_J| +1 \right)\sign(N_J)}-1}{e^{\sign(N_J)
	\lambda\Delta E}-1}
	+ \xi \delta_{N_J,0} \left( \cosh{(\lambda \Delta E)} -\tanh\left(\frac{\Delta_{\text{min}}}{2k_BT}\right) \sinh{(\lambda \Delta E)} \right)\right].
	\end{aligned}
		\label{eq:omega_ql}
	\end{equation}
	Using Eqs.~(\ref{eq:transport_coeff_omega}) and (\ref{eq:omega_ql}), 
	we finally obtain the multiterminal transport coefficients
	\begin{equation}
	\begin{aligned}
		&G_{\alpha\beta} = \left( \delta_{\alpha\beta}\Gamma_\alpha - \frac{\Gamma_\alpha\Gamma_\beta}{\Gamma_{\text{tot}}} \right)\frac{e^2}{4k_BT\cosh^2(\frac{\Delta_{\text{min}}}{2k_BT})}\left(1+|N_J|\right), \\
		&S_{\alpha\beta} =  \delta_{\alpha\beta} \frac{1}{eT}\left(\Delta_{\text{min}} +\frac{\Delta E}{2}N_J \right), \\
		&K_{\alpha\beta} =\left( \delta_{\alpha\beta}\Gamma_\alpha - \frac{\Gamma_\alpha\Gamma_\beta}{\Gamma_{\text{tot}}} \right) k_B\left(\frac{\Delta E}{k_BT}\right)^2
		\begin{cases}
		\hfill \frac{1}{12}e^{-|\Delta_{\text{min}}|/k_BT}|N_J|\left(|N_J|+1\right)\left(|N_J|+2\right) \hfill &\text{if $N_J \neq 0$}, \\

	\hfill \dfrac{1}{e^{\Delta E/k_BT}+4\cosh^2(\Delta_{\text{min}}/2k_BT)}
	\hfill &\text{if $N_J = 0$.}
		\end{cases}
	\end{aligned}
	\label{eq:GSKmatrices}
	\end{equation}
	\end{widetext}
	We have computed $G_{\alpha\beta}$ and $S_{\alpha\beta}$ setting $\xi=0$, since the term proportional to $\xi$ in Eq.~(\ref{eq:omega_ql}) only yields minor corrections that make these quantities more ``smooth'' as a function of $\mu$; the calculation of $K_{\alpha\beta}$ instead requires a non null value of $\xi$.
	Eqs.~(\ref{eq:GSKmatrices}) exhibit a number of interesting features. First of all,  
	$G_{\alpha\beta}$ shows peaks as a function of the electrochemical potential $\mu$   
	every time $\Delta_{\text{min}}=0$, namely when $\mu = \tilde{\epsilon}(N)$.
	For example, the $N$-th peak corresponds to $\mu$ equal to the $N$-th dominant transition energy,
\begin{equation}
\mu=\mu_N\equiv (N-1) \Delta E+(2N-1)E_C.
\label{eq:muN}
\end{equation}
  We set $E_p = (p-1)\Delta E$, for $p=1,2,\dots$,
	and therefore the separation between two nearby peaks 
	is given by $\Delta E+2E_C$. 
	Due to the factor $\cosh^{-2}(\frac{\Delta_{\text{min}}}{2k_BT})$
	in $G_{\alpha\beta}$,
	these peaks have a bell shape with amplitude of the order of $k_BT$. 
	On the other hand, the thermal conductance $K_{\alpha\beta}$ has plateaus of width 
	$2\Delta E$ around $\mu_N$ \cite{bib:zianni_2007}, 
corresponding to the second line 
	of the expression of $K_{\alpha\beta}$ for $N_J=0$. 
	For $N_J\ne 0$, 
	the thermal conductance $K_{\alpha\beta}$ is then exponentially suppressed 
	due to the term $e^{-|\Delta_{\text{min}}|/k_BT}$.
	The local thermopower $S_{\alpha\alpha}$ vanishes 
	at the values $\mu_N$ where the electrical and thermal conductances 
	$G_{\alpha\alpha}$ and $K_{\alpha\alpha}$ exhibit a maximum. 
	$S_{\alpha\alpha}$ has a linear 
	dependence on $\mu$ with slope $dS_{\alpha\alpha}/d\mu=-1/eT$,
	with jumps when either $N_{\rm min}$ or $N_J$ change by one. 
	Therefore we have, for $E_C\gg \Delta E$, main oscillations 
	of period  $\Delta E+2E_C$, and a fine structure with spacing
	$\Delta E$ \cite{bib:beenakker2}. 
	We note that the fine structure is present also for 
	$G_{\alpha\beta}$ and $K_{\alpha\beta}$, but in these cases 
        the amplitude of the fine structure oscillations is
	exponentially small.

	\subsection{Two-terminal system}
	\label{2terms}

	In the two-terminal case, the matrices
	$G_{\alpha\beta}$, $S_{\alpha\beta}$, and $K_{\alpha\beta}$ 
	reduce to the familiar transport coefficients, namely the electrical 
	conductance $G=G_{22}$, the thermopower $S=S_{22}$, and 
	the thermal conductance $K=K_{22}$.   
	From Eqs.~(\ref{eq:GSKmatrices}) we recover the formulas
	for $G$ and $S$ well-known in literature~\cite{bib:beenakker2},
	while our expression for $K$ coincides for $N_J=0$ 
	with the result of Ref.~\onlinecite{bib:zianni_2007}, but also provides
	the fine structure oscillations for $N_J\ne 0$. 
	Although such oscillations are not appreciable in $K$ as a function of $\mu$, we will see below that they give rise to a visible fine structure in $ZT$.

	As we have shown in Sec.~\ref{chap:eff_pow_def}, within the linear response regime the relevant quantities to characterize
	the performance of a thermoelectric device are the power factor
	$Q$ and the figure of merit $ZT$.
	From the expressions of the transport coefficients in Eqs.~(\ref{eq:GSKmatrices}), 
	specified for the two-terminal case, 
	one can compute $Q$ and $ZT$ analytically within the quantum limit.
	The obtained expressions are given below, and compared with a 
	numerical calculation performed using the kinetic equations. 

\subsubsection{Power factor}

	Let us start by studying the power factor $Q$. 
	Within the quantum limit, we find that
	\begin{equation}
	Q = \frac{\gamma  \left( 1+|N_J| \right) }{4k_BT^3\cosh^2{(\Delta_{\text{min}}/2k_BT)}} 
	\left( \Delta_{\text{min}}
	+ \frac{N_J}{2}\Delta E \right)^2,
	\label{eq:p_max_ql}
	\end{equation}
	where we have defined $\gamma \equiv \Gamma_1\Gamma_2/(\Gamma_1+\Gamma_2)$.
	\begin{figure}[!tb]
		\centering
		\includegraphics[width = 1.0 \columnwidth]{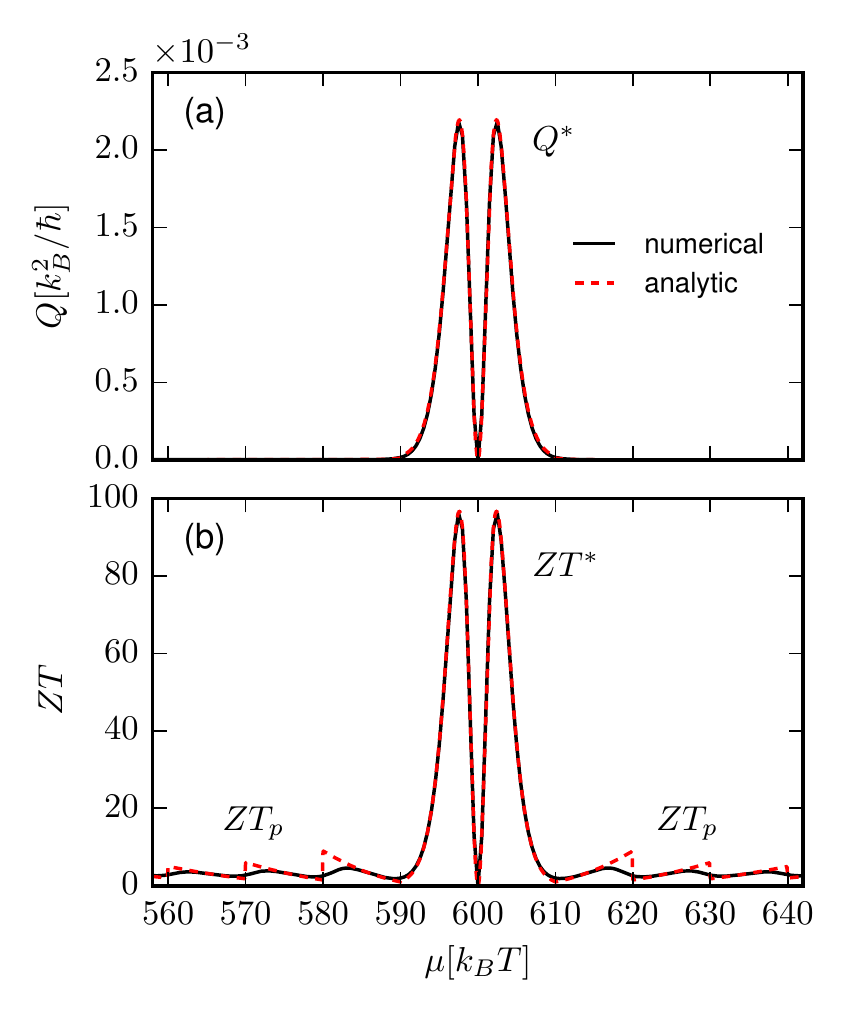}
		\caption{(Color online) Power factor $Q$ (a) and figure of merit $ZT$ (b) are plotted as a function of the electrochemical potential $\mu$.
		For both quantities the analytical quantum limit [given by Eqs.~(\ref{eq:p_max_ql}) and (\ref{eq:zt_ql})] is plotted as a red dashed curve, while the numerically calculated result is plotted as a black solid curve.
		All curves are calculated at $E_C = 50\,k_BT$, $\Delta E=10\,k_BT$, and $\hbar\Gamma_1(p) = \hbar\Gamma_2(p) = (1/100)\,k_BT$.} 
		\label{fig:p_max_ql}
	\end{figure}
	As for $G$, the power factor $Q$ is dominated by a fast decrease, 
	given by the term $\cosh^{-2}{(\Delta_{\text{min}}/2k_BT)}$, thus becoming vanishingly 
	small within a few $k_BT$ around $\mu = \mu_N$ [see Fig.~\ref{fig:p_max_ql}(a)].
	In fact the fine structure, given by the terms with $N_J$, is not visible in Fig.~\ref{fig:p_max_ql}(a) due to the 
	rapid suppression given by the 
$\cosh^{-2}{(\Delta_{\text{min}}/2k_BT)}$ term in Eq.~(\ref{eq:p_max_ql}). 
	Differently from $G$, the power factor vanishes at $\mu = \mu_N$,
	due to the fact that in this point the thermopower $S=0$. 
	So as $\mu$ moves away from $\mu_N$, 
	$Q$ increases quadratically due to the linear growth of the thermopower with $\mu$, 
	and then it rapidly decreases within a few $k_BT$ due to 
the $\cosh^{-2}(\Delta_{\text{min}}/2k_BT)$ term. 
	Hence there are two symmetric peaks around 
	$\mu = \mu_N$, within a few $k_BT$. 
	These double peaks are the dominant feature of Fig.~\ref{fig:p_max_ql}(a)
	and identify the optimal values of $\Delta_{\text{min}}$ (and consequently of 
$\mu$) to obtain the absolute maximum power $P_{\text{peak}}$,
namely when the power factor $Q$ is maximum, $Q=Q^*$.
From Eq.~(\ref{eq:p_max_ql}), we obtain that $Q$ is maximum 
for values $\Delta_{\text{min}}^*$ of $\Delta_{\text{min}}$ such that 
		\begin{equation}
			\frac{\Delta_{\text{min}}^*}{2k_BT} = 
	\coth\left(\frac{\Delta_{\text{min}}^*}{2k_BT}\right).
		\end{equation}
	The numerical solution is $\Delta_{\text{min}}^* \simeq \pm 2.40k_BT$, 
	which corresponds to $\mu = \mu_N \pm 2.40k_BT$. 
	This result does not depend on any energy scale of the system except for $k_BT$, and coincides with the non-interacting single-level case (see App.~\ref{oneEL}). 
	The value $Q^*$ of $Q$ in these points is
	\begin{equation}
   Q^* \simeq 0.44 \frac{\gamma k_B}{T},
   \label{eq:q_ql}
  \end{equation}
	so that the peaks of the power factor only depend on $\gamma$ and on the 
	reference temperature. 
	In conclusion, if we want to extract maximum power from this system, 
	we must choose $\mu = \mu_N \pm 2.40k_BT$.
	We will now show that also $ZT$ reaches a maximum at these 
	same values of the electrochemical potential confirming that
	these are the optimal values for heat to work conversion in the quantum limit
	linear response regime.

\subsubsection{Figure of merit}

	Let us now study the figure of merit $ZT$ in the quantum limit. 
	To obtain a more manageable
	analytical expression, we compute $K$ from the function 
	$\Omega[\lambda]$ expanded to the first order in $\xi$
	[this corresponds to approximating $K$ with a constant plateau 
	when $N_J=0$, namely 
	$4\cosh^2(\Delta_{\text{min}}/2k_BT)$ is neglected with respect to
	$e^{\Delta E/k_BT}$ in the last line of Eqs.~(\ref{eq:GSKmatrices})].
	We then obtain
	\begin{equation}
		ZT = 
		\begin{dcases}
			\frac{1}{4}\left(\frac{\Delta_{\text{min}}}{\Delta E} \right)^2 
	\frac{e^{\Delta E/k_BT}}{\cosh^2\left(\frac{\Delta_{\text{min}}}{2k_BT}\right)} \hfill &\text{if $N_J=0$}, \\
			\frac{3|N_J|}{2+|N_J|}\left( 1- 2\frac{|\Delta_{\text{min}}|}{\Delta E|N_J|} \right)^2 \hfill &\text{if $N_J \neq 0$,}
		\end{dcases}
		\label{eq:zt_ql}
	\end{equation}
	which implies that the behavior of $ZT$ is different for the two cases $|\mu - \mu_N| < \Delta E$ ($N_J=0$)
	and $|\mu - \mu_N| > \Delta E$ ($N_J\neq 0$). 
	In the former case $K$ exhibits a plateau, so that $ZT$ is 
	directly proportional to $Q$ and therefore it has the same double peak structure 
	at $\mu = \mu_N \pm 2.40k_BT$.
	This is clearly shown in Fig.~\ref{fig:p_max_ql}(b) where $ZT$ is plotted as a function of $\mu$. 

	The value of $ZT$ in these points is
		\begin{equation}
			ZT^* \approx 0.44\frac{e^{\Delta E/k_BT}}{\left(\Delta E / k_BT\right)^2}.
			\label{eq:zt_star}
		\end{equation}
This results has been obtained also in Ref. \onlinecite{Tsaousidou2010}. Eq.~(\ref{eq:zt_star}) shows that in the limit $\Delta E/ k_BT \rightarrow \infty$, 
	we have that $ZT \rightarrow \infty$. 
	For example, for $\Delta E = 6 k_BT$, we reach $ZT^* \approx 5$; 
	for $\Delta E = 10 k_BT$, we reach $ZT^* \approx 97$, and so on. 
	This is consistent with Mahan and Sofo's observation \cite{bib:mahan} that a narrow transmission function yields $ZT \rightarrow \infty$. Furthermore, these peaks in $ZT$ correspond to peaks in $Q$, so in these points we can maximize $P_{\text{max}}$ and $\eta(P_{\text{max}})$ simultaneously.  
	Instead, when $N_J \neq 0$, $ZT$ has a discontinuity every time $\mu = \epsilon(N,p)$ with $p \neq N$, which means with a $\Delta E$ spacing.
	This fine structure is the origin of the saw-tooth oscillations of 
	Fig.~\ref{fig:p_max_ql}(b). The value of $ZT$ in each $\mu = \epsilon(N,p)$ is given by
	\begin{equation}
		ZT_p = 
		\begin{dcases}
			3\frac{|N-p| + 1}{|N-p| -1} \hfill &\text{if $|N-p| \geq 2$, } \\
			1 \hfill &\text{if $|N-p| = 1$. }
		\end{dcases}
	\end{equation}
	The height of these peaks, as opposed to $ZT^*$, has no dependance on the parameters of the system.
	The highest peak is obtained for $|N-p| = 2$, where
	$	ZT_{p = N\pm 2} = 9.$
	For values of $p$ distant from $N$, the height of the peak decreases 
	to an asymptotic value of $ZT_{\infty} = 3$.

\subsection{Comparison with a non-interacting QD}
\label{sec:comparisonintnonint}

Here we compare the efficiency of an interacting QD (with $2E_C > \Delta E$) with the efficiency of a non-interacting QD ($E_C=0$) that has the same energy spacing $\Delta E$; the comparison is performed within the linear response quantum limit for a two-terminal setup. The generating function in Eq.~(\ref{eq:omega_ql}) cannot be used in the case $E_C=0$ since it requires $2E_C >\Delta E \gg k_BT$. The generating function for $E_C=0$ will be denoted as $\Omega_{\text{NI}}[\lambda]$ (where NI stands for ``non interacting'') and calculated as follows.
As we can see from Eq.~(\ref{eq:omega_def}), we must compute ${\cal P}[\exp{\{\lambda(\epsilon(N,p)-\mu)\}}]$ using the definition of ${\cal P}$ given in Eq.~(\ref{eq:o_def}). When $E_C=0$, $\epsilon(N,p) = E_p$, so the transition energies correspond to the energy levels of the QD, and they do not depend on the number of electrons in the QD. As a consequence, there is no dependance on $N$ in the argument of ${\cal P}$, so we can explicitly perform the sum over $N$ in Eq.~(\ref{eq:o_def}) yielding the following expression: 
\begin{equation}
		{\cal P}_{\text{NI}}[x(p)] \equiv \sum\limits_{p=1}^\infty \frac{1}{4\cosh^2\left[(E_p-\mu)/2k_BT\right]} x(p).
		\label{eq:o_ni_ql}
\end{equation}
Each term in the series in Eq.~(\ref{eq:o_ni_ql}), as a function of $\mu$, is a bell-shaped function centered around $E_p$ of width set by $k_BT$. Therefore, within the quantum limit we can restrict the sum over $p$ to the three bell-shaped functions closest to $\mu$, namely $p = N_{\text{min}},N_{\text{min}}\pm 1$ ($N_{\text{min}}$, defined as in Eq.~(\ref{eq:n_min}), is such that $E_{N_{\text{min}}}$ is the closest energy level to $\mu$). This approximation allows us to find 
\begin{multline}
	\Omega_{\text{NI}}[ \lambda ] 
	= -\ln\left[4\cosh^2\left(
	\frac{\Delta_{\text{min}}}{2k_B T}\right)\right] 
	+ \lambda\Delta_{\text{min}} \\
	+\ln\left[1+2 \xi \cosh{\left(\lambda \Delta E- \frac{\Delta_{\text{min}}}{k_BT}\right)} \right],
		\label{eq:omega_ql_ni}
\end{multline}
where $\Delta_{\text{min}}$ in the non-interacting case reduces to the distance between $\mu$ and the nearest energy level: 
$\Delta_{\text{min}}=E_{N_{\text{min}}} - \mu$. 

\begin{figure}[!tb]
	\centering
	\includegraphics[width=1\columnwidth]{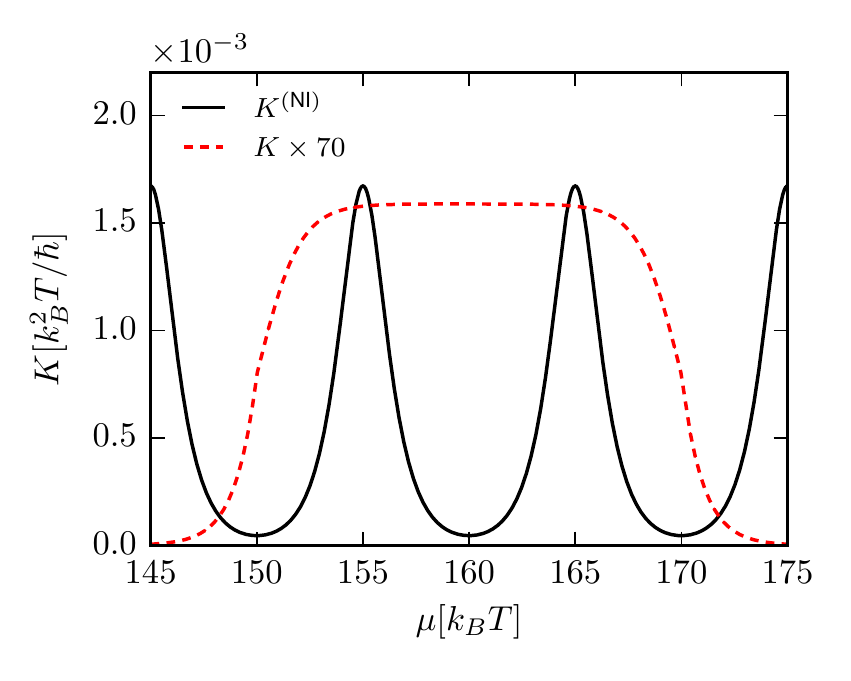}
	\caption{(Color online) Comparison between $K^{(\text{NI})}= K_{22}^{(\text{NI})}$ and $K=K_{22}$, given respectively by Eqs.~(\ref{eq:k_ql_ni}) and (\ref{eq:GSKmatrices}), plotted as a function of $\mu$. Both cases have been computed with the parameters used in Fig.~\ref{fig:p_max_ql}, except for setting $E_C=0$ in the non-interacting case. The interacting thermal conductance has been multiplied by a factor $70$. In particular, its maximum value is half the minimum of $K^{(\text{NI})}$.}
	\label{fig:k_int_nonint}
\end{figure}

Let us now compare Eq.~(\ref{eq:omega_ql_ni}) with its interacting counterpart, Eq.~(\ref{eq:omega_ql}), setting $N_J=0$, thus neglecting the fine structure. We notice that the two expressions are identical when $\xi =0$; this implies that $G_{\alpha\beta}$ and $S_{\alpha\beta}$ are equal in the interacting and non-interacting case [in Eqs.~(\ref{eq:GSKmatrices}) $G_{\alpha\beta}$ and $S_{\alpha\beta}$ are calculated at $\xi=0$], while $K_{\alpha\beta}$ is different in the two cases, since it is determined by the term proportional to $\xi$ in Eqs.~(\ref{eq:omega_ql}) and (\ref{eq:omega_ql_ni}). We find that
\begin{multline}
	K_{\alpha\beta}^{(\text{NI})} = 2k_B \left( \delta_{\alpha\beta}\Gamma_\alpha - \frac{\Gamma_\alpha\Gamma_\beta}{\Gamma_{\text{tot}}} \right) \left(\frac{\Delta E}{k_BT}\right)^2 e^{-\Delta E/k_BT} \\ \times
	\frac{2\xi + \cosh{\left(\Delta_{\text{min}}/k_BT\right)}}{1 + 2\xi\cosh{\left(\Delta_{\text{min}}/k_BT\right)}}.
	\label{eq:k_ql_ni}
\end{multline}
A comparison between the interacting and non-interacting thermal conductances is plotted for a two-terminal system in Fig.~\ref{fig:k_int_nonint}, using Eqs.~(\ref{eq:GSKmatrices}) and (\ref{eq:k_ql_ni}) at equal $\Delta E=10k_BT$.
In this figure the dominant transition energy of the interacting system is located,
see Eq.~(\ref{eq:muN}), at $\mu=\mu_{N=2}=160\,k_BT$, while the energy levels of the non-interacting system are chosen as $E_p = (p-1)10k_BT$ so that in both cases $G$ has a peak in $\mu=160\,k_BT$. We have verified that a numerical calculation is in very good agreement with Eq.~(\ref{eq:k_ql_ni}) using the parameters of Fig.~\ref{fig:k_int_nonint}. As we can see from Fig.~\ref{fig:k_int_nonint}, $K^{(\text{NI})}= K^{(\text{NI})}_{22}$ and $K=K_{22}$ are very different. While $K$ has a plateau of width $2\Delta E$ centered in $\mu_{N=2}$, $K^{(\text{NI})}$ has a minimum in $\mu_{N=2}$, and reaches a maximum value for $\mu$ between two energy levels. Furthermore, by comparing Eq.~(\ref{eq:k_ql_ni}) with the last line of Eqs.~(\ref{eq:GSKmatrices}), we see that $K^{(\text{NI})} \approx 2 \cosh{\left(\Delta_{\text{min}}/k_BT\right)}K$, so the minimum of $K^{(\text{NI})}$, occurring at $\Delta_{\text{min}}=0$, is twice the maximum of $K$, and $K^{(\text{NI})}$ increases exponentially respect to $K$ as $\Delta_{\text{min}}$ increases. 

Intuitively, the striking difference between the two models can be explained as follows. As discussed in App.~\ref{oneEL}, if we consider a single energy level QD the thermal conductance vanishes ($K=0$) since $K$ is computed at zero charge current and charge and heat currents are proportional in this case. However, $K$ can be finite when at least two energy levels are available, and gets bigger by increasing the flux of electrons tunneling at different energies. Now, Coulomb interaction produces a correlation between electrons tunneling at different energies. Namely, if one electron enters the QD the electrostatic energy increases by $2E_C$, preventing other electrons from entering the QD at any other energy level. Therefore, until that electron tunnels out of the QD, all other processes are blocked: this is a manifestation of Coulomb blockade. On the contrary, in the non-interacting case all tunneling events are independent. This correlation is thus responsible for suppressing simultaneous tunneling through different energy levels in the interacting case, which results in a suppression of $K$.

So in general $K$ is much smaller than $K^{(\text{NI})}$. As a consequence of these observations, we expect $ZT$ to be suppressed in the non-interacting case.

By setting $\xi=0$ in Eq.~(\ref{eq:k_ql_ni}), we find 
	\begin{equation}
		ZT_{\text{NI}} = \frac{1}{8}\left(\frac{\Delta_{\text{min}}}{\Delta E} \right)^2 \frac{e^{\Delta E/k_BT}}{\cosh^2\left(\frac{\Delta_{\text{min}}}{2k_BT}\right)\cosh\left(\frac{\Delta_{\text{min}}}{k_BT}\right)}.
		\label{eq:zt_ni}
	\end{equation}
Comparing Eq.~(\ref{eq:zt_ni}) with Eq.~(\ref{eq:zt_ql}), we see that, for $N_J=0$, $ZT = 2\cosh{\left( \Delta_{\text{min}}/k_BT \right)}ZT_{\text{NI}}$, so $ZT_{\text{NI}}$ is exponentially suppressed as $\Delta_{\text{min}}$ increases. Given this suppression, the maximum of $ZT_{\text{NI}}$ occurs at $\Delta_{\text{min}}^* \approx \pm 1.36k_BT$, corresponding to  
\begin{equation*}
	ZT_{\text{NI}}^* \approx \frac{1}{13.8} \frac{e^{\Delta E/k_BT}}{\left(\Delta E/k_BT\right)^2}.
\end{equation*}
This value is approximately 6 times smaller than $ZT^*$ for the interacting case, see Eq.~(\ref{eq:zt_star}). 

Furthermore, since $K$ is ``flat'' around the dominant transition energies in the interacting case, the peak power $P_{\text{peak}}$ and the figure of merit $ZT^*$ are reached at the same electrochemical potential, $\Delta_{\text{min}}^* \approx \pm 2.40k_BT$, so $ZT(P_{\text{peak}}) = ZT^*$. Instead in the non-interacting case these two quantities are not simultaneously maximized, due to the strong dependance of $K^{(\text{NI})}$ on $\mu$, so we have that
\begin{equation*}
	ZT_{\text{NI}}(P_{\text{peak}}) \approx \frac{1}{25.3} \frac{e^{\Delta E/k_BT}}{\left(\Delta E/k_BT\right)^2},
\end{equation*}
which is approximately 11 times smaller than $ZT(P_{\text{peak}})$, see Eq.~(\ref{eq:zt_star}). 

In conclusion, within the linear response quantum limit, an interacting QD (with $2E_C> \Delta E$) has a considerably higher $ZT$ with respect to a non-interacting QD both at peak efficiency, and at peak power, while having the same power factor ($G$ and $S$ being equal).

At last we will study how these two models violate the Wiedemann-Franz law, which states that for macroscopic ordinary metals the Lorenz ratio $L = K/GT$ is a constant equal to the Lorenz number $L_0 = (k_B/e)^2(\pi^2/3)$. Using Eqs.~(\ref{eq:GSKmatrices}) and (\ref{eq:k_ql_ni}), we find that
\begin{equation}
L = \frac{L_0}{\pi^2}\left(\frac{\Delta E}{k_BT}\right)^2
\begin{cases}
	\hfill |N_J|(|N_J|+2) \hfill &\text{if $N_J \neq 0$}, \\
	\hfill 12 e^{-\frac{\Delta E}{k_BT}}\cosh^2 \frac{\Delta_{\text{min}}}{2k_BT}   \hfill &\text{if $N_J = 0$,}
\end{cases}
 \label{eq:l_i}
\end{equation}
and
\begin{equation}
 L_{\text{NI}} =  L_0 \left(\frac{\Delta E}{k_BT}\right)^2 \frac{24}{\pi^2} e^{-\frac{\Delta E}{k_BT}}\cosh^2 \frac{\Delta_{\text{min}}}{2k_BT}\cosh\frac{\Delta_{\text{min}}}{k_BT},
 \label{eq:l_ni}
\end{equation}
where $L$ refers to the interacting case, and the expression of $K$, for simplicity, has been computed at $\xi=0$.
In both cases the Wiedemann-Franz law is strongly violated: at $\Delta_{\text{min}}=0$, the Lorenz ratio is exponentially smaller  than $L_0$ thanks to $(\Delta E/k_BT)^2\exp{(-\Delta E/k_BT)}$ (this has been noticed in Ref. \onlinecite{Tsaousidou2010} for the interacting model). In both cases the Lorenz ratio exponentially increases with $\Delta_{\text{min}}$. In the non-interacting model the exponent is $2\Delta_{\text{min}}/(k_BT)$ (twice the interacting case), and the maximum value, achieved at $\Delta_{\text{min}}=\Delta E/2$, is of the order of $L_{\text{NI}} \approx (\Delta E/k_BT)^2L_0$. Interestingly, in the interacting case [Eq.~(\ref{eq:l_i})], when $|\Delta_{\text{min}}|>\Delta E$, i.e. $N_J\neq 0$, we find plateaus whose height increases with $N_J$.

\subsection{Three-terminal system}
\label{3term}

In this Section we consider the case of a three-terminal system, which allows to study the non-local transport coefficients and the influence of an additional terminal on the thermoelectric performance of the system. We focus on the simplest case when the couplings to the reservoirs are energy independent, that is, the rates $\Gamma_\alpha$ do not depend on $p$, and we will consider an equidistant QD spectrum. Analytical expressions in the quantum limit for the power and the efficiency at maximum power can be obtained also in this case on the basis of the expressions written in  Ref.~\onlinecite{Mazza2014} and of Eqs.~(\ref{eq:GSKmatrices}) for the transport coefficients. All considerations made in this Section are valid for both interacting and non-interacting systems.

In a three-terminal setup with time reversal symmetry we have nine independent coefficients: three electrical conductances $G_{22}$, $G_{33}$, and $G_{23}$, three thermopowers $S_{22}$, $S_{33}$, and $S_{23}$, and three thermal conductances $K_{22}$, $K_{33}$, and $K_{23}$. According to the expressions in Eqs.~(\ref{eq:GSKmatrices}), valid in the quantum limit when the couplings to the leads are independent of energy, one finds that the local and non-local electrical (thermal) conductances are characterized by peaks (plateaus) located in the same positions as for the two-terminal case. The two local electrical (thermal) conductances $G_{22}$, $G_{33}$, ($K_{22}$, $K_{33}$), can have different heights if the coupling to reservoirs $2$ and $3$ are different ($\Gamma_2\ne\Gamma_3$), while the non-local conductances, $G_{23}$ and $K_{23}$, are negative. On the other hand, the local thermopowers $S_{22}$ and $S_{33}$ are equal and exhibit the same fine structure as in the two-terminal case. Moreover, the non-local thermopower $S_{23}$ vanishes, as a consequence of energy independent tunneling rates.

The power and efficiency of a three-terminal system are defined in Eqs.~(\ref{eq:p_def}) and (\ref{eq:eta_def}). For definiteness, let's consider $T_3 \geq T_2 \geq T_1$, so $\Delta T_3 \geq \Delta T_2 \geq 0$.  In general, Carnot's efficiency cannot be written only in terms of the temperatures $T_1$, $T_2$, and $T_3$, but it depends on the details of the system \cite{Mazza2014}. Nonetheless, if we fix the temperature of the hottest and coldest reservoir, that is, $T_3$ and $T_1$, it can be shown that $\eta_C \leq \eta_C^{(2)}$, where $\eta_C^{(2)} = 1 - T_1/T_3$ is the Carnot efficiency of the two-terminal system; the equal sign can be achieved when two reservoirs have the same temperatures. So adding a third terminal at an intermediate temperature cannot increase the maximum efficiency beyond the two terminal Carnot's efficiency.

Also in the three terminal case the efficiency at maximum power cannot go beyond the linear response Curzon-Ahlborn efficiency 
$\eta_{CA} = \eta_C^{(2)}/2$. Our aim is to maximize the power $P$ with respect to $\Delta \mu_2$ and $\Delta \mu_3$, at given temperature differences. Then we will consider a fixed value of $\Delta T_3$, and we will study the maximum power $P_{\text{max}}$ and the efficiency at maximum power $\eta(P_{\text{max}})$ varying $T_2$ between the fixed $T_1$ and $T_3$. These calculations can be performed by writing the currents in terms of the temperature and electrochemical potential differences through the Onsager matrix $L_{ij}$ \cite{Mazza2014}; in turn, $L_{ij}$ can be related to the transport coefficients. By rewriting Eqs.~(\ref{eq:transport_coeff_omega}) as
\begin{equation}
	\begin{aligned}
		&G_{\alpha\beta} = M_{\alpha\beta} \mathcal{G}, \\
		&S_{\alpha\beta} = \delta_{\alpha\beta} \mathcal{S}, \\
				&K_{\alpha\beta} = M_{\alpha\beta}\mathcal{K},
	\end{aligned}
	\label{eq:gsk_3t}
\end{equation}
	where
\begin{equation}
		M_{\alpha\beta} = \delta_{\alpha\beta}\Gamma_\alpha - \frac{\Gamma_\alpha\Gamma_\beta}{\Gamma_{\text{tot}}},
\end{equation}
and by defining 
\begin{align}
	&\mathcal{Z}T = \frac{\mathcal{G}\mathcal{S}^2T}{\mathcal{K}}, &\Gamma_{ij} = \Gamma_i + \Gamma_j,
\end{align}
we can write the currents as
\begin{equation}
	\begin{pmatrix}
		J^c_2/\Gamma_2 \\ J^c_3/\Gamma_3 
	\end{pmatrix} = 
	\frac{\mathcal{G}}{e\Gamma_{\text{tot}}}
	\begin{pmatrix}
		 \Gamma_{13} & \Gamma_{13} & -\Gamma_3 & -\Gamma_3 \\
		-\Gamma_2 & -\Gamma_2 & \Gamma_{12} & \Gamma_{12} 
	\end{pmatrix}
	\begin{pmatrix}
		\Delta \mu_2 \\ e\mathcal{S}\Delta T_2 \\ \Delta \mu_3 \\ e\mathcal{S}\Delta T_3
	\end{pmatrix}, 
\end{equation}
\begin{multline}
	\begin{pmatrix}
		J^h_2/\Gamma_2 \\ J^h_3/\Gamma_3
	\end{pmatrix} 
	= 
	\frac{\mathcal{K}}{\Gamma_{\text{tot}}}
	\begin{pmatrix}
		 \Gamma_{13} & \Gamma_{13} & -\Gamma_3 & -\Gamma_3 \\
		-\Gamma_2 & -\Gamma_2 & \Gamma_{12} & \Gamma_{12} 
	\end{pmatrix}
	\begin{pmatrix}
		0  \\ \Delta T_2 \\ 0 \\ \Delta T_3
	\end{pmatrix}
	\\ +
	\mathcal{S}T
	\begin{pmatrix}
		J^c_2/\Gamma_2 \\ J^c_3/\Gamma_3 
	\end{pmatrix}.
\end{multline}
Note that the quantities $\mathcal{G}$, $\mathcal{S}$, $\mathcal{K}$ and $\mathcal{Z}T$ only depend on the properties of the QD, which can be interacting or non-interacting, and on the reference electrochemical potential $\mu$; they do not depend on the number of reservoirs nor on the tunneling rates.

The electrochemical potential differences that maximize $P$ at given reservoir temperatures can be written as
\begin{equation}
	\Delta \mu_\alpha = -\frac{1}{2}e\mathcal{S}\Delta T_\alpha.
\end{equation}
Inserting these expressions into $P$ yields
\begin{multline}
		P_{\text{max}} = \frac{1}{4}\frac{\mathcal{Q}}{\Gamma_{\text{tot}}} \times \\
		\left[ \Gamma_1\Gamma_2\Delta T_2^2 + \Gamma_1\Gamma_3\Delta T_3^2 +\Gamma_2\Gamma_3\left( \Delta T_3-\Delta T_2 \right)^2 \right],
		\label{eq:pmax_3t}
\end{multline}
where $\mathcal{Q} = \mathcal{G}\mathcal{S}^2$. The maximum power is thus an always positive quantity and, in the same manner as in the two-terminal case, see Eq.~(\ref{eq:p_max_q}), it is proportional to $\mathcal{Q}$ and quadratic in the temperature differences. Furthermore, the properties of the QD and the chosen $\mu$ are all contained in the $\mathcal{Q}$ term, while the coupling to the reservoirs and the temperature differences are separately accounted for in the term between square parentheses in Eq.~(\ref{eq:pmax_3t}). The efficiency at maximum power instead is given by
\begin{widetext}
\begin{multline}
	\eta\left(P_{\text{max}}\right) = \frac{\eta_C^{(2)}}{2}\frac{\mathcal{Z}T}{\mathcal{Z}T+2} \left[\Gamma_1\Gamma_2\Delta T_2^2 + \Gamma_1\Gamma_3\Delta T_3^2 +\Gamma_2\Gamma_3\left( \Delta T_3-\Delta T_2 \right)^2 \right] \times \\
	\times
	\begin{dcases}
		\frac{1}{\Gamma_1\Delta T_3\left( \Gamma_2\Delta T_2 + \Gamma_3 \Delta T_3 \right)}  \hfill&\text{if $\frac{\Gamma_3}{\Gamma_1+\Gamma_3}\Delta T_3 \leq \Delta T_2 \leq \Delta T_3$},  \\
		\frac{1}{\Gamma_3\Delta T_3\left[ \Gamma_1\Delta T_3 + \Gamma_2 \left(\Delta T_3-\Delta T_2\right) \right]}  \hfill&\text{if  $0\leq \Delta T_2 \leq \frac{\Gamma_3}{\Gamma_1+\Gamma_3}\Delta T_3$}.
	\end{dcases}
	\label{eq:eta_pmax_3t}
\end{multline}
\end{widetext}
It is interesting to notice that in this equation 
the term before square parenthesis, which does not depend on $\Gamma_\alpha$, is exactly equal to the two terminal efficiency at maximum power, see Eq.~(\ref{eq:eta_p_max_zt}), since $\mathcal{Z}T = ZT$ for a two terminal system.
The remaining part of Eq.~(\ref{eq:eta_pmax_3t}) takes into account the particular temperatures and couplings to the three terminals. Furthermore the efficiency at maximum power, also in this 3 terminal system, only depends on $\mathcal{Z}T$. There are three limiting cases we will first study: $\Delta T_2 = 0$, $\Delta T_2 = \Delta T_3$ and $\Delta T_2 = \Delta T_3\Gamma_3/(\Gamma_1+\Gamma_3)$.

If $\Delta T_2=0$, $T_1=T_2$, so we have one hot reservoir at temperature $T_3$ and two cold ones at the same temperature. In this case we obtain
\begin{align}
	&P_{\text{max}} = \frac{1}{4}\mathcal{Q} \Delta T_3^2 \frac{\left( \Gamma_1 + \Gamma_2\right)\Gamma_3}{\left( \Gamma_1 + \Gamma_2\right) +\Gamma_3}, \label{Pmax3}\\
	&\eta\left(P_{\text{max}}\right) = \frac{\eta_C}{2}\frac{\mathcal{ZT}}{\mathcal{ZT}+2}. \label{etamax3}
\end{align}
Note that for a two-terminal system the maximum power, obtained by inserting Eq.~(\ref{eq:gsk_3t}) into Eq.~(\ref{eq:p_max_q}), is given by
\begin{equation}
\label{Pmax2}
	P_{\text{max}}^{(2)} = \frac{1}{4}\mathcal{Q}\Delta T_2^2\frac{\Gamma_1\Gamma_2}{\Gamma_1+\Gamma_2}.
\end{equation}
Comparing Eq.~(\ref{Pmax2}) with Eq.~(\ref{Pmax3}), and Eq.~(\ref{eq:eta_p_max_zt}) with Eq.~(\ref{etamax3}), we notice that the 3 terminal system is formally equivalent to a 2-terminal system with temperature difference $\Delta T_3$, with tunneling rate $\Gamma_1+\Gamma_2$ instead of $\Gamma_1$, and $\Gamma_3$ instead of $\Gamma_2$.

If $\Delta T_2=\Delta T_3$, $T_2=T_3$, so we have two hot reservoir at temperature $T_3$ and one cold reservoir at temperature $T_1$. In this case we obtain
\begin{align}
	&P_{\text{max}} = \frac{1}{4}\mathcal{Q} \Delta T_3^2 \frac{\Gamma_1\left( \Gamma_2 + \Gamma_3\right)}{\Gamma_1 + \left(\Gamma_2 +\Gamma_3\right)}, \\
	&\eta\left(P_{\text{max}}\right) = \frac{\eta_C}{2}\frac{\mathcal{Z}T}{\mathcal{Z}T+2}.
\end{align}
As in the previous limiting case, this system behaves like a 2-terminal with temperature difference $\Delta T_3$ and with tunneling rate $\Gamma_2 +\Gamma_3$ instead of $\Gamma_2$.

If $\Delta T_2 = \Delta T_3\Gamma_3/(\Gamma_1+\Gamma_3)$, reservoir 2 has an intermediate temperature such that $J^h_2=0$. In fact this specific value of $\Delta T_2$ distinguishes the two regimes where $J^h_2> 0$ and $J^h_2< 0$. In this case we obtain
\begin{align}
	&P_{\text{max}} = \frac{1}{4}\mathcal{Q} \Delta T_3^2 \frac{\Gamma_1\Gamma_3}{\Gamma_1 + \Gamma_3}, \\
	&\eta\left(P_{\text{max}}\right) = \frac{\eta_C}{2}\frac{\mathcal{Z}T}{\mathcal{Z}T+2}.
\end{align}
As in the other two limiting cases, this system behaves like a 2-terminal system, where reservoir 2 has been removed; this is to be expected because at this particular temperature, no heat flows through the second reservoir.

\begin{figure}[!b]
	\centering
	\includegraphics[width=1\columnwidth]{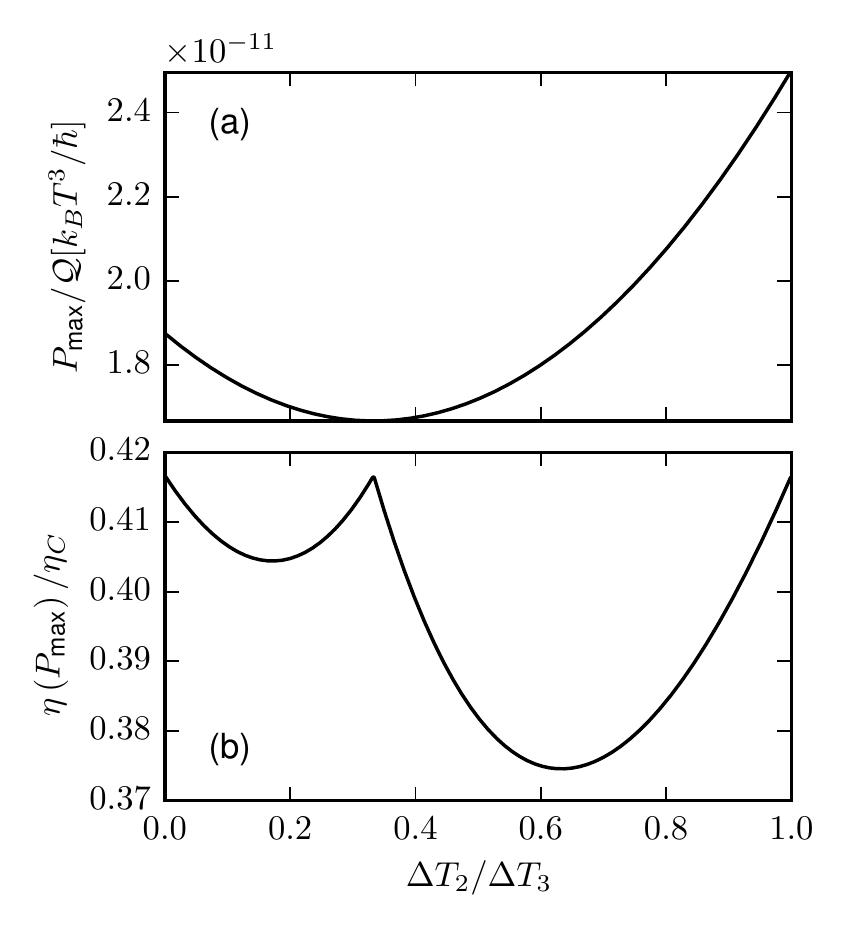}
	\caption{(a) Maximum power, Eq.~(\ref{eq:pmax_3t}), normalized to $\mathcal{Q}$ and (b) efficiency at maximum power, Eq.~(\ref{eq:eta_pmax_3t}), normalized to Carnot's efficiency, plotted as a function of $\Delta T_2/\Delta T_3$. Curves computed with $\mathcal{Z}T=10$, $\Delta T_3/T = 10^{-4}$, $\hbar\Gamma_1 = 0.02k_BT$, $\hbar\Gamma_2 = \hbar\Gamma_3 = 0.01k_BT$. }
	\label{fig:pow_eff_3t}
\end{figure}
According to Eq.~(\ref{eq:eta_pmax_3t}), all values of $\Delta T_2$ other than the three cases discussed above decrease the efficiency at maximum power with respect to the two-terminal case.
The maximum power instead, at given tunneling rates, is increased with respect to the two-terminal case. In fact the maximum power is formally equal to that of a two-terminal system, coupled to the same QD, with increased tunneling rates. So if we have fixed values of $\Gamma_1$, $\Gamma_2$ and $\Gamma_3$, we achieve the largest maximum power by choosing $\Delta T_2=0$ if $\Gamma_1 < \Gamma_3$ and $\Delta T_2 = \Delta T_3$ if $\Gamma_1 > \Gamma_3$. In Fig.~\ref{fig:pow_eff_3t}, $P_{\text{max}}$ and $\eta\left(P_{\text{max}}\right)$ from Eqs. (\ref{eq:pmax_3t}) and (\ref{eq:eta_pmax_3t}) are plotted as a function of $\Delta T_2/\Delta T_3$, choosing $\hbar\Gamma_1 = 0.02k_BT$ and $\hbar\Gamma_2 = \hbar\Gamma_3 = 0.01k_BT$.
As we can see in panel (a), the power is maximum when $\Delta T_2 = \Delta T_3$
(this result is expected since $\Gamma_1 > \Gamma_3$). Furthermore the three maxima in panel (b) correspond to the three limiting cases previously studied, where $\eta\left(P_{\text{max}}\right)$ reaches the two-terminal performance.

\section{Non-linear response}
\label{chap:beylinear}
The linear response theory describes correctly the thermoelectric properties of bulk materials in most experimental conditions.
However, as discussed for instance in Ref.~\onlinecite{bib:whitney}, non-linear effects are important in nanoscopic setups, since the temperature difference is applied across very small elements of the order of tens or hundreds of nanometers.
As far as heat-to-work conversion is concerned, there is a practical reason to consider the non-linear response, namely efficiency and power output may increase with increasing temperature difference.
Furthermore, for systems with time-reversal symmetry 
the efficiency at maximum power can overcome the limit of $\eta_C/2$ only beyond the linear response \footnote{Note, however, that the Curzon-Ahlborn limit can be overcome also within linear response when time-reversal symmetry is broken, see Refs.~\onlinecite{Benenti2011,Balachandran2013,Brandner2013,Brandner2013b,Yamamoto2016}.}.

In this Section we will consider a two-terminal QD system and discuss the numerical results obtained solving the kinetic equations (\ref{eq:kinetic_def}) as discussed in Sec.~\ref{theory}.
We will focus our discussion on the thermoelectric properties, and on the efficiency and power produced by a QD-based heat engine.
Let us define the charge current $J^c\equiv J^c_2=-J^c_1$, thanks to charge current conservation, and 
the average reservoir temperature $\bar{T} = (T_1+ T_2)/2$, which determines the typical thermal energy scale of the system beyond linear response (all energies will be given in units of $k_B\bar{T}$).
Furthermore, $\Delta\mu\equiv \Delta\mu_2=eV$, with $V$ applied voltage, $\Delta T\equiv\Delta T_2$, and assume equidistant energy levels with spacing given by $\Delta E$.
In order to describe the potential drop between the QD and the two reservoirs, we will assume that the set of energy levels is shifted as $E_p(V)=E_p+(1-\theta_0) eV$, where $0\leq\theta_0\leq 1$ is the fraction of potential $V$ that drops over the tunnel barrier which couples reservoir 2 to the QD.

Regarding charge transport, we recall that in the linear response, within the quantum limit, the conductance exhibits peaks occurring at the dominant transition energies, i.e. when $\mu=\mu_N$, of width given by $k_B\bar{T}$.
By applying a finite voltage bias $V$, in the absence of a temperature difference ($\Delta T=0$), the differential conductance, defined as
\begin{equation}
G=\left( \frac{\partial J^c}{\partial V} \right) ,
\end{equation}
exhibits the typical Coulomb diamond structure, with visible excited states, as a function of $V$ and $\mu$, see for instance
Ref.~\onlinecite{Vandersypen}.

\subsection{Non-linear Seebeck and Peltier coefficients}
\label{sec-SP}
In the non-linear regime the thermopower (Seebeck coefficient) can be defined as follows
\begin{equation}
	S = -\at{\frac{V}{\Delta T }}{J^c = 0}, 
	\label{esse}
\end{equation}
i.e. as the ratio between the \textit{thermovoltage} $V$ that develops as a result of a finite $\Delta T$ applied, at open circuit ($J^c=0$).
\begin{figure}[!htb]
\centering
\includegraphics[width=1\columnwidth]{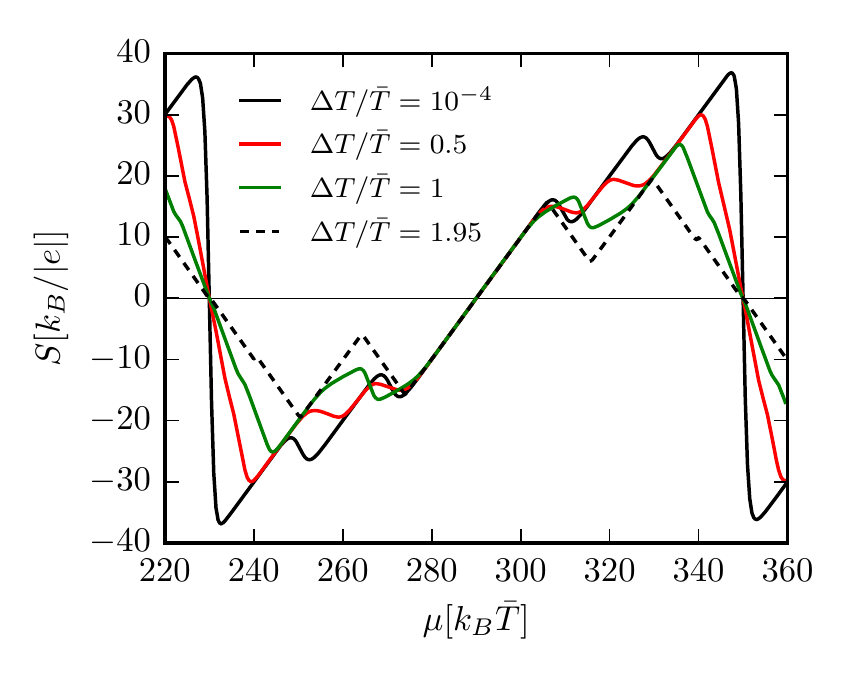}
\caption{(Color online) Non-linear thermopower $S$ plotted as a function of $\mu$ for various values of $\Delta T/\bar{T}$, and $\Delta E = 20k_B\bar{T}, E_C = 50k_B\bar{T}, \hbar\Gamma_L(p) = \hbar\Gamma_R(p) = 0.01k_B\bar{T}, \theta_0=1/2$.}
\label{fig-S}
\end{figure}
In Fig.~\ref{fig-S}, $S$ is plotted as a function of $\mu$ for various values of $\Delta T/\bar{T}$.
The black solid curve (calculated for $\Delta T/\bar{T}=10^{-4}$) is the linear-response reference that is well approximated by the expression given in Eqs.~(\ref{eq:GSKmatrices}).
As discussed in Sec.~\ref{qr}, the black solid curve presents main oscillations of period  $\Delta E+2E_C$, and a fine structure with a $\Delta E$ spacing~\cite{bib:beenakker2}.
Since we have chosen an equidistant energy spectrum, all curves share a number of features with the linear-response reference.
Namely, i) $S$ crosses zero with positive slope at the main transition energies $\mu_N$ and is periodic with periodicity $\Delta E+2E_C$ (in Fig.~\ref{fig-S} $\mu=290$ $k_B\bar{T}$ corresponds [see Eq.~(\ref{eq:muN}))] to $\mu_{N=3}$); ii) in the range of $\mu$ considered, $S$ is antisymmetric with respect to $\mu=290$ $k_B\bar{T}$; iii) $S$ vanishes for values of $\mu$ in the middle points between two dominant transitions $\mu_N$ and $\mu_{N+1}$ (in Fig.~\ref{fig-S} such points are located at $\mu=230$ $k_B\bar{T}$ and $\mu=350$ $k_B\bar{T}$).
Moreover, since we set $\theta_0=1/2$, the linear increase of $S$ for $\mu\simeq\mu_N$ does not depend on the ratio $\Delta T/\bar{T}$, i.e. it is well described by the linear-response proportionality coefficient $1/(|e|\bar{T})$.
Interestingly, such features (except for the fine structure oscillations) can be understood in terms of a non-interacting model (see App.~\ref{oneEL}), which also explains the reduction of the negative slope of $S$ at the middle points as the ratio $\Delta T/\bar{T}$ increases.

Let us now discuss the behavior of $S$ when departing from the linear response regime.
Fig.~\ref{fig-S} shows that for all values of $\Delta T$ the thermopower deviates from the linear-response curve only for $\mu$ above 310 $k_B\bar{T}$ (or below 270 $k_B\bar{T}$).
A sharp departure already occurs at $\Delta T/\bar{T}=0.5$ (red curve).
This can be understood from the fact that, for $\mu>310$ $k_B\bar{T}$, $S$ is of the order of $15 ~k_B/|e|$ which corresponds to a value of the thermovoltage ($V=-7.5~k_B\bar{T}/|e|$) such that $|eV|\gg k_B T$ \footnote{We have also checked that we are well beyond the linear response for the electric conductance.}.
Note that $\mu=310$ $k_B\bar{T}$ roughly corresponds to the first step of the fine structure in linear response.
In particular, while the first step hardly moves by increasing $\Delta T$ from its position in linear response, the second step, occurring at $\mu=330$ $k_B\bar{T}$ in the linear response, shifts to a smaller value for increasing $\Delta T/\bar{T}$, eventually disappearing or merging with the first step. This behavior may be attributed to the combination of the following two effects.
On one hand, the thermovoltage $V$, which determines the transport energy window, depends on $\mu$ and increases with $\Delta T$ according to the definition (\ref{esse}).
On the other hand, an increase of $\Delta T/\bar{T}$ moves the lowest temperature ($T_1$) towards absolute zero, thus sharpening the Fermi distribution function $f_1(E)$. This last effect is also responsible for the sharpening of $S(\mu)$ as $\Delta T/\bar{T}$ increases. Note furthermore that the extremal values of $S$ decrease as $\Delta T/\bar{T}$ increases.

Let us now consider the non-linear Peltier coefficient defined, for
a given voltage $V$, as
\begin{equation}
	\Pi = \at{\frac{J^h_2}{J^c}}{\Delta T= 0}.
\end{equation}
Our aim is to assess the failure of the Onsager reciprocity relation $\Pi=TS$, which holds in the linear response regime.
Beyond linear response, for a single-level non-interacting QD, one finds a ``corrected'' reciprocity relation, namely $\Pi+V/2=\bar{T} S$ in the case where $\theta_0=1/2$ (see App.~\ref{oneEL}).
\begin{figure}[!t]
\centering
\includegraphics[width=1\columnwidth]{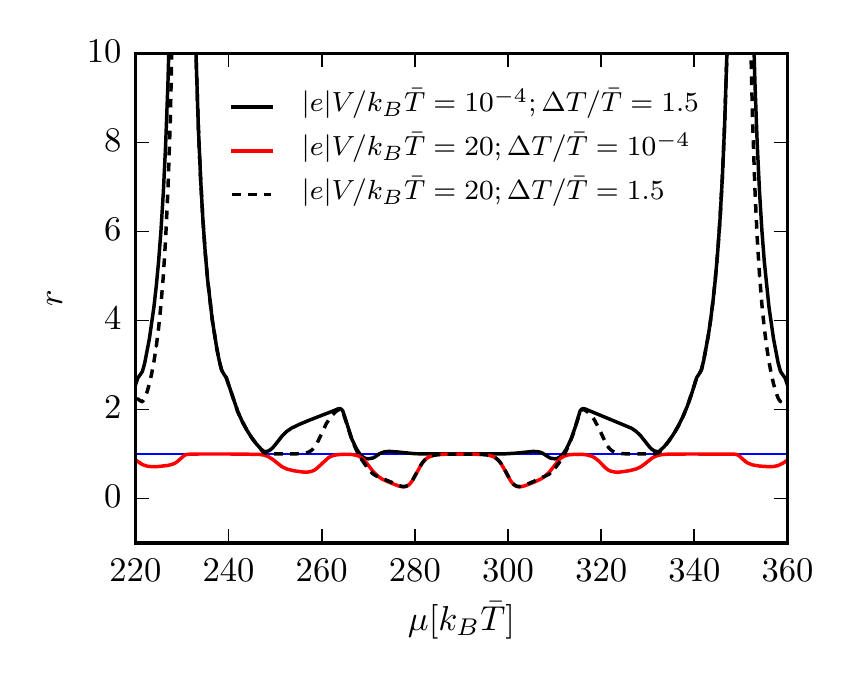}
\caption{(Color online) Ratio $r=(\Pi+V/2)/(\bar{T} S)$ plotted as a function of $\mu$ for various values of 
$|e|V/k_B T$ (for the Peltier coefficient) and of 
$\Delta T/\bar{T}$ (for the thermopower), for the same parameter values 
as in Fig.~\ref{fig-S}. The blue thin line is the reference $r=1$.}
\label{figPiS}
\end{figure}
To single out the effect of interactions in a multi-level QD, in Fig.~\ref{figPiS} we plot the ratio $r=(\Pi+V/2)/(\bar{T} S)$ as a function of $\mu$ for various values of $\Delta T/\bar{T}$ and $|e|V/(k_B\bar{T})$ ($\Delta T/\bar{T}$ is used to compute $S$, while $|e|V/k_BT$ to compute $\Pi$).
Fig.~\ref{figPiS} shows that the ratio $r$ departs significantly from 1, the linear response result (blue thin line), only far enough from the dominant transition energy $\mu_{N=3}=290$ $k_B\bar{T}$.
In particular, when $\Pi$ is in the linear response regime and $S$ is not (black solid curve) the strong deviations occurring for $\mu$ around the middle points between dominant transition energies, namely $\mu=230$ $k_B\bar{T}$ and $\mu=350$ $k_B\bar{T}$, can be explained by a two-level non-interacting model (see App.~\ref{oneEL}).
However, the deviations occurring in the range of values of $\mu$ between 250 $k_B\bar{T}$ and 330 $k_B\bar{T}$ can be imputed to interaction effects.
In the opposite case, where $S$ is in the linear response and $\Pi$ is not (red curve), the deviations of $r$ from 1 are entirely due to interaction effects and $r$ takes values between 0 and 1 in the entire range of values of $\mu$.
When both $S$ and $\Pi$ are beyond linear response the two behaviors discussed above coexist giving rise to the black dashed curve \footnote{The black dashed curve is actually equal to the product of the solid black and the red curve.
Indeed, due to the Onsager reciprocal relations in this product the 
linear response Peltier $\Pi$ (which appears in the ratio
$r$ for the solid black curve) simplifies with the linear response 
$\bar{T}S$ (which appears in the ratio
$r$ for the red curve).}.

\subsection{Efficiency and output power}
\label{blEP}
In this Section we consider the efficiency for heat-to-work conversion and output power in a two terminal system. Specifying Eq.~(\ref{eq:eta_def}) to a two terminal system where $\Delta T>0$, we have that
\begin{equation}
	\eta = \frac{P}{J^h_2} ,
\end{equation}
where $P>0$ is the output power, defined in Eq.~(\ref{eq:p_def}), and $J^h_2>0$ is the heat current absorbed by the system. Apart from the system's parameters, $\eta$ depends on $V$ and $\Delta T$.

\begin{figure}[!htb]
\centering
\includegraphics[width=1\columnwidth]{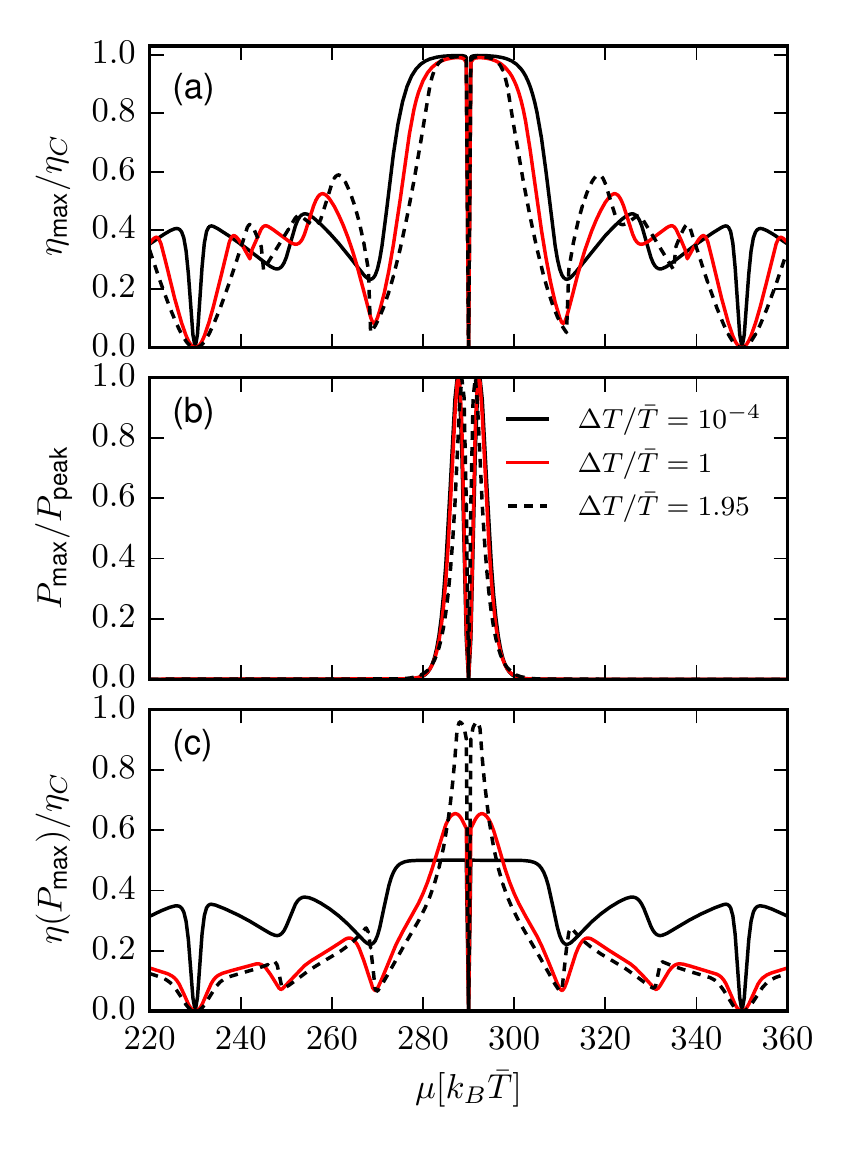}
\caption{(Color online) (a) Maximum efficiency, normalized to Carnot's efficiency, (b) maximum output power $P_{\text{max}}$, normalized to its peak value $P_{\text{peak}}$, and (c) efficiency at maximum power $\eta(P_{\text{max}})$, normalized to $\eta_{C}$, plotted as a function of $\mu$ for various values of $\Delta T/\bar{T}$, for the same parameter values as in Fig.~\ref{fig-S}.}
\label{figEtaPmax}
\end{figure}

Let us first consider the maximum efficiency $\eta_{\text{max}}$, obtained by maximizing the efficiency $\eta$ with respect to the applied voltage $V$, at given $\Delta T$.
$\eta_{\text{max}}$ is plotted, normalized to $\eta_{C}$, in Fig.~\ref{figEtaPmax}(a) as a function of $\mu$ for different values of $\Delta T/\bar{T}$.
All plots show pairs of peaks close to $\mu_N$, whose maximum is very close to $\eta_C$, and secondary peaks of smaller height.
The solid black curve, relative to the linear response regime 
($\Delta T/\bar{T}=10^{-4}$), is related through  Eq.~(\ref{eq:eta_max_zt}) to the plot of $ZT$ [Fig.~\ref{fig:p_max_ql}(b)]\footnote{Notice that in Fig~\ref{figEtaPmax} $\Delta E=20\,k_BT$, while in Fig.~\ref{fig:p_max_ql} $\Delta E=10\,k_BT$. } discussed in Sec.~\ref{2terms}.
For the black curve a pair of maxima approaching $\eta_C$ occur at $\mu=\mu_N\pm 2.40 k_B\bar{T}$, while $\eta_{\text{max}}$ vanishes at the dominant transition energies (and at the middle points between two dominant transition energies).
Moreover, a fine structure of secondary peaks, with spacing $\Delta E$, appears for intermediate values of $\mu$.
Moving away from the linear response, the main observation is that an increase of $\Delta T/\bar{T}>0.1$ produces only quantitative changes to the curves.
As shown in Fig.~\ref{figEtaPmax}(a), the main peaks of $\eta_{\text{max}}$ are still approximately located at $\mu=\mu_N\pm 2.40 k_B\bar{T}$ and approaching the Carnot efficiency, while the peaks' width reduces slightly with increasing $\Delta T$, and the fine structure of the secondary peaks gets simply distorted.

Another important quantity in heat-to-work conversion is the maximum output power generated $P_{\text{max}}$, which is obtained by maximizing the output power with respect to the applied voltage $V$.
It turns out that $P_{\text{max}}$ exhibits pairs of peaks approximately located at $\mu=\mu_N\pm 2.40 k_B\bar{T}$ whose height increases approximately quadratically with $\Delta T$, as long as $\Delta T/\bar{T}$ is not to close to 2.
Interestingly, Fig.~\ref{figEtaPmax}(b) shows that the maximum output power $P_{\text{max}}$, when normalized to its peak value $P_{\text{peak}}$, only very weakly depends on the ratio $\Delta T/\bar{T}$.
In particular, $P_{\text{max}}/P_{\text{peak}}$ is well approximated by the linear-response result, whose analytical expression is obtained by substituting Eq.~(\ref{eq:p_max_ql}) into Eq.~(\ref{eq:p_max_q}).

The efficiency at maximum power $\eta(P_{\text{max}})$ can now be calculated by taking, for each value of $\mu$, the value of $V$ which maximizes the power.
$\eta(P_{\text{max}})$ is plotted, as a function of $\mu$, in Fig.~\ref{figEtaPmax}(c) for various values of the ratio $\Delta T/\bar{T}$.
By increasing such ratio starting from the linear response [solid black curve, related to the plot of $ZT$ in Fig.~\ref{fig:p_max_ql}(b) through Eq.~(\ref{eq:eta_p_max_zt})] one finds that the peak values, again occurring approximately at $\mu=\mu_N\pm 2.40 k_B\bar{T}$, increase well above 
$\eta_C/2$ (the upper limit for the linear response).
On the contrary, the efficiency at maximum power for values of $\mu$ away from $\mu_N$, relative to the fine structure, decreases with increasing $\Delta T/\bar{T}$ beyond the linear response, but is only slightly different moving from $\Delta T/\bar{T}=1$ (red curve) to $\Delta T/\bar{T}=1.95$ (black dashed curve).

\begin{figure}[!tb]
\centering
\includegraphics[width=1\columnwidth]{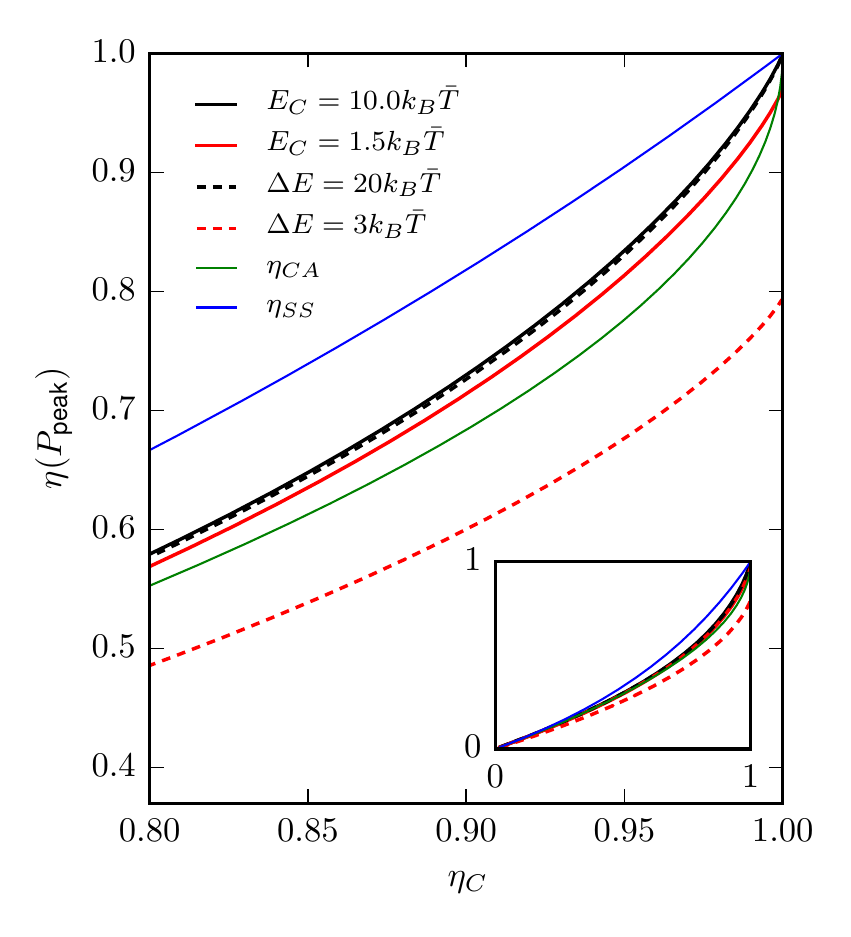}
\caption{(Color online) Efficiency at maximum power $\eta(P_{\text{peak}})$ plotted as a function of $\eta_C$ for various values of $E_C$, with $\Delta E=0$ (solid curves) and for various values of $\Delta E$, with $E_C=0$ (dashed curves). Thin solid curves represent the CA and SS efficiencies, see text.
Tunneling rates are $\hbar\Gamma_1(p) = \hbar\Gamma_2(p) = 0.01k_B\bar{T}$.
The inset shows the same curves on the entire range $\eta_C\in [0,1]$.}
\label{figEtaPmaxEtaC}
\end{figure}
It is now interesting to compare the peak values of the efficiency at maximum power with various reference values, such as the Curzon-Alhborn (CA) efficiency~\cite{bib:curzon} $\eta_{\text{CA}}=1-\sqrt{1-\eta_C}$ and the Schmiedl-Seifert (SS) efficiency~\cite{SchmiedlSeifert2008} $\eta_{\text{SS}}=\eta_C/(2-\eta_C)$ \footnote{This is actually the upper bound of the Schmiedl-Seifert efficiency,
see Refs.~\onlinecite{SchmiedlSeifert2008,bib:esposito2}.}.
To do so we calculate the peak power, i.e. maximizing the power with respect to $V$ and $\mu$, and plot the corresponding efficiency as a function of $\eta_C$ (determined by the temperature difference) in various situations, see Fig.~\ref{figEtaPmaxEtaC}.
In particular, we consider the case of a QD with one doubly degenerate level with a finite charging energy (solid thick curves) and the case of a QD with two non-degenerate levels and zero charging energy (dashed curves).
The parameters are chosen such that the two situations can be compared, namely $\Delta E=2E_C$, i.e. the differential conductance consists of two peaks separated by the same electrochemical potential.
Figure~\ref{figEtaPmaxEtaC} (inset) shows the following general feature for small $\eta_C$, i.e. in the linear response regime: $\eta(P_{\text{peak}})$ increases linearly with $\eta_C$  with slope determined by the value of $ZT$ [see Eq.~(\ref{eq:eta_p_max_zt})]. 
In particular, the two black curves (relative to $E_C=10$ $k_B\bar{T}$ and to $\Delta E=20$ $k_B\bar{T}$) virtually coincide, and are equal to the one for a single non-interacting level QD [see App.~\ref{oneEL} and the first line of Eqs.~(\ref{eq:eta_pmax_1liv})], since the parameters are such that $k_B\bar{T}\ll E_C, \Delta E$, where the transport is mostly accounted for by a single energy level.
On the contrary, the two red curves (relative to $E_C=1.5$ $k_B\bar{T}$ and to $\Delta E=3$ $k_B\bar{T}$) differ by a large extent, with the interacting case (solid red curve) exhibiting larger efficiency at maximum power with respect to the associated non-interacting case (dashed red curve).
Note that the efficiency at maximum power relative to the interacting case goes beyond the CA efficiency, when $\eta_C$ is larger than about 0.5, for all values of $E_C$ between 1.5 $k_B\bar{T}$ and 10 $k_B\bar{T}$.
Finally, we find that the SS efficiency is never overcome.
\begin{figure}[!t]
\centering
\includegraphics[width=1\columnwidth]{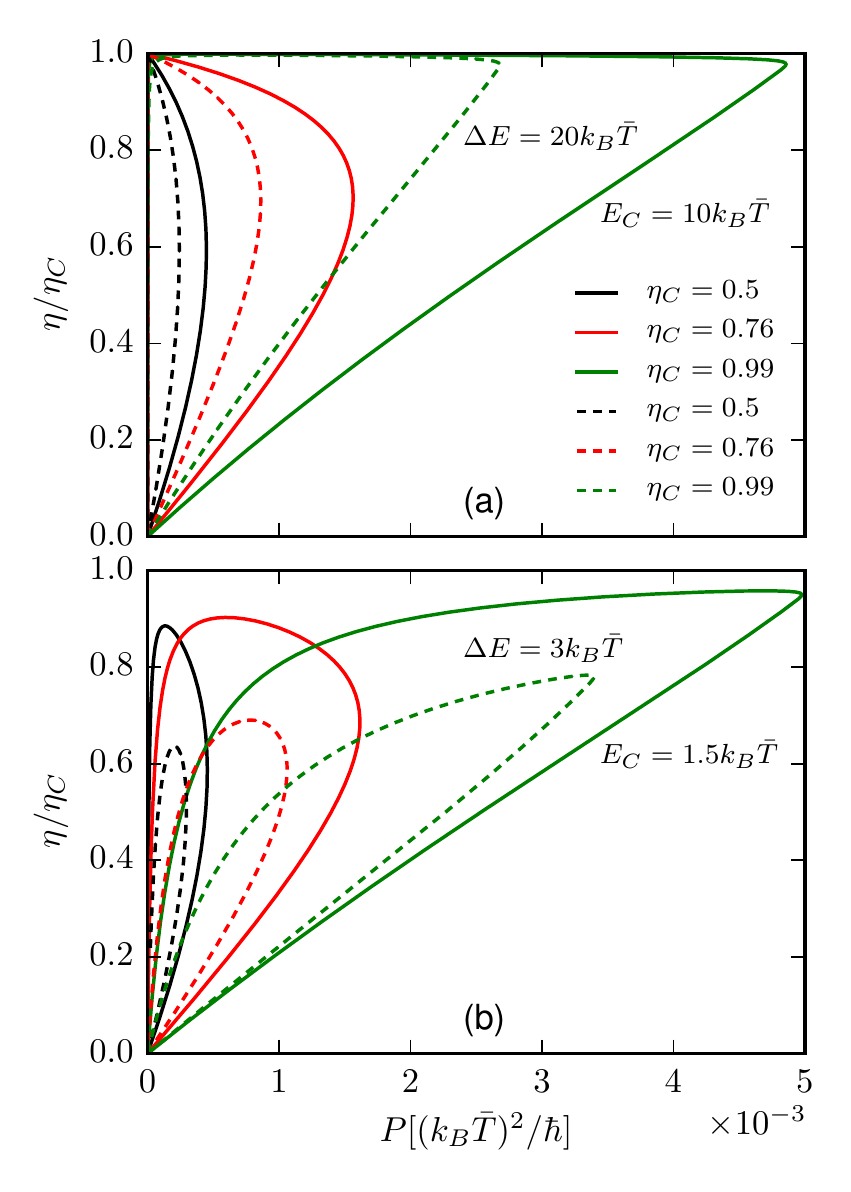}
\caption{(Color online) Correlation between efficiency and output power calculated for a few points of the plots in Fig.~\ref{figEtaPmaxEtaC}, both for doubly degenerate interacting case (solid curves), and non-degenerate non-interacting case (dashed curves). Curves obtained by increasing the value of $V$ from zero to the thermovoltage value, corresponding to the open-circuit situation, for various values of $\eta_C$. Panel (a) refers to $\Delta E=20$ $k_B\bar{T}$ (with $E_C=0$, dashed curves) and to $E_C=10$ $k_B\bar{T}$ (with $\Delta E=0$, solid curves), while panel (b) refers to $\Delta E=3$ $k_B\bar{T}$ (with $E_C=0$, dashed curves) and to $E_C=1.5$ $k_B\bar{T}$ (with $\Delta E=0$, solid curves).}
\label{figEtaP}
\end{figure}

To complete the analysis, we show the correlation between efficiency and power corresponding to a few points (i.e. a few values of $\eta_C$) in the curve of Fig.~\ref{figEtaPmaxEtaC}.
More precisely, Fig.~\ref{figEtaP} shows how the value of the power $P$ and the efficiency $\eta$ evolve by increasing the applied voltage $V$ from zero (where both $P$ and $\eta$ vanish) to the thermovoltage (where $P$ vanishes as a consequence of the fact that the charge current vanishes) \footnote{For a given 
value of $\eta_C$, we first find the values $\mu=\bar{\mu}$ and
$V=\bar{V}$ leading to the peak value $P_{\text{peak}}$ for power. 
We then set $\mu=\bar{\mu}$ and vary the voltage $V$ from zero to the
thermovoltage.}. 
In particular, panel (a) refers to the case $\Delta E=20$ $k_B\bar{T}$ (dashed curves) and $E_C=10$ $k_B\bar{T}$ (solid curves), while panel (b) refers to the case $\Delta E=3$ $k_B\bar{T}$ (dashed curves) and $E_C=1.5$ $k_B\bar{T}$ (solid curves).
We checked that in the linear response (when $\eta_C\ll 1$) the power reaches its maximum when the efficiency is nearly equal to $\eta_C/2$.
By increasing $\eta_C$, for all the curves in the figure, both maximum power and efficiency at maximum power increase.
For large values of $\eta_C$ (green curves), the efficiency remains close to the Carnot efficiency when $V$ is increased beyond the point of maximum power.
The general feature is that in the interacting case (solid curves) the power is much larger than in the associated non-interacting situation (dashed curves).
When $\Delta E$ and $E_C$ are of the same order as $k_B\bar{T}$ [Fig.~\ref{figEtaP}(b)], both the maximum power and the efficiency at maximum power are increased in the interacting case as compared with the non-interacting case.
A remarkable property of both regimes discussed in panels (a) and (b) 
is that in the strongly nonlinear regime, 
the maximum power is obtained for values where the efficiency 
is high and close to the maximum efficiency. 

We finally discuss a peculiarity of the maximum power output considering a doubly degenerate energy level in the interacting case \footnote{The effect of degenerate energy levels on
the electron and thermal conductance and on thermopower of
a QD was considered in Ref.~\onlinecite{bib:zianni_2008}.} (as we will point out, some considerations are also valid in the non degenerate case).
\begin{figure}[!t]
\centering
\includegraphics[width=1\columnwidth]{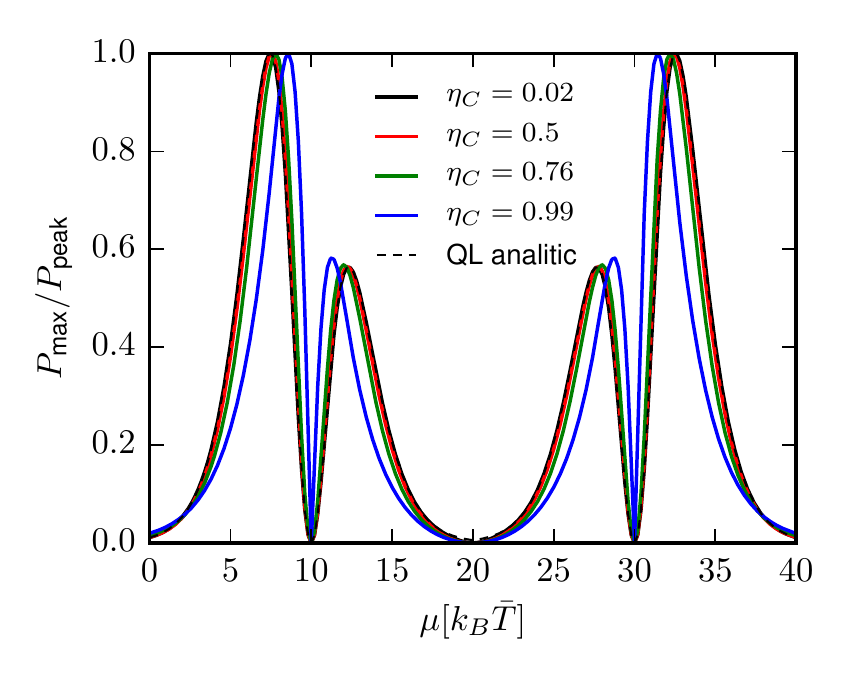}
\caption{(Color online) Maximum power, normalized to the peak value, plotted as a function of $\mu$ for different values of $\eta_C$ with $E_C=10$ $k_B\bar{T}$, $\Delta E=0$ and considering a single doubly degenerate energy level. Surprisingly, not only the linear-response curve (black) is very well approximated by the analytic expression in Eq.~(\ref{Pdegen}) (black dashed curve) valid in the quantum limit.
Tunneling rates are $\hbar\Gamma_1(p) = \hbar\Gamma_2(p) = 0.01k_B\bar{T}$.}
\label{figPdeg}
\end{figure}
Fig.~\ref{figPdeg} shows the maximum power, normalized with respect to the peak value, as a function of $\mu$ for different values of $\eta_C$ including the linear-response case (parameter values are the same as for Fig.~\ref{figEtaPmaxEtaC}, with $E_C=10$ $k_B\bar{T}$).
The first remarkable feature is that the two peaks around a dominant transition energy have different heights (here $\mu_{N=1}=10$ $k_B\bar{T}$ and $\mu_{N=2}=30$ $k_B\bar{T}$); in the absence of degeneracy, within the quantum limit all the peaks have the same height (see Fig.~\ref{fig:p_max_ql}).
More precisely, see Eq.~(\ref{Pdegen}), the external peaks, displaced of $2.53k_B\bar{T}$ from $\mu_N$ are higher with respect to the internal peaks, displaced of $2.32k_B\bar{T}$ from $\mu_N$, whose height is almost equal to the non-degenerate (interacting) case.
The second feature is that all curves, apart from the case of very large $\eta_C$ (blue curve), are well approximated by the linear-response quantum limit expression (black dashed curve in Fig.~\ref{figPdeg})
\begin{equation}
P_{\text{max}} = \frac{\sqrt{2}\gamma}{16k_B\bar{T}} \left(\frac{\Delta T}{\bar{T}}\right)^2 \frac{\Delta_{\text{min}}^2}{\cosh{(\frac{\Delta_{\text{min}}}{2k_B\bar{T}} \pm \frac{\log{2}}{2})} \cosh{(\frac{\Delta_{\text{min}}}{2k_B\bar{T}})}}.
\label{Pdegen}
\end{equation}
The plus (minus) sign in Eq.~(\ref{Pdegen}) is to be taken when $N_{\text{min}}$ is even (odd). 
Eq.~(\ref{Pdegen}) allows us to find that level degeneracy gives rise to an 
enhancement of output power of about 1.77 times respect to the non-degenerate case, independently of the parameter values (see also Ref.~\onlinecite{kuo2016}).
Eq.~(\ref{Pdegen}) also makes clear that the origin of the asymmetry for
$\Delta_{\text{min}}\to -\Delta_{\text{min}}$ and of the difference in 
peaks' height is the term $\pm\log 2/2$ occurring in the 
presence of degeneracy. The case of $E_C=1.5$ $k_B\bar{T}$ (not shown), which is not within the quantum limit, produces a much more asymmetric behavior. We have verified that also in the non degenerate case the analytic formula derived in the linear response regime [see Eq.~(\ref{eq:p_max_ql})] well describes the maximum power also beyond the linear response regime.

\section{Conclusions}
\label{conc}
We have studied the thermoelectric properties of a multi-level 
interacting QD in the sequential tunneling regime, in a multi-terminal setup,
both in the linear response regime and beyond.
In particular, we have
\begin{itemize}
	\item generalized the sequential tunneling method put forward by Beenakker in Refs.~\onlinecite{bib:beenakker1} and \onlinecite{bib:beenakker2} to a multi-terminal configuration and set the range of validity of the expressions for the charge and heat currents in the linear response regime;
	\item found simple analytic formulas for the multi-terminal transport coefficients in the low temperature limit;
	\item found simple analytic formulas for the power factor $Q$ and the figure of merit $ZT$ in the low temperature limit for a two-terminal setup;
	\item found that $Q$ and $ZT$ can be simultaneously maximized for suitable values of the electrochemical potential;
	\item found that Coulomb interactions can dramatically enhance $ZT$ by suppressing the thermal conductance;
	\item found that both the interacting an non-interacting models strongly violate the Wiedemann-Franz law;
	\item found analytic expressions for the maximum power and for the efficiency at maximum power  in a three terminal setup;
	\item investigated the nonlinear Seebeck and Peltier coefficients in a two-terminal setup, identifying features of the breakdown on the Onsager reciprocity relation;
	\item computed numerically the maximum efficiency, the maximum power, and the efficiency at maximum power in the non-linear regime, finding optimal system parameters for heat-to-work conversion such that the efficiency at maximum power can go beyond Curzon-Alhborn's efficiency;
	\item compared the case of a doubly degenerate level with interaction and the case of two non-degenerate levels without interaction finding that the interacting case enhances the power output and, especially when charging energy and level spacing are of the order of the thermal energy, it increases the efficiency at maximum power that can go beyond Curzon-Alhborn's efficiency;
	\item found that the non-linear maximum power is well approximated by the analytic linear response expression;
	\item found that QDs with degenerate energy levels and Coulomb interactions achieve higher efficiency and output power than non-degenerate QDs; in particular the maximum power is enhanced almost of a factor 2;
	\item calculated the transport coefficients for a non-linear, non-interacting QD with 1 and 2 energy levels (App.~\ref{oneEL});
	\item found approximate analytic expressions for the maximum power and efficiency at maximum power for a non-linear, non-interacting QD with 1 energy level (App.~\ref{oneEL}).
\end{itemize}
The multi-terminal formalism developed in this paper and the 
expressions we have obtained for charge and heat currents, transport coefficients, power and efficiency could be used to design and analyze experimental data. Extensions of the 
studies presented in this paper could include level spacings different 
from the equidistant and regimes beyond the 
quantum limit. Finally, a comprehensive description of the thermoelectric 
properties and performance of a QD should assess the role of quantum 
coherence going beyond the sequential tunneling limit and the relevance 
of phonon contribution to heat transport.

\section*{Acknowledgements}
We would like to acknowledge fruitful discussions with Stefano Roddaro.
This work has been supported by the SNS internal projects ``Thermoelectricity in nanodevices'' and ``Non-equilibrium dynamics of one-dimensional quantum systems: From synchronization to many-body localization'', by the EU project ThermiQ and by the COST Action MP1209 ``Thermodynamics in the quantum regime''.
G.B. acknowledges support by INFN through the project ``QUANTUM''.

	\begin{appendix}

	\section{The kinetic equations always allows a non-trivial solution.}
	\label{app:KE}
	To prove this statements, let us consider the kinetic equations as written in Eq.~(\ref{eq:kinetic_a_b_def}). This equation is made up of a sum over $p$ of the following expression:
	\begin{multline}
			(\delta_{n_p,1} - \delta_{n_p,0})\left[ P\left( \{n_i\}, n_p=0 \right)A_{\widetilde{N},p}\right.\\
	\left. - P\left( \{n_i\}, n_p=1 \right)B_{\widetilde{N},p}  \right],
	\end{multline}
	which is a function of a generic configuration $\{n_i\}$. Let us sum this expression for the two particular configurations: $(\{n_i\}, n_p=1)$ and $(\{n_i\}, n_p=0)$. When $n_p=1$, the first term in round parenthesis gives a plus sign, while when $n_p=0$, it gives a minus sign. The rest of the expression does not depend on $n_p$, since $\widetilde{N} = \sum\nolimits_{i\neq p} n_i $, thus the sum over the above configurations exactly vanishes. Now let us go back to Eq.~(\ref{eq:kinetic_a_b_def}), and let us sum over the two configurations $(\{n_i\}, n_k=1)$ and $(\{n_i\}, n_k=0)$, $k$ being a given index. 
	According to the argument given above, the term in the sum where $p=k$ vanishes, and we obtain
	\begin{multline}
		\sum\limits_{n_k=0,1}  \sum\limits_{p\neq k} (\delta_{n_p,1} - \delta_{n_p,0})\Big[ P\left( \{n_i\}, n_p=0 \right)A_{\widetilde{N},p}  \\
		- P\left( \{n_i\}, n_p=1 \right)B_{\widetilde{N},p}  \Big]	= 0.
	\end{multline}
	Thus by summing over a given occupation number $n_k=0,1$ we have removed the case $p=k$ in the sum over $p$. If we now sum over all occupation numbers, we will remove all terms from the sum, yielding zero \footnote{By summing over all occupation numbers except for a fixed $n_p$, we could demonstrate that the LBEs derive from the kinetic equations.}. The sum over all occupation numbers $\{n_i\}$ is the sum of all $2^L$ equations of the kinetic equations: this demonstrates that any equation in the kinetic equations is linearly dependent from the other ones. 
Thus the matrix $M$ defined 
by the kinetic equations has a null space of 
dimension at least 1, since we demonstrated that the rows of $M$ are not 
linearly independent. Furthermore, if we perform the same sum over all occupation numbers
to the time dependent kinetic equations, given in Eq.~(\ref{eq:kinetic_def}), 
we find that:
	\begin{equation}
		\frac{\partial}{\partial t} \left( \sum\limits_{\{n_i\}} P \right) = 0.
	\end{equation}
	This is an obvious but important property that says that the probability normalization does not depend on time.

\section{The DBEs are not consistent in general}
	\label{app:DBE1}
	In general, if $E_C \neq 0$, the DBEs are not consistent. This means that no set of $P(\{n_i\})$ exists that can simultaneously satisfy all the DBEs. In App.~(\ref{app:DBE2}) we will discuss which conditions guarantee their consistency within the linear response regime.

	Here let us demonstrate their inconsistency in the special case $L=2$. Let us start from Eq.~(\ref{eq:det_bal}) and consider non null temperature and electrochemical potential differences. We will show that these equations form an over-complete set for $P(\{n_i\})$ that in general does not allow any non null solution. Since $L=2$, we have $2^L = 4$ unknown probabilities, and the number of DBEs is $ 2^{L-1}2 = 4$. In this case, the DBEs can be represented in matrix form as follows
	\begin{equation}
		M_D  \vec{P} \equiv
		\begin{pmatrix}
			A_{0,1} & -B_{0,1} & 0 & 0 \\
			0 & 0 & A_{1,1} & -B_{1,1} \\
			A_{0,2} & 0 & -B_{0,2} & 0 \\
			0 & A_{1,2} & 0 & -B_{1,2} 
		\end{pmatrix} 
		\begin{pmatrix}
			P_{00} \\ P_{01} \\ P_{10} \\ P_{11}
		\end{pmatrix}
		=
		\begin{pmatrix}
			0 \\ 0 \\ 0 \\ 0
		\end{pmatrix},
	\end{equation}
	where $P_{n_2 n_1} \equiv P(n_2,n_1)$ and the coefficients are defined in Eqs.~(\ref{eq:a_def}) and (\ref{eq:b_def}).
	In order to show that this linear algebra problem does not allow a non null solution, we will show that the determinant $\det (M_D)$ is in general not zero.
	We have
	\begin{multline}
		\det (M_D)	 
		=\Gamma_{\text{tot}}(1)\Gamma_{\text{tot}}(2)\left( A_{0,2}A_{1,1} - A_{0,1}A_{1,2} \right) \\
		+\Gamma_{\text{tot}}(1) A_{0,2}A_{1,2}\left( A_{0,1} - A_{1,1} \right) 
	\\
		+\Gamma_{\text{tot}}(2) A_{0,1}A_{1,1}\left( A_{1,2} - A_{0,2} \right),
	\end{multline}
	where $\Gamma_{\text{tot}}(p) = \sum_\alpha \Gamma_\alpha(p)$.
	It is pretty clear that since $\Gamma_\alpha(p)$, $E_p$ and $E_C$ are arbitrary, this determinant cannot be in general zero. For instance, choosing a two terminal system with 
	$\Gamma_1(p) = \Gamma_2(p) = \Gamma$, $\Delta \mu_2 = 4k_BT$, $E_1=1/2k_BT$, $E_2=3/2k_BT$, $\mu = \Delta T_2 = 0$
	we obtain $\text{det}(M_D)/\Gamma^4 \simeq 0.20$. 
	It is interesting to notice that at equilibrium $(\Delta \mu_2 = \Delta T_2 = 0)$, the DBEs are all exactly satisfied by the gran canonical distribution.
	Furthermore, in the non-interacting limit $E_C = 0$, the coefficients $A_{\widetilde{N},p}$ and $B_{\widetilde{N},p}$ depend only on $p$, so that we can drop the $\widetilde{N}$ argument. Thus $\text{det}\left( M_D \right) = A_{2}A_{1}B_{1}B_{2} - A_{1}A_{2}B_{2}B_{1} \equiv 0$.

	The proof can be extended to any number $L$ of levels as follows.
We rewrite the DBEs, Eq.~(\ref{eq:det_bal}), as
        \begin{equation}
                \ln{P\left( \{n_i\}, n_p=0 \right)} - \ln{P\left( \{n_i\}, n_p=1 \right)} = \ln{\frac{B_{\widetilde{N},p}}{A_{\widetilde{N},p}}}.
        \end{equation}
This equation has the same form as the LDBEs (\ref{eq:ldbe_delta})
which we will consider in App.~\ref{app:DBE2}, where the unknown probabilities are replaced by the logarithm of the probabilities, and where
\begin{equation}
	\delta_p(\widetilde{N}) = \ln{\frac{B_{\widetilde{N},p}}{A_{\widetilde{N},p}}}.
\end{equation}

As we shall show in App.~\ref{app:DBE2}, these equations are consistent 
if $\delta_p(\widetilde{N})$ satisfies property (\ref{eq:ldbe_property}).
This is the case if $E_C = 0$, since $A_{\widetilde{N},p}$ and  
$B_{\widetilde{N},p}$ are then independent of $\widetilde{N}$.
Then condition (\ref{eq:ldbe_property}) is trivially fulfilled with $c=0$
and the DBEs are consistent.

To summarize, as we will see in Appendix \ref{app:DBE2}, the DBEs are not consistent in general, but they are if we set $E_C = 0$ or $\Delta \mu_\alpha = \Delta T_\alpha = 0$ for all $\alpha$. Furthermore, within the linear response regime, they are valid in many more cases, for example if the tunneling rates are proportional, i.e. $\Gamma_\alpha(p) = k_\alpha \Gamma_1(p)$ for $\alpha=2,3,\dots,{\cal N}$ .

\section{Conditions of validity of the linearized DBEs}
\label{app:DBE2}
	In this Appendix we will assess under which conditions the linearized DBEs are consistent. 
We start by rewriting Eq.~(\ref{DBErel}) in the following way:
\begin{equation}
	\psi(\{n_i\},n_p=0) - \psi(\{n_i\},n_p=1) = \delta_p(\widetilde{N}),
	\label{eq:ldbe_delta}
\end{equation}
where
\begin{multline}
	\delta_p(\widetilde{N}) \equiv -\frac{1}{k_BT}\sum_{\alpha}\frac{\Gamma_\alpha(p)}{\Gamma_{\text{tot}}(p)} \\
  \times \left[ \left(E_p+(2\widetilde{N}+1)E_C-\mu \right) \frac{\Delta T_\alpha}{T} +\Delta\mu_\alpha \right].
	\label{eq:delta_def}
\end{multline}
We will prove that the linearized DBEs, written in Eq.~(\ref{eq:ldbe_delta}), are consistent if
$\delta_p(\widetilde{N})$ satisfies the property
\begin{equation} 
	\delta_p(N) - \delta_p(M) = c (N-M),
	\label{eq:ldbe_property}
\end{equation}
where $c$ is a constant that does not depend on $p$, $N$ or $M$. This statement will be explicitly proven for a two energy level system, then it will be extended to $L$ energy levels by induction.

Property (\ref{eq:ldbe_property}) is in general satisfied if the tunneling rates are proportional, i.e. $\Gamma_\alpha(p) = k_\alpha \Gamma_1(p)$, or if $E_C = 0$, or if $\Delta T_\alpha = 0$ for all $\alpha$. Furthermore, in a three terminal system, property (\ref{eq:ldbe_property}) is satisfied also if $\Delta T_2 = 0$ and $\Gamma_3(p) = k \Gamma_{\text{tot}}(p)$, or if  $\Delta T_3 = 0$ and $\Gamma_2(p) = k \Gamma_{\text{tot}}(p)$. These conditions can be generalized to 
a generic ${\cal N}$-terminal system by requiring that  
\begin{equation}
	E_C\sum\limits_\alpha \frac{\Gamma_\alpha(p)}{\Gamma_{\text{tot}}(p)}\Delta T_\alpha
\end{equation}
is independent of $p$.

\subsection{Two energy level system ($L=2$)}
We will prove that if property (\ref{eq:ldbe_property}) is satisfied, then the linearized DBEs allow a solution. Eq. (\ref{eq:ldbe_delta}) represents a linear algebra problem for $\psi(\{n_i\})$. For $L=2$, $\psi(\{n_i\}) = \psi(n_2,n_1) \equiv \psi_{n_2,n_1}$.  Let $\vec{\psi}$ be the vector $(\psi_{00},\psi_{01},\psi_{10},\psi_{11})$ and let $B$ be the corresponding matrix such that Eq.~(\ref{eq:ldbe_delta}) can be written as
\begin{equation}
	B \vec{\psi} = \\
	\begin{pmatrix}
		1 & -1 & 0 & 0 \\
		0 & 0 & 1 & -1 \\
		1 & 0 & -1 & 0 \\
		0 & 1 & 0 & -1 \\
	\end{pmatrix}
	\begin{pmatrix}
		\psi_{00} \\ \psi_{01} \\ \psi_{10} \\ \psi_{11}
	\end{pmatrix}
	=
	\begin{pmatrix}
		\delta_1(0) \\ \delta_1(1) \\ \delta_2(0) \\ \delta_2(1) 
	\end{pmatrix}
	\equiv
	\vec{\delta}.
	\label{eq:b2_example}
\end{equation}
Matrix $B$ has a null space of dimension 1, generated by $\vec{\psi} = (1,1,1,1)$, thus it is not invertible. 
This vector actually represents the equilibrium distribution: when $\Delta T_\alpha = \Delta \mu_\alpha =0$, $\delta_p(\widetilde{N}) = 0$, so $\vec{\psi} = (1,1,1,1)$ satisfies Eq.~(\ref{eq:b2_example}). Since $P = P_{eq} (1+\psi)$, $P = 2P_{eq}$. Normalizing the probabilities yields $P \equiv P_{eq}$, so we have demonstrated that the equilibrium distribution is in fact given by the grand canonical distribution.
Eq. (\ref{eq:b2_example}) will allow a solution if and only if vector $\vec{\delta}$ belongs to the image of matrix $B$. The dimension of the image of matrix $B$ is 3, so there is a one dimensional space orthogonal to the image of $B$ that cannot be obtained by linear combinations of $B$'s columns. Vector 
\begin{equation}
	\vec{v_0} = 
		\begin{pmatrix}
		1 \\ -1 \\ -1 \\ 1
	\end{pmatrix}
\end{equation}
is orthogonal to the columns of $B$, so it is a generator of this one dimensional space. Thus a solution $\vec{\psi}$ will exist if and only if vector $\vec{\delta}$ only belongs to the image of $B$, thus it cannot have a projection on $\vec{v_0}$. The projection is zero when
\begin{equation}
\label{v0q}
	0=\vec{v_0}\cdot \vec{\delta} = \delta_1(0) - \delta_1(1) - \delta_2(0) + \delta_2(1)
\end{equation}
which is satisfied using property (\ref{eq:ldbe_property}).

\begin{figure}[!t]
	\centering	
	\includegraphics[width=1\columnwidth]{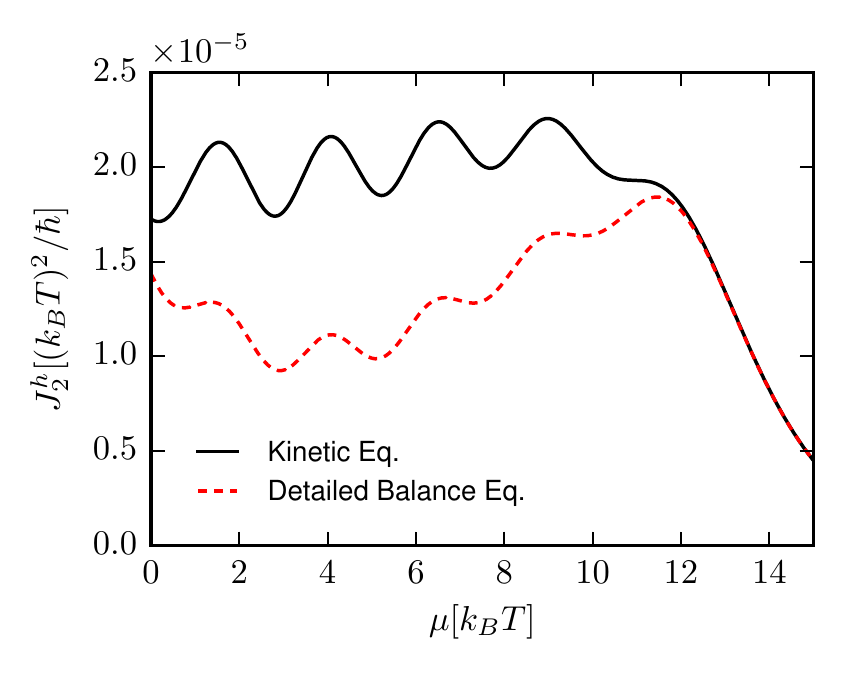}
	\caption{(Color online) 
		Heat current computed using Eq.~(\ref{eq:jh_general}) and using non equilibrium probabilities found solving exactly the kinetic equations (black solid line) and a subset of independent DBEs (red dashed line), as a function of the electrochemical potential $\mu$. The parameters used are: $\Delta T/T = 10^{-4}$, $\Delta\mu = 0$, $E_C = k_BT$, 5 equidistant energy levels with $\Delta E = 0.2 k_BT$,
$\hbar\Gamma_1(p) = (5)^{p-1} k_BT$, and $\hbar\Gamma_2(p) = (0.01)^{p-1} k_BT$.
}
\label{fig:iq_different}
\end{figure}
As we have shown in Sec.~\ref{chap:linear}, Eq.~(\ref{eq:lin_jc}) 
for the linearized charge current is correct in general, instead 
the linearized heat current given in Eq.~(\ref{eq:lin_jh}) is correct 
only if the linearized DBE are consistent. 
We verified numerically all these statements computing exactly, for small temperature and electrochemical potential differences, the charge and heat currents using both the DBEs and the kinetic equations.
As demonstrated by Fig.~\ref{fig:iq_different}, which shows the heat current as a function of $\mu$ for a choice of tunneling rates that are not proportional to each other,
a particular subset of DBEs leads to an incorrect result (red dashed curve) which differs from the one obtained using the kinetic equations (black solid curve). On the other hard, there is no difference when plotting the charge current using the same parameters as in Fig.~\ref{fig:iq_different}.

\subsection{$L$ energy level system}
We extend the previous demonstration to a generic system with $L$ energy levels. We will use the following notation and conventions:
\begin{itemize}
	\item $\{n_i\}$ indicates a generic set of occupation numbers, conventionally ordered the following way: $\{n_L,n_{L-1},\dots,n_2,n_1\}$.
	\item $(n_p=1,\{n_i\})$ is a shorthand notation for: $\{n_L,\dots,n_{p+1},1,n_{p-1},\dots,n_1\}$.
	\item $(n_p=0,\{n_i\})$ is a shorthand notation for: $\{n_L,\dots,n_{p+1},0,n_{p-1},\dots,n_1\}$.
	\item $(n_p=1,\{0\})$ is a shorthand notation for: $\{0,\dots,0,1,0,\dots,0\}$, where the $1$ is relative to $n_p=1$.
	\item $(n_p=0,\{0\})$ is equivalent to: $\{0,\dots,0\}$.
	\item $\vec{\psi}_i$ is the component of vector $\vec{\psi}$ with index $i$.
	\item Each linearized DBE in Eq.~(\ref{eq:ldbe_delta}) can be uniquely defined by specifying $p$ and $\{n_i\}$, so $[p|\{n_i\}]$ is a shorthand notation to indicate that given equation.
	\item Let $A$ be a matrix written in block form, for example:
	\begin{equation}
		A = \left(
		\begin{array}{c|c}
			B  & 0 \\
			0 &  C \\
			D & E
	\end{array} \right),
	\end{equation}
	where $B$, $C$, $D$ and $E$ are matrices. By ``block row'' we intend the set of rows in the block. For example, the first block row of $A$ is $(B,0)$, the second block row of $A$ is $(0,C)$, and so on.
\end{itemize}
In order to represent the linearized DBE, given in Eq.~(\ref{eq:ldbe_delta}), in matrix form, we will put the $2^L$ unknowns $\psi(\{n_i\})$ into a vector as follows:
$	\vec{\psi}_{I(\{n_i\})} \equiv \psi(\{n_i\})$,
where
$	I(\{n_i\}) = \sum\limits_{i=1}^{L} n_i 2^{i-1}$.

Let us define $B(L)$ as the $L2^{L-1} \times 2^L$ matrix such that Eq.~(\ref{eq:ldbe_delta}) can be represented as follows:
\begin{equation}
B(L) \vec{\psi} = \vec{\delta}(L),
\label{eq:ldbe_matrix}
\end{equation}
where $\vec{\delta}(L)$ is a vector containing the various values of $\delta_p(\widetilde{N})$. Each row of $B(L)$ contains zeros except for a $+1$ and a $-1$ in the position corresponding to $\psi(n_p=0,\{n_i\})$ and $\psi(n_p=1,\{n_i\})$. Let's notice that $B(L)$ has the following property:
\begin{equation}
	B(L) 
	\begin{pmatrix}
		1 \\
		\vdots \\
		1
	\end{pmatrix}
	 = 
		\begin{pmatrix}
		0 \\
		\vdots \\
		0
	\end{pmatrix}.
	\label{eq:b_1_0_property}
\end{equation}
This property is trivial since, as we just discussed, each row of $B(L)$ only has a $+1$ and a $-1$.

\subsubsection{Recursive decomposition}
We want to write $B(L)$ in terms of $B(L-1)$. We will thus order the rows in $B(L)$, and accordingly the elements of $\vec{\delta}(L)$, starting from those LDBE $[p|\{n_i\}]$ where $p < L$, then we will add the ones involving level $L$. In particular, we will first put equations $[p|(n_L=0,\{n_i\})]$ with $p<L$, then we will add equations $[p|(n_L=1,\{n_i\})]$ with $p<L$, and at last we will add equations $[L|\{n_i\}]$, which are the ones involving level $L$. For example, $B(2)$ will be written as in Eq.~(\ref{eq:b2_example}).

This ordering allows us to relate $B(L)$ to $B(L-1)$. The first set of equations $[p|(n_L=0,\{n_i\})]$, with $p<L$, represents all possible LDBE between elements of $\vec{\psi}$ with $n_L=0$, so they are equivalent to the LDBE for $L-1$ levels. Since $n_L=0$, according to our indexing convention, these rows involve all indexes of $\vec{\psi}_i$ with $0 \leq i < 2^{L-1}$, which corresponds to the first half of vector $\vec{\psi}$. Instead, the equations $[p|(n_L=1,\{n_i\})]$, with $p<L$, represents all possible LDBE between elements of $\vec{\psi}$ with $n_L=1$, so also these equations are equivalent to the LDBE for $L-1$ levels. Since $n_L=1$, according to our indexing convention, these rows involve all indexes of $\vec{\psi}_i$ with $i \geq 2^{L-1}$, which corresponds to the second half of vector $\vec{\psi}$. At last, equations $[L,\{n_i\}]$ relate components of $\psi$ where only the occupation number $n_L$ is changed. Since $I(n_L=1,\{n_i\}) = I(n_L=0,\{n_i\}) + 2^{L-1}$, these equations relate indexes of $\vec{\psi}$ that are distant $2^{L-1}$, thus they mix elements between the first and second half of $\vec{\psi}$. We will thus have the following block representation:
\begin{equation}
B(L) = \\
	\left(
	\begin{array}{c|c}
		B(L-1)  & 0 \\
		0 &  B(L-1) \\
		I_{d}(L-1) & -I_{d}(L-1)
	\end{array}
	\right),
	\label{eq:b_decomp}
\end{equation}
where $I_{d}(L)$ is the $2^L \times 2^L$ identity matrix. 
Since we trivially have that 
$B(1) = (1, -1)$, 
Eq.~(\ref{eq:b_decomp}) can be used to define $B(L)$ recursively, yielding a precise row ordering. 
For example, as we can see in Eq.~(\ref{eq:b2_example}), $B(2)$ can be obtained by applying Eq.~(\ref{eq:b_decomp}) to $B(1)$.

Also $\vec{\delta}(L)$ allows a decomposition in terms of $\vec{\delta}(L-1)$. The first block row in Eq.~(\ref{eq:b_decomp}), $(B(L-1),0)$, corresponds to equations $[p|(n_L=0,\{n_i\})]$ with $p < L$. Recalling Eq.~(\ref{eq:delta_def}), we will notice that $\delta_p(\widetilde{N})$ depends on level $L$ only through: 
\begin{equation}
	\widetilde{N}(L) = \sum\limits_{i\neq p}^L n_i = \sum\limits_{i\neq p}^{L-1} n_i = \widetilde{N}(L-1),
\end{equation}
since $n_L = 0$. 
So in the first block row we have that $\vec{\delta}(L) = \vec{\delta}(L-1)$. The second block row, $(0,B(L-1))$, corresponds to equations $[p|(n_L=1,\{n_i\})]$ with $p < L$. So this time the presence of level $L$ will change the values of $\widetilde{N}$ the following way:
\begin{equation}
	\widetilde{N}(L) = \sum\limits_{i\neq p}^L n_i = 1 + \sum\limits_{i\neq p}^{L-1} n_i = 1 +\widetilde{N}(L-1),
\end{equation}
since $n_L = 1$. So in the second block row $\vec{\delta}(L)$ is given by $\vec{\delta}(L-1)$ replacing $\widetilde{N}$ with $\widetilde{N} + 1$. Using property (\ref{eq:ldbe_property}), we have that:
\begin{equation}
	\delta_p(\widetilde{N} + 1) = \delta_p(\widetilde{N}) + c,
\end{equation}
where $c$ is a constant that does not depend on $p$ or $\widetilde{N}$. So in the second block row we will have that $\vec{\delta}(L) = \vec{\delta}(L-1) + c$. The last row $(I_d(L-1),-I_d(L-1))$ in Eq.~(\ref{eq:b_decomp}) corresponds to equations $[L|\{n_i\}]$, which involve level $L$. So let us define $\vec{\partial}(L)$ as the vector that corresponds to $\vec{\delta}(L)$ in the third block row. 

Using these observations and Eq.~(\ref{eq:b_decomp}), we can rewrite Eq.~(\ref{eq:ldbe_matrix}) using the following decomposition:
\begin{multline}
	\left(
	\begin{array}{c|c}
		B(L-1)  & 0 \\
		0 &  B(L-1) \\
		I_{d}(L-1) & -I_{d}(L-1)
	\end{array}
	\right) 
	\begin{pmatrix}
		\vec{\psi}(n_L = 0, \{n_i\}) \\
		\vec{\psi}(n_L = 1, \{n_i\})
	\end{pmatrix} \\ =
	\begin{pmatrix}
		\vec{\delta}(L-1) \\
		\vec{\delta}(L-1) + c \\
		\vec{\partial}(L)
	\end{pmatrix}.
	\label{eq:ldbe_matrix_decomp}
\end{multline}
This decomposition will be the main tool to perform a demonstration by induction.

\subsubsection{The independent part of \texorpdfstring{$B(L)$}{B(L)}}
Among the $L2^{L-1}$ rows of $B(L)$, we will explicitly show how to extract $2^L-1$ linearly independent rows; we will later show that all other rows can be obtained by linear combinations. Let's define $\widetilde{B}(L)$ as the matrix containing only the rows of $B(L)$ corresponding to equations $[p|\{n_i\}]$ with $n_i = 0$ for $i<p$; these are $2^L - 1$ rows. We will now prove by induction that all the rows of $\widetilde{B}(L)$ are linearly independent.

 Since $\widetilde{B}(L)$ is made by selecting rows from $B(L)$, we can decompose it in terms of $\widetilde{B}(L-1)$ just like we did for $B(L)$ in Eq.~(\ref{eq:b_decomp}). An interesting characteristic of $\widetilde{B}(L)$ is that it only has one equation in the block $[L|\{n_i\}]$ involving $\psi(0,0,\dots,0)$ and $\psi(1,0,\dots,0)$. So the decomposition becomes:
\begin{equation}
\widetilde{B}(L) = \\
	\left(
	\begin{array}{c|c}
		\widetilde{B}(L-1)  & 0 \\
		0 &  \widetilde{B}(L-1) \\
		1,0,\dots,0 & -1,0,\dots,0
	\end{array}
	\right).
	\label{eq:b_tilde_decomp}
\end{equation}
We can use Eq.~(\ref{eq:b_tilde_decomp}) to define $\widetilde{B}(L)$ recursively noticing that $\widetilde{B}(1) = B(1) =  (1, -1)$.

Now we can prove by induction over $L$ the following statement: all the rows of $\widetilde{B}(L)$ are linearly independent. This statement is obvious for $L=1$ and $L=2$. We will now assume that our statement is valid for $L-1$, and we will show that this implies that it is valid for $L$. The first block row in Eq.~(\ref{eq:b_tilde_decomp}) contains rows that are linearly independent by hypothesis. The same is true for the second block row. Also putting together the first and second block row yields $2^L -2$ independent rows, since the space generated by the respective linear combinations are clearly orthogonal thanks to the block decomposition. So we have to show that the last row cannot be written as a linear combination of previous rows. We will make a proof by contradiction. Let's assume that the last row $\vec{v} = (1,0,\dots,0|-1,0,\dots,0)$ can be written as a linear combination of a vector belonging to the space generated the first block row, $\vec{w_0}$, and of one belonging to the second block row, $\vec{w_1}$:
\begin{equation}
	\vec{v} = \vec{w_0} + \vec{w_1}.
	\label{eq:v_absurd}
\end{equation}
If we define $\vec{n} = (1,\dots,1|0,\dots,0)$, we will have that:
\begin{equation}
\begin{aligned}
	&\vec{n} \cdot \vec{w_0} = 0, \\
	&\vec{n} \cdot \vec{w_1} = 0.
\end{aligned}
\end{equation}
The first equation is zero because, as we have shown in Eq.~(\ref{eq:b_1_0_property}), $B(L)$ has the property that $B(L) (1,\dots,1)^T = 0$, so it is true also for $\widetilde{B}(L)$. The second equation is zero because $\vec{n}$ has zeroes on all non null components of $\vec{w_1}$. So multiplying each sides of Eq.~(\ref{eq:v_absurd}) by $\vec{n}$ yields
$\vec{n} \cdot \vec{v} = \vec{n} \cdot (\vec{w_0} + \vec{w_1}) =0$,
which is absurd because $\vec{n} \cdot \vec{v}=1$. Thus we have completed the proof by induction.

\subsubsection{Consistency of the linearized DBEs}
We will now show that if property (\ref{eq:ldbe_property}) is valid, the linearized DBEs are consistent and allow a solution for $\psi$ given by a one dimensional space; the unique solution can then be found by imposing the normalization condition on the probabilities. We will thus prove by induction over $L$ the following statement: if $\delta_p(N)$ satisfies property (\ref{eq:ldbe_property}), all equation in the linear algebra system in Eq.~(\ref{eq:ldbe_matrix_decomp}) can be written as linear combinations of the equations involving the rows of $\widetilde{B}(L)$. Since there are $2^L-1$ independent rows in $\widetilde{B}(L)$, by virtue of Rouch\'e-Capelli's theorem we will have a solution given by a $2^L - (2^L-1) = 1$ dimensional space. This statement is trivially true for $L=1$, since $B(1) = \widetilde{B}(1)$, and we have proved it explicitly for $L=2$, so we will show that if it's valid for $L-1$, then it's valid for $L$.

Let's consider the subset of rows of Eq.~(\ref{eq:ldbe_matrix_decomp}) where $B(L) = \widetilde{B}(L)$. Among these equations, we will denote by ``set 0'' those belonging to the first block row, and ``set 1'' those belonging to the second block row. As we can see from Eq.~(\ref{eq:b_tilde_decomp}), this leaves out one equation in the last block row, which is:
\begin{equation}
	\psi(0,0,\dots,0) - \psi(1,0,\dots,0) = \delta_L(0).
	\label{eq:last_row_b_tilde}
\end{equation}
By induction hypothesis, all equations in the first block row of Eq.~(\ref{eq:ldbe_matrix_decomp}) can be written as linear combinations of set 0. Furthermore, if $\delta_p(N)$ satisfies property (\ref{eq:ldbe_property}), also $\delta_p(N) + c$ satisfies the same property, so by induction hypothesis, also all equations in the second block row of Eq.~(\ref{eq:ldbe_matrix_decomp}) can be written as linear combinations of set 1. So any row in the first and second block row in Eq.~(\ref{eq:ldbe_matrix_decomp}) can be written as linear combinations of those equations involving $\widetilde{B}(L)$. 

We are left to show that equations in the last block row of Eq.~(\ref{eq:ldbe_matrix_decomp}) can be written as linear combinations of equations involving $\widetilde{B}(L)$. Thanks to what we have just demonstrated, this is is equivalent to showing that the equations in the last block row are linear combinations of the first two block rows of Eq.~(\ref{eq:ldbe_matrix_decomp}), and of Eq.~(\ref{eq:last_row_b_tilde}). The equations of the last block row of Eq.~(\ref{eq:ldbe_matrix_decomp}) are of the form
\begin{equation}
	\psi(n_L=0,\{n_i\}) - \psi(n_L=1,\{n_i\}) = \delta_L(\widetilde{N}).
	\label{eq:ldbe_last_block}
\end{equation}
Given any fixed set of $\{n_i\}$, which fixes $\widetilde{N} = \sum_{i\neq L}n_i$, we will now show how to obtain Eq.~(\ref{eq:ldbe_last_block}) by relating $\psi(n_L=0,\{n_i\})$ to $\psi(n_L=1,\{n_i\})$ only using the equations of the first two block rows of Eq.~(\ref{eq:ldbe_matrix_decomp}) and Eq.~(\ref{eq:last_row_b_tilde}). We can first relate $\psi(n_L=0,\{n_i\})$ to $\psi(n_L=0,\{0\})$ using the equations in the first block row $\widetilde{N}$ times, each time changing a non null $n_i$ to zero. We will obtain the following result:
\begin{equation}
	\psi(n_L=0,\{n_i\}) - \psi(n_L=0,\{0\}) = - \sum\limits_{i=1}^{\widetilde{N}} \delta_{p_i}(\widetilde{N} - i),
\end{equation}
where $p_i$ is the energy level whose occupation number has been changed to zero at step $i$; the argument $\widetilde{N}-i$ is given by the fact that at each step we use a linearized DBE where $\sum_{i\neq p_i} n_i$ decreases by one. Using Eq.~(\ref{eq:last_row_b_tilde}) we obtain
\begin{equation}
	\psi(n_L=0,\{n_i\}) - \psi(n_L=1,\{0\}) = \delta_L(0) - \sum\limits_{i=1}^{\widetilde{N}} \delta_{p_i}(\widetilde{N} - i).
\end{equation}
Now we can relate $\psi(n_L=1,\{0\})$ to $\psi(n_L=1,\{n_i\})$ using the equations of the second block row to set each non null $n_i$ back to $1$. If we do this in the exact same order as we did in the previous step, we will obtain
\begin{multline}
	\psi(n_L=0,\{n_i\}) - \psi(n_L=1,\{n_i\})  \\
	= \sum\limits_{i=1}^{\widetilde{N}} \delta_{p_i}(i) + \delta_L(0) - \sum\limits_{i=1}^{\widetilde{N}} \delta_{p_i}(\widetilde{N} - i).
\end{multline}
The ``new'' term has the same set of $\{p_i\}$, and the argument $i$ is given by the fact that $\sum_{i\neq p_i} n_i$ increases by one at each step, and it starts from $1$ since now $n_L=1$. Eq. (\ref{eq:ldbe_last_block}) is of the same form, therefore the equations on the last block row can be expressed as linear combinations of $\widetilde{B}(L)$ if and only if they represent the same equation, which requires:
\begin{equation}
	 \delta_L(\widetilde{N}) - \delta_L(0) = \sum\limits_{i=1}^{\widetilde{N}} \left[ \delta_{p_i}(i) - \delta_{p_i}(\widetilde{N}-i) \right].
\end{equation}
Using property (\ref{eq:ldbe_property}), this last equation reduces to $\widetilde{N}=\widetilde{N}$, concluding the proof by induction. 

\section{Non-interacting QD: Non-linear effects}
\label{oneEL}

In this Appendix we will study the transport coefficients, maximum and peak efficiency and efficiency at peak power of a two-terminal non-interacting QD.
We will first consider a single-level QD and then a two-levels QD (with energy-independent tunneling rates).

Let us first consider a QD with a single energy level $E_1$. In this case, the kinetic equations with $E_C=0$ reduces to a single simple equation, so it is possible to compute the currents. Using the two-terminal notation introduced in Secs.~\ref{2terms} and \ref{chap:beylinear} we obtain
\begin{equation}
	\begin{aligned}
		&J^c = e\gamma(f_2 - f_1), \\
		&J^h_2 =  ( \bar{\Delta}_\text{min} - \theta_0 eV)\gamma(f_2 - f_1),
	\end{aligned}
	\label{eq:curr_1liv}
\end{equation}
where
\begin{equation}
\begin{aligned}
	&f_1 = f\left(\frac{\bar{\Delta}_{\text{min}} +(1-\theta_0)eV}{k_BT_1} \right), \\
	&f_2 = f\left(\frac{\bar{\Delta}_{\text{min}} -\theta_0 eV}{k_BT_2} \right), \\
	&\bar{\Delta}_{\text{min}} = E_1 - \mu,
\end{aligned}
\end{equation}
with the Fermi function $f(x)=[1+\exp(x)]^{-1}$. Note that the same result could be obtained using the Landauer-B\"uttiker scattering formalism with a narrow in energy single-level QD transmission probability. Using Eq.~(\ref{eq:curr_1liv}), we can compute
\begin{multline}
	G \equiv \at{\frac{\partial J^c}{\partial V}}{\Delta T=0} = \\
	\frac{e^2\gamma}{4k_BT}\left[ \frac{\theta_0}{\cosh^2\left(\frac{\bar{\Delta}_{\text{min}} - \theta_0 eV}{2k_BT}\right)} + \frac{1-\theta_0} {\cosh^2\left(\frac{\bar{\Delta}_{\text{min}} + (1-\theta_0) eV}{2k_BT}\right)} \right] .
\end{multline}
Note that the value of $\theta_0$ can be determined by measuring $G$.

We can also compute $S$ using Eq.~(\ref{eq:curr_1liv}). The condition $J^c=0$ is satisfied when the arguments of the two Fermi distributions $f_1$ and $f_2$
are equal, so we obtain
\begin{equation}
	S \equiv -\at{\frac{V}{\Delta T}}{J^c=0} = \frac{1}{eT^*} \bar{\Delta}_{\text{min}},
\end{equation}
where
\begin{equation}
	T^* = \theta_0 T_1 + (1-\theta_0)T_2
\end{equation}
is the average reservoir temperature, weighed with $\theta_0$. As we can see, the slope of $S(\mu)$ beyond the linear response regime is strongly determined by $\theta_0$. In the linear response regime $T^* \simeq T$, where $T$ is the average reservoirs' temperature, so once again we obtain the result we obtained in the quantum limit linear response regime of an interacting QD neglecting the fine structure oscillations, due to other energy levels.

\begin{figure}[!htb]
	\centering
	\includegraphics[width=1\columnwidth]{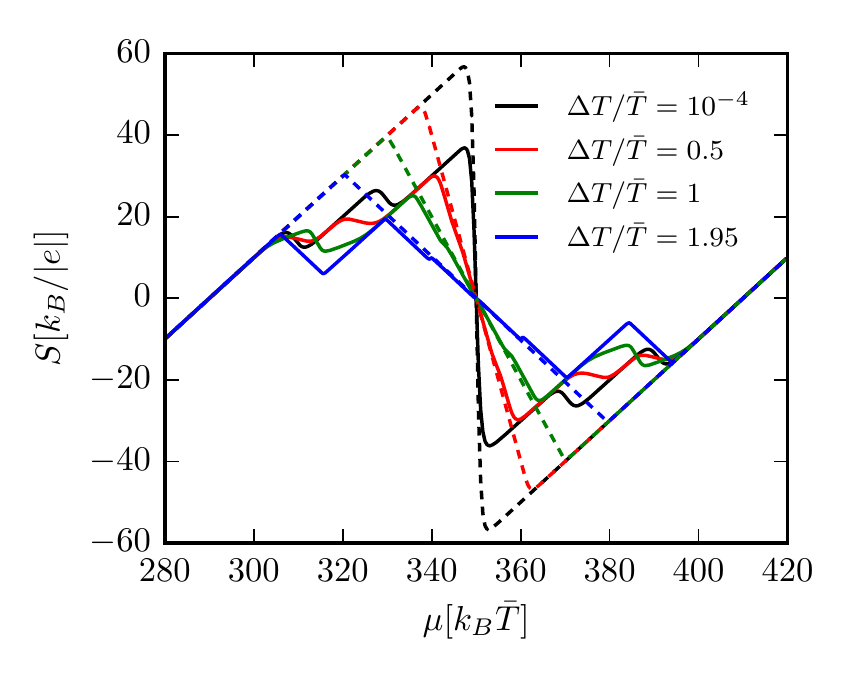}
	\caption{(Color online) Thermopower $S$ as a function of $\mu$ for a two-level non-interacting QD with energies $E_1=290k_B\bar{T}$ and $E_2=410k_B\bar{T}$ (dashed lines) and of a multi-level interacting QD (solid lines), considered in Sec.~\ref{chap:beylinear}, with the same parameters used in Fig.~\ref{fig-S}.}
	\label{S2levels}
\end{figure}

It is now interesting to study the influence of a second energy level on $S$. In Fig.~\ref{S2levels} we show the comparison between the thermopower of a non-interacting two energy level QD (dashed lines) and the multilevel interacting QD considered in Fig.~\ref{fig-S} (solid lines), for different values of $\Delta T$. For the sake of comparison, we choose the value for the two energy levels $E_1$ and $E_2$ to match the two dominant transition energies of the multilevel interacting QD, in particular $E_1 = \mu_{N=3} = 290$ $k_B\bar{T}$ and $E_2 = \mu_{N=4} = 410$ $k_B\bar{T}$.
On one hand, the non-interacting thermopower shows a linear dependence on $\mu$ when $\mu$ is close to $E_1$ or $E_2$, whose slope is independent of $\Delta T$ (only because $\theta_0=1/2$) and is the same as the interacting QD when $|\Delta_{\text{min}}| < \Delta E$. On the other, when $\mu$ is close to the middle point $\mu^* = (E_1+E_2)/2$ between $E_1$ and $E_2$, the thermopower depends linearly on $\mu$ with a negative slope ($S\simeq \alpha\delta\mu$, with $\delta\mu$ being a small displacement with respect to $\mu^*$) which depends on the temperatures as 
\begin{equation}
\alpha= \frac{1}{e\Delta T} \frac{T_1 \cosh^2 (\frac{\Delta \varepsilon}{4k_BT_1})-T_2 \cosh^2 (\frac{\Delta \varepsilon}{4k_BT_2})}{\theta_0 T_1 \cosh^2 (\frac{\Delta \varepsilon}{4k_BT_1})+(1-\theta_0)T_2 \cosh^2 (\frac{\Delta \varepsilon}{4k_BT_2})} ,
\label{eq:alpha_def}
\end{equation}
where $\Delta\varepsilon =E_2-E_1$. Eq.~(\ref{eq:alpha_def}) has been found by expanding the charge current around $\mu=\mu^*$. Once again, the slope of the interacting and non-interacting QDs are the same in this region. In between instead the fine structure oscillations of the interacting QD create a substantial difference that causes the maximum value of $S$ to decrease.

At last we can compute the Peltier coefficient for a single energy level QD using Eq.~(\ref{eq:curr_1liv}):
\begin{equation}
	\Pi \equiv \at{\frac{J^h_2}{J^c}}{\Delta T=0} = \frac{\bar{\Delta}_{\text{min}}}{e} - \theta_0 V.
	\label{eq:pi_1liv}
\end{equation}
Surprisingly, we would have obtained the same result also without setting $\Delta T=0$. On the other hand, it strongly depends on $V$: increasing the voltage shifts $\Pi$ by $\theta_0 V$, so measuring this shift would allow to measure $\theta_0$. Furthermore, if we compute $\Pi$ in the linear response regime, setting $V=0$ in Eq.~(\ref{eq:pi_1liv}), we can explicitly verify that $\Pi = TS$ is respected. If we set $\theta_0 = 0$, we see that $\Pi = T^*S$, even beyond the linear response regime, for arbitrary temperature and voltage differences.
Interestingly, in the presence of two levels the Peltier coefficient beyond linear response $\Pi$ still satisfies the relation $(\Pi+\theta_0V)=TS_{\text{lin}}$, where $S_{\text{lin}}$ is the linear-response thermopower (black dashed curve in Fig.~\ref{S2levels}), so that $(\Pi+\theta_0V)$ does not depend on $V$.

The linearization of $G$ and $S$ for a single-level non-interacting QD yields the same results obtained for a multilevel interacting QD, in the quantum limit linear response regime, if we restrict $|\Delta_{\text{min}}| < \Delta E$. We could expect this result since considering a single energy level is intuitively equivalent to sending $\Delta E$ and $E_C$ to infinity in the interacting QD.
On the contrary, a single level QD model cannot be used to estimate $K$ since it predicts $K=0$ [in fact, from Eq.~(\ref{eq:curr_1liv}) we have that $J^h_2 \propto J^c$, so $K$, caluclated at $J^c=0$, vanishes].

We now want to study the power and efficiency of a single-level QD. Considering $\Delta T = T_2-T_1>0$ and inserting Eq.~(\ref{eq:curr_1liv}) into the definitions in Eqs.~(\ref{eq:p_def}) and (\ref{eq:eta_def}), we can write the power and efficiency as
\begin{align}
&	P = -V J^c = -\gamma eV (f_2 - f_1), \label{eq:one_level_p} \\
&	\eta = \frac{P}{J_2^h} = \frac{eV}{\theta_0 eV - \bar{\Delta}_{\text{min}}}. \label{eq:one_level_eta}
\end{align}
We will consider a fixed temperature difference $\Delta T = T_2-T_1>0$, and a variable $V$ such that the system behaves as a heat engine ($P>0$). The power is positive when $V \in [0,V_{\text{stop}}]$, where $V_{\text{stop}}$ is the non zero voltage that creates a null charge current. Imposing this condition we find
\begin{equation}
	V_{\text{stop}} = \frac{\bar{\Delta}_{\text{min}}}{-e} \frac{\eta_C}{1 - \theta_0 \eta_C}.
\end{equation}
Without loss of generality, we can specify our analysis to the region where $\mu < E_1$, $E_1$ being the energy of the single energy level; thus $\bar{\Delta}_{\text{min}}>0$, so $V_{\text{stop}}$ is positive, and our system will behave as a heat engine when
\begin{equation} 
	0 \leq V \leq V_{\text{stop}} = \frac{\bar{\Delta}_{\text{min}}}{-e} \frac{\eta_C}{1 - \theta_0 \eta_C}.
	\label{eq:one_level_v_range}
\end{equation}

Let us now discuss the peak efficiency of the system. We have to maximize Eq.~(\ref{eq:one_level_eta}) with respect to $\bar{\Delta}_{\text{min}}$ and $V$, at fixed $\Delta T>0$, respecting $\bar{\Delta}_{\text{min}} >0$ and Eq.~(\ref{eq:one_level_v_range}). Since $\eta$ is a growing function of $V$ for $\bar{\Delta}_{\text{min}} > 0$, $\eta$ will be maximum when computed at the highest allowed voltage, $V_{\text{stop}}$. Inserting $V_{\text{stop}}$ in Eq.~(\ref{eq:one_level_eta}) yields:
\begin{equation}
	\eta_{\text{max}} = \eta_C.
\end{equation} 
Thus a single level QD in the sequential tunneling regime always achieves $\eta_{\text{max}} =\eta_C$, regardless of the the temperature difference, $\theta_0$ and the distance between $\mu$ and $E_1$ \cite{bib:linke,bib:humphrey}. 
Using Eq.~(\ref{eq:one_level_p}) we can compute the power when $\eta$ is maximum, i.e. when $V = V_{\text{stop}}$: this yields $P=0$. These results agrees with the expectation that Carnot's efficiency is reached when the heat exchange is ``reversible'', thus when the power is vanishingly small. 

Let us now study the efficiency at peak power, $\eta(P_{\text{peak}})$. $P_{\text{peak}}$ is obtained by maximizing the power with respect to $\bar{\Delta}_{\text{min}}$ and $V$, at fixed $\Delta T>0$, imposing $\bar{\Delta}_{\text{min}}>0$ and Eq.~(\ref{eq:one_level_v_range}). By imposing this request we obtain two coupled equations that cannot be solved analytically \cite{esposito2009}. However, the Fermi function is always evaluated when the argument is positive, and we approximate it with its exponential tail. By doing so, we obtain
\begin{equation}
\begin{aligned}
	&\eta(P_{\text{peak}}) = \eta_C \frac{\eta_C}{\eta_C - (1-\eta_C)\ln{(1-\eta_C)}}, \\
	&\begin{multlined}
			P_{\text{peak}} = \gamma \bar{e} k_B \Delta T  \\
			\times \frac{\eta_C}{\left[1+\bar{e}(1-\eta_C)^{1-1/\eta_C}\right]\left[ \bar{e} + (1-\eta_C)^{1/\eta_C} \right] },
	\end{multlined}	
\end{aligned}
	\label{eq:eta_pmax_1liv}
\end{equation}
where $\bar{e}$ is Napier's constant. 
These equations provide an approximate expression of $P_{\text{peak}}$ and $\eta(P_{\text{peak}})$ for a single level QD, valid for any reservoir temperatures.
Note how $\eta(P_{\text{peak}})$ only depends on $\eta_C$, while $P_{\text{peak}}$ depends on both $T_1$ and $T_2$ through $\eta_C$ and $\Delta T$. Note $\eta(P_{\text{peak}}) \rightarrow \eta_C/2$ as $\eta_C \rightarrow 0$, as expected from the fact that $ZT\to\infty$ for a narrow transmission probability \cite{bib:mahan}.
\begin{figure}[!tb]
	\centering
	\includegraphics[width=1\columnwidth]{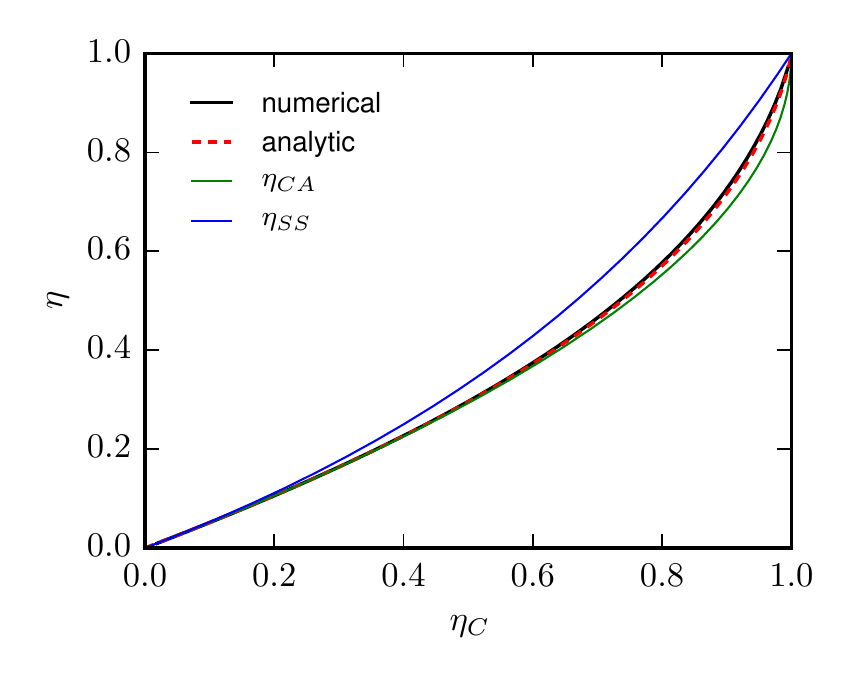}
	\caption{(Color online) Comparison between $\eta(P_{\text{peak}})$ computed numerically and using Eq.~(\ref{eq:eta_pmax_1liv}). Also $\eta_{CA}$ and $\eta_{SS}$ are displayed.}
	\label{fig:eta_pmax_1liv}
\end{figure}
As we can see in Fig.~\ref{fig:eta_pmax_1liv}, there is a good agreement between $\eta(P_{\text{peak}})$ given in Eq.~(\ref{eq:eta_pmax_1liv}) and a numerical calculation: the analytic expression slightly underestimates $\eta(P_{\text{peak}})$. Furthermore, we can see that the efficiency at peak power goes beyond $\eta_{CA}$, while it remains under $\eta_{SS}$. We have also verified that $P_{\text{peak}}$ given in Eq. (\ref{eq:eta_pmax_1liv}) is in good agreement with the numerical calculation. 

To further assess the validity of the approximate analytical formulas (\ref{eq:eta_pmax_1liv}), we expand them around $\eta_C =0$ and $\eta_C=1$, and compare the obtained results with the exact expansions known in these limiting 
cases \cite{esposito2009}. An expansion of Eq.~(\ref{eq:eta_pmax_1liv})
around $\eta_C=0$ yields
\begin{equation}
\begin{aligned}
	&\eta(P_{\text{peak}}) = \frac{\eta_C}{2} + \frac{\eta_C^2}{8} + \frac{7}{96}\eta_C^3 + O(\eta_C^4), \\
	&P_{\text{peak}} = \gamma k_BT \frac{\bar{e}^2}{(1+\bar{e}^2)^2} \left(\frac{\Delta T}{T}\right)^2 + O\left(\left(\Delta T/T\right)^3\right),
\end{aligned}
\end{equation}
where $T$ is the average temperature in the linear response regime. 
Our result for $\eta(P_{\text{peak}})$ has to be compared with 
exact expansion of Ref.~\onlinecite{esposito2009},
\begin{equation}
	\eta(P_{\text{peak}}) = \frac{\eta_C}{2} + \frac{\eta_C^2}{8} + \frac{7}{96}(1+0.0627)\eta_C^3 + O(\eta_C^4).
\end{equation}
\begin{figure}[!ht]
\centering
\includegraphics[width=1\columnwidth]{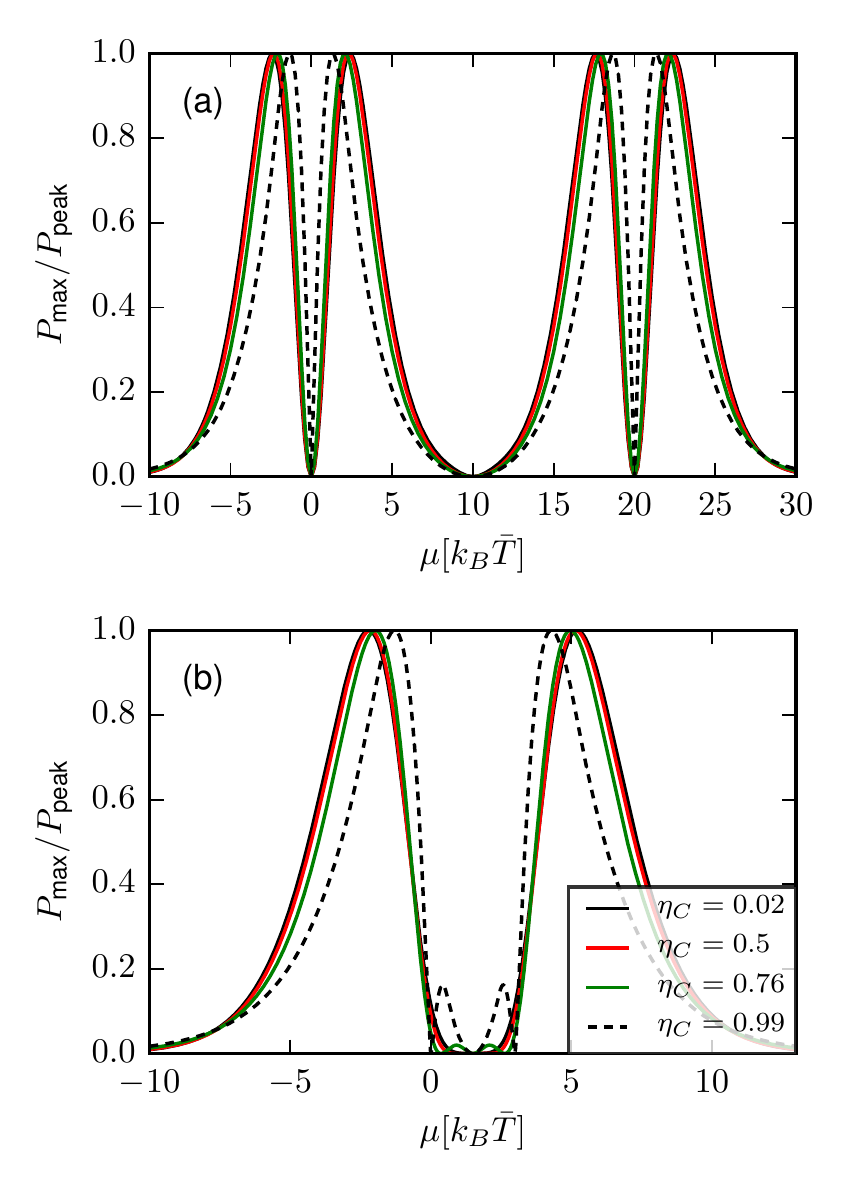}
\caption{(Color online) Maximum power, normalized to the peak value, plotted as a function of $\mu$ for different values of $\eta_C$ with $\Delta E=20 \,k_B\bar{T}$ and $E_C=0$ (top panel) and with $\Delta E=3\,k_B\bar{T}$ and $E_C=0$ (bottom panel). Tunneling rates are $\hbar\Gamma_1(p) = \hbar\Gamma_2(p) = 0.01k_B\bar{T}$.}
\label{figPdegNI}
\end{figure}
As we can see the first two orders are exactly the same, while the third order only has $\approx 6\%$ correction, confirming that Eq.~(\ref{eq:eta_pmax_1liv}) slightly underestimates the exact result. The expansion of $P_{\text{peak}}$ instead confirms that in the linear response regime the peak power depends on $\Delta T^2$. Furthermore, evaluating the coefficient numerically yields $P_{\text{peak}} \approx 0.105 \gamma k_BT (\Delta T/T)^2$, which is in very good agreement with the result obtained in Eq. (\ref{eq:q_ql}) for a multilevel QD in the quantum limit linear response regime (note that $P_{\text{peak}} = Q^*\Delta T^2/4$). 

An expansion of Eq.~(\ref{eq:eta_pmax_1liv}) around $\eta_C=1$ yields
\begin{equation}
	\eta(P_{\text{peak}}) \approx  1 + (1-\eta_C)\ln{(1-\eta_C)},
\end{equation}
\begin{equation}
	P_{\text{peak}} \approx \frac{\gamma k_B\Delta T}{1+\bar{e}}\left( 1 + \frac{\bar{e}}{1+\bar{e}}(1-\eta_C)\ln{(1-\eta_C)} \right), \label{eq:p_peak_1liv}
\end{equation}
to be compared with the exact 
expansion \cite{esposito2009}
\begin{equation}
	\eta(P_{\text{peak}}) \approx 1 + \frac{ (1-\eta_C)\ln{(1-\eta_C)} }{1.278} .
\end{equation}
There is a good agreement between the two formulas, and we can see that Eq.~(\ref{eq:eta_pmax_1liv}) slightly underestimates $\eta(P_{\text{peak}})$. 
We stress that the expression of $P_{\text{peak}}$ for $\eta_C\approx 1$, Eq.~(\ref{eq:p_peak_1liv}), shows that the peak power is proportional to $\Delta T$, as opposed to $\Delta T^2$ as in the linear response regime. Furthermore the peak power approaches its maximum value given by $P_{\text{peak}} = \gamma k_B\Delta T/(1+\bar{e})$ when $\eta_C=1$ . 

At last, we have computed numerically the maximum power $P_{\text{max}}$ in the case of a two-level non-interacting QD. $P_{\text{max}}$ is computed by maximizing the power only over $V$; the peak of $P_{\text{max}}$ as a function of $\mu$ will thus yield $P_{\text{peak}}$. In Fig.~\ref{figPdegNI} $P_{\text{max}}/P_{\text{peak}}$ is plotted as a function of $\mu$ for $\Delta E = 20\,k_BT$ and $\Delta E =2\,k_BT$ respectively in panel (a) and (b). Each curve corresponds to a different value of $\eta_C$, starting from the linear-response case (black solid curve) to the extremely non-linear behavior (black dashed curve) at $\eta_C = 0.99$. In panel (a), representing the quantum limit, all curves nearly coincide with the linear-response one (apart for very large values of $\eta_C$).
For each value of $\eta_C$ the height of the four peaks is equal, contrary to the case where interaction is present (see Fig~\ref{figPdeg}). This is to be expected, since the two energy levels do not influence each other in the quantum limit in the absence of interaction, so the behavior is essentially dictated by a single level. In the bottom panel instead, representing a case away from the quantum limit, the external peaks are much higher with respect to the internal peaks, especially for small values of $\eta_C$. 
This is due to the fact that $\Delta E$ is of the order of $k_B\bar{T}$ and
therefore we cannot consider a single energy level at a time, and this produces the peak asymmetry.

\end{appendix}


\begin{thebibliography}{90}

\bibitem{bib:dresselhaus} 
	M. S. Dresselhaus, G. Chen, M. Y. Tang, R. G. Yang, H. Lee, D. Z. Wang, Z. F. Ren, J. P. Fleurial, and P. Gogna,
\href {http://dx.doi.org/10.1002/adma.200600527}{Adv. Mater. {\bf 19}, 1043 (2007)}.

\bibitem{bib:snyder} 
G. J. Snyder and E. S. Toberer, 
\href{http://www.nature.com/nmat/journal/v7/n2/full/nmat2090.html}{Nat. Mater. {\bf 7}, 105 (2008)}.

\bibitem{bib:vineis} 
	C. J. Vineis, A. Shakouri, A. Majumdar, and M. G. Kanatzidis, 
\href{http://onlinelibrary.wiley.com/doi/10.1002/adma.201000839/abstract}{Adv. Mater. {\bf 22}, 3970 (2010)}.
	
\bibitem{PRreview} 
G. Benenti, G. Casati, K. Saito, and R. Whitney, 
\href{https://arxiv.org/abs/1608.05595}{arXiv:1608.05595}.

\bibitem{Hochbaum} 
A. I. Hochbaum, R. Chen, R. D. Delgado, W. Liang, E. C. Garnett, 
M. Najarian, A. Majumdar, and P. Yang,
\href{http://www.nature.com/nature/journal/v451/n7175/abs/nature06381.html}{Nature {\bf 451}, 163 (2008)}.

\bibitem{bib:hicks}  
L. D. Hicks, T. C. Harman, and M. S. Dresselhaus, 
\href{http://aip.scitation.org/doi/10.1063/1.110207}{Appl. Phys. Lett. 
{\bf 63}, 3230 (1993)}.

\bibitem{bib:mahan} 
G. D. Mahan and J. O. Sofo, 
\href{https://www.ncbi.nlm.nih.gov/pmc/articles/PMC38761/}{Proc. Natl. Acad. 
Sci. USA {\bf 93}, 7436 (1996)}.

\bibitem{bib:beenakker1} C. W. J Beenakker, 
\href{http://journals.aps.org/prb/abstract/10.1103/PhysRevB.44.1646}{Phys. Rev. B {\bf 44}, 1646 (1991)}.

\bibitem{bib:beenakker2} C. W. J. Beenakker and A. A. M. Staring, 
\href{http://journals.aps.org/prb/abstract/10.1103/PhysRevB.46.9667}{Phys. Rev. B {\bf 46}, 9667 (1992)}.

\bibitem{fazio2001}
D. Boese, and R. Fazio,
\href{https://doi.org/10.1209/epl/i2001-00559-8}{Europhys. Lett. {\bf 56}, 576 (2001)}.

\bibitem{andreev2001}
A. V. Andreev, and K. A. Matveev,
\href{https://doi.org/10.1103/PhysRevLett.86.280}{Phys. Rev. Lett. {\bf 86}, 280 (2001)}.

\bibitem{turek2002}
M. Turek, and K. A. Matveev,
\href{https://doi.org/10.1103/PhysRevB.65.115332}{Phys. Rev. B {\bf 65}, 115332 (2002)}.

\bibitem{koch2004}
J. Koch, F. von Oppen, Y. Oreg, and E. Sela, 
\href{https://doi.org/10.1103/PhysRevB.70.195107}{Phys. Rev. B {\bf 70}, 195107 (2004)}.

\bibitem{kubala2006}
B. Kubala, and J. K{\"o}nig,
\href{https://doi.org/10.1103/PhysRevB.73.195316}{Phys. Rev. B {\bf 73}, 195316 (2006)}.

\bibitem{bib:zianni_2007} X. Zianni, 
\href{http://journals.aps.org/prb/abstract/10.1103/PhysRevB.75.045344}{Phys. Rev. B {\bf 75}, 045344 (2007)}.

\bibitem{jukka2008}
B. Kubala, J. K\"{o}nig, and J. Pekola,
\href{https://doi.org/10.1103/PhysRevLett.100.066801}{Phys. Rev. Lett. {\bf 100}, 066801 (2008)}.

\bibitem{bib:zianni_2008} 
X. Zianni,
\href{http://journals.aps.org/prb/abstract/10.1103/PhysRevB.78.165327}{Phys. Rev. B {\bf 78}, 165327 (2008)}.

\bibitem{jacquet2009}
P. A. Jacquet,
\href{https://doi.org/10.1007/s10955-009-9697-1}{J. Stat. Phys. {\bf 134}, 709 (2009)}.

\bibitem{costi2010}
T. A. Costi, and V. Zlati\'{c},
\href{https://doi.org/10.1103/PhysRevB.81.235127}{Phys. Rev. B {\bf 81}, 235127 (2010)}.

\bibitem{billings2010}
G. Billings, A. D. Stone, and Y. Alhassid,
\href{https://doi.org/10.1103/PhysRevB.81.205303}{Phys. Rev. B {\bf 81}, 205303 (2010)}.

\bibitem{mani2011}
P. Mani, N. Nakpathomkun, E. A. Hoffmann, and H. Linke,
\href{https://dx.doi.org/10.1021/nl202258f}{Nano Lett. {\bf 11}, 4679 (2011)}.

\bibitem{wohlman2012}
O. Entin-Wohlman, and A. Aharony,
\href{https://doi.org/10.1103/PhysRevB.85.085401}{Phys. Rev. B {\bf 85}, 085401 (2012)}.

\bibitem{rejec2012}
T. Rejec, R. \v{Z}itko, J. Mravlje, and Anton Ram\v{s}ak,
\href{https://doi.org/10.1103/PhysRevB.85.085117}{Phys. Rev. B {\bf 85}, 085117 (2012)}.

\bibitem{sanchezPRB2013}
R. L\'{o}pez, and David S\'{a}nchez,
\href{http://doi.org/10.1103/PhysRevB.88.045129}{Phys. Rev. B {\bf 88}, 045129 (2013)}.

\bibitem{kennes2013}
D. M. Kennes, and V. Meden,
\href{https://doi.org/10.1103/PhysRevB.87.075130}{Phys. Rev. B {\bf 87}, 075130 (2013)}.

\bibitem{dutt2013}
P. Dutt, and K. Le Hur, 
\href{https://doi.org/10.1103/PhysRevB.88.235133}{Phys. Rev. B {\bf 88}, 235133 (2013)}.

\bibitem{bjorn2013}
R. S\'{a}nchez, B. Sothmann, A. N. Jordan, and M. B\"uttiker,
\href{https://doi.org/10.1088/1367-2630/15/12/125001}{New J. Phys. {\bf 15}, 125001 (2013)}.

\bibitem{muralidharan2013}
B. Muralidharan, and M. Grifoni,
\href{https://doi.org/10.1103/PhysRevB.88.045402}{Phys. Rev. B {\bf 88}, 045402 (2013)}.

\bibitem{Sanchez2014} 
M. A. Sierra and D. S\'{a}nchez,
\href{http://journals.aps.org/prb/abstract/10.1103/PhysRevB.90.115313}{Phys. Rev. B {\bf 90}, 115313 (2014)}.

\bibitem{zimbovskaya2016}
N. A. Zimbovskaya,
\href{http://dx.doi.org/10.1063/1.4922907 }{J. Chem. Phys. {\bf 142}, 244310 (2016)}.

\bibitem{karol2006} 
M. Krawiec and K. I. Wysoki\'{n}ski,
\href{http://journals.aps.org/prb/abstract/10.1103/PhysRevB.73.075307}{Phys. Rev. B {\bf 73}, 075307 (2006)}.

\bibitem{murphy2008}  
P. Murphy, S. Mukerjee, and J. Moore,
\href{http://dx.doi.org/10.1103/PhysRevB.78.161406}{Phys. Rev. B {\bf 78}, 161406 (2008)}.

\bibitem{dubi2009} 
Y. Dubi and M. Di Ventra,
\href{https://doi.org/10.1103/PhysRevB.79.081302}{Phys. Rev. B {\bf 79}, 081302(R) (2009)}.

\bibitem{esposito2009}  
M. Esposito, K. Lindenberg, and C. Van den Broeck,
\href{http://dx.doi.org/10.1209/0295-5075/85/60010}{Eurphys. Lett. {\bf 85}, 60010 (2009)}.

\bibitem{swirkowicz2009} 
R. \'Swirkowicz, M. Wierzbicki, and J. Barna\'s,
\href{http://dx.doi.org/10.1103/PhysRevB.80.195409}{Phys. Rev. B {\bf 80}, 195409 (2009)}.

\bibitem{kuo2009}
D. M.-T. Kuo,
\href{http://dx.doi.org/10.1143/JJAP.48.125005}{Jap. J. App. Phys. {\bf 48}, 125005 (2009)}.

\bibitem{leijnse2010} 
M. Leijnse, M. R. Wegewijs, and K. Flensberg,
\href{http://dx.doi.org/10.1103/PhysRevB.82.045412}{Phys. Rev. B {\bf 82}, 045412 (2010)}.

\bibitem{liu2010} 
J. Liu, Q. F. Sun, and X. C. Xie,
\href{http://dx.doi.org/10.1103/PhysRevB.81.245323}{Phys. Rev. B {\bf 81}, 245323 (2010)}.

\bibitem{imry2010} 
O. Entin-Wohlman, Y. Imry, and A. Aharony,
\href{http://journals.aps.org/prb/abstract/10.1103/PhysRevB.82.115314}{Phys. Rev. B {\bf 82}, 115314 (2010)}.

\bibitem{zianni2010} 
X. Zianni,
\href{http://dx.doi.org/10.1103/PhysRevB.82.165302}{Phys. Rev. B {\bf 82}, 165302 (2010)}.

\bibitem{nakpathomkun2010} 
N. Nakpathomkun, H. Q. Xu, and H. Linke,
\href{http://dx.doi.org/10.1103/PhysRevB.82.235428}{Phys. Rev. B {\bf 82}, 235428 (2010)}.

\bibitem{wierzbicki2010}
M. Wierzbicki, and R. \'{S}wirkowicz,
\href{https://doi.org/10.1088/0953-8984/22/18/185302}{J. Phys.: Condens. Matter {\bf 22}, 185302 (2010)}.

\bibitem{kuo2010}
D. M.-T. Kuo, and Y.-C. Chang,
\href{https://doi.org/10.1103/PhysRevB.81.205321}{Phys. Rev. B {\bf 81}, 205321 (2010)}.

\bibitem{buttiker2011} 
R. S\'{a}nchez and M. B\"{u}ttiker,
\href{https://doi.org/10.1103/PhysRevB.83.085428}{Phys. Rev. B {\bf 83}, 085428 (2011)}.

\bibitem{karlstrom2011} 
O. Karlstr\"om, H. Linke, G. Karlstr\"om, and A. Wacker,
\href{http://dx.doi.org/10.1103/PhysRevB.84.113415}{Phys. Rev. B {\bf 84}, 113415 (2011)}.

\bibitem{liu2011}
Y.-S. Liu, D.-B. Zhang, X.-F. Yang, and J.-F. Feng, 
\href{https://doi.org/10.1088/0957-4484/22/22/225201}{Nanotechnology {\bf 22}, 225201 (2011)}.

\bibitem{sothmann2012} B. Sothmann,  and M. B{\"u}ttiker, 
\href{http://dx.doi.org/10.1209/0295-5075/99/27001}{Europhys. Lett. {\bf 99}, 27001 (2012)}.

\bibitem{trocha2012} 
P. Trocha, and J. Barna\'s,
\href{http://dx.doi.org/10.1103/PhysRevB.85.085408}{Phys. Rev. B {\bf 85}, 085408 (2012)}.

\bibitem{muralindharan2012} 
B. Muralidharan and M. Grifoni,
\href{http://dx.doi.org/10.1103/PhysRevB.85.155423}{Phys. Rev. B {\bf 85}, 155423 (2012)}.

\bibitem{buttiker2013} 
A. N. Jordan, B. Sothmann, R. S\'{a}nchez, and M. B\"{u}ttiker,
\href{http://journals.aps.org/prb/abstract/10.1103/PhysRevB.87.075312}{Phys. Rev. B {\bf 87}, 075312 (2013)}.

\bibitem{kuo2013}
D. M.-T. Kuo, and Y.-C. Chang,
\href{https://doi.org/10.1088/0957-4484/24/17/175403}{Nanotechnology {\bf 24}, 175403 (2013)}.

\bibitem{weymann2013}
I. Weymann, and J. Barna\'{s},
\href{https://doi.org/10.1103/PhysRevB.88.085313}{Phys. Rev. B {\bf 88}, 085313 (2013)}.

\bibitem{Mazza2014}
F. Mazza, R. Bosisio, G. Benenti, V. Giovannetti, R. Fazio, and F. Taddei,
\href{http://dx.doi.org/10.1088/1367-2630/16/8/085001}{New J. Phys. {\bf 16}, 085001 (2014)}. 

\bibitem{agarwal2014}
A. Agarwal, and B. Muralidharan,
\href{http://dx.doi.org/10.1063/1.4888859}{Appl. Phys. Lett. {\bf 105}, 013104 (2014)}.

\bibitem{donsa2014}
S. Donsa, S. Andergassen, and K. Held,
\href{https://doi.org/10.1103/PhysRevB.89.125103}{Phys. Rev. B {\bf 89}, 125103 (2014)}.

\bibitem{tseng2015} 
Y.-C. Tseng, D. M.-T. Kuo, Y.-C. Chang, and C.-W. Tsai,
\href{https://arxiv.org/abs/1504.06082v2}{arXiv:1504.06082}. 

\bibitem{taylor2015} 
E. Taylor and D. Segal,
\href{http://dx.doi.org/10.1103/PhysRevB.92.125401}{Phys. Rev. B {\bf 92}, 125401 (2015)}.

\bibitem{szukiewicz2016} 
B. Szukiewicz, U. Eckern, and K. I. Wysoki\'{n}ski,
\href{http://dx.doi.org/10.1088/1367-2630/18/2/023050}{New. J. Phys. {\bf 18}, 023050 (2016)}.

\bibitem{perroni2016}
C. A. Perroni, D. Ninno, and V. Cataudella, 
\href{http://doi.org/10.1088/0953-8984/28/37/373001}{J. Phys.: Condens. Matter {\bf 28}, 373001 (2016)}.

\bibitem{de2016}  
B. De, and B. Muralidharan,
\href{http://journals.aps.org/prb/abstract/10.1103/PhysRevB.94.165416}{Phys. Rev. B {\bf 94}, 165416 (2016)}.

\bibitem{kuo2016}
D. M. T. Kuo, C.-C. Chen, and Y.-C. Chang,
\href{https://arxiv.org/abs/1701.04515}{arXiv:1701.04515}.

\bibitem{rossello2017}
G. Rossell\'{o}, R. L\'{o}pez, and R. S\'{a}nchez,
\href{https://arxiv.org/abs/1702.03110}{arXiv:1702.03110}.

\bibitem{sothmann2015}  
B. Sothmann, R. S\'anchez, and A. N. Jordan, 
\href{http://dx.doi.org/10.1088/0957-4484/26/3/032001}{Nanotechnology {\bf 26}, 032001 (2015)}.

\bibitem{molenkamp1992} 
L. W. Molenkamp, Th. Gravier, H. van Houten, O. J. A. Buijk, M. A. A. Mabesoone, and C. T. Foxon,
\href{https://doi.org/10.1103/PhysRevLett.68.3765}{Phys. Rev. Lett. {\bf 68}, 3765 (1992)}.

\bibitem{staring1993} 
A. A. M. Staring, L. W. Molenkamp, B. W. Alphenaar, H. van Houten, O. J. A. Buyk, M. A. A. Mabesoone, C. W. J. Beenakker, and C. T. Foxon, 
\href{http://dx.doi.org/10.1209/0295-5075/22/1/011}{Europhys. Lett. {\bf 22}, 57 (1993)}.

\bibitem{dzurak1993}
A. S. Dzurak, C. G. Smith, M. Pepper, D. A. Ritchie, J. E. F. Frost, G. A. C. Jones, and D. G. Hasko,
\href{http://dx.doi.org/10.1016/0038-1098(93)90819-9}{Solid State Commun. {\bf 87}, 1145 (1993)}.

\bibitem{dzurak1998}
A. S. Dzurak, C. G. Smith, C. H. W. Barnes, M. Pepper, L. Mart\'in-Moreno, C. T. Liang, D. A. Ritchie, and G. A. C. Jones,
\href{http://dx.doi.org/10.1016/S0921-4526(98)00115-X}{Physica B {\bf 249-251}, 281 (1998)}.

\bibitem{godijn1999} 
S. F. Godijn, S. M\"{o}ller, H. Buhmann, L. W. Molenkamp, and S. A. van Langen,
\href{https://doi.org/10.1103/PhysRevLett.82.2927}{Phys. Rev. Lett. {\bf 82}, 2927 (1999)}.

\bibitem{harman2002}
T. C. Harman, P. J. Taylor, M. P. Walsh, and B. E. LaForge,
\href{http://doi.org/10.1126/science.1072886}{Science {\bf 297}, 2229 (2002)}.

\bibitem{llaguno2003}
M. C. Llaguno, J. E. Fischer, A. T. Johnson, and J. Hone,
\href{http://dx.doi.org/10.1021/nl0348488}{Nano Lett. {\bf 4}, 45 (2003)}.

\bibitem{scheibner2005}
R. Scheibner, H. Buhmann, D. Reuter, M. N. Kiselev, and L. W. Molenkamp,
\href{https://doi.org/10.1103/PhysRevLett.95.176602}{Phys. Rev. Lett. {\bf 95}, 176602 (2005)}.

\bibitem{pogosov2006} 
A. G. Pogosov, M. V. Budantsev, R. A. Lavrov, A. E. Plotnikov, A. K. Bakarov, 
A. I. Toropov, and J. C. Portal,
\href{http://dx.doi.org/10.1134/S002136400603009X}{JETP Lett. {\bf 83}, 122 (2006)}.

\bibitem{reddy2007}
P. Reddy, S.-Y. Jang, R. A. Segalman, and A. Majumdar,
\href{http://dx.doi.org/10.1126/science.1137149}{Science {\bf 315}, 1568 (2007)}.

\bibitem{scheibner2007}
R. Scheibner, E. G. Novik, T. Borzenko, M. K\"{o}nig, D. Reuter, A. D. Wieck, H. Buhmann, and L. W. Molenkamp,
\href{https://doi.org/10.1103/PhysRevB.75.041301}{Phys. Rev. B {\bf 75}, 041301(R) (2007)}

\bibitem{scheibner2008} 
R. Scheibner, M. K{\"o}nig, D. Reuter, A. D. Wieck, C. Gould, H. Buhmann, and L. W. Molenkamp,   
\href{http://dx.doi.org/10.1088/1367-2630/10/8/083016}{New J. Phys. {\bf 10}, 083016 (2008)}.

\bibitem{svensson2012}
S. F. Svensson, A. I. Persson, E. A. Hoffmann, N. Nakpathomkun, H. A. Nilsson, H. Q. Xu, L. Samuelson, and H. Linke, 
\href{http://doi.org/10.1088/1367-2630/14/3/033041}{New J. Phys. {\bf 14}, 033041 (2012)}.

\bibitem{svensson2013} 
S. F. Svensson, E. A. Hoffmann, N. Nakpathomkun, P. M. Wu, H. Q. Xu, H. A. Nilsson, D. S\'{a}nchez, V. Kashcheyevs, and H. Linke,
\href{http://dx.doi.org/10.1088/1367-2630/15/10/105011}{New J. Phys. {\bf 15}, 105011 (2013)}.

\bibitem{thierschmann2013}
H. Thierschmann, M. Henke, J. Knorr, L. Maier, C, Heyn, W. Hansen, H. Buhmann, and L. W. Molenkamp,
 \href{https://doi.org/10.1088/1367-2630/15/12/123010}{New J. Phys. {\bf 15}, 123010 (2013)}.

\bibitem{matthews2014} 
J. Matthews, F. Battista, D. S\'{a}nchez, P. Samuelsson, and H. Linke,
\href{https://doi.org/10.1103/PhysRevB.90.165428}{Phys. Rev. B {\bf 90}, 165428 (2014)}.

\bibitem{glattli2015} 
B. Roche, P. Roulleau, T. Jullien, Y. Jompol, I. Farrer, D. A. Ritchie,
and D. C. Glattli,
\href{http://www.nature.com/articles/ncomms7738}{Nature Commun. {\bf 6}, 6738 (2015)}.

\bibitem{hartmann2015} 
F. Hartmann, P. Pfeffer, S. H\"{o}fling, M. Kamp, and L. Worschech
\href{http://journals.aps.org/prl/abstract/10.1103/PhysRevLett.114.146805}{Phys. Rev. Lett. {\bf 114}, 146805 (2015)}.

\bibitem{molenkamp2015} 
H. Thierschmann, R. S\'{a}nchez, B. Sothmann, F. Arnold, C. Heyn, W. Hansen, H. Buhmann, and L. W. Molenkamp,
\href{http://www.nature.com/nnano/journal/v10/n10/full/nnano.2015.176.html}{Nature Nanotech. {\bf 10}, 854 (2015).}

\bibitem{svilansPhysE2016}
A. Svilans, A. M. Burke, S. F. Svensson, M. Leijnse, and H. Linke,
\href{http://dx.doi.org/10.1016/j.physe.2015.10.007}{Physica E {\bf 82}, 34 (2016)}.

\bibitem{svilans2016}
A. Svilans, M. Leijnse, and H. Linke,
\href{http://dx.doi.org/10.1016/j.crhy.2016.08.002}{C. R. Phys. {\bf 17}, 1096 (2016)}.

\bibitem{bib:yvon} J. Yvon, Proceedings of the International Conference on Peaceful Uses of Atomic Energy (United Nations, New York, 1955).

\bibitem {bib:chambadal} P. Chambadal, \textit{Les centrales nucleaires} (Armand Colin, Paris, 1957).

\bibitem{bib:novikov} I. I. Novikov, 
\href{http://www.sciencedirect.com/science/article/pii/0891391958902444}{J. Nuclear Energy II {\bf 7}, 125 (1958)}.

\bibitem{bib:curzon} F. Curzon and B. Ahlborn, 
\href{http://dx.doi.org/10.1119/1.10023}{Am. J. Phys. {\bf 43}, 22 (1975)}.

\bibitem{bib:broeck} C. Van den Broeck, 
\href{http://journals.aps.org/prl/abstract/10.1103/PhysRevLett.95.190602}{Phys. Rev. Lett. {\bf 95}, 190602 (2005)}.

\bibitem{Sanchez2013}
D. S\'{a}nchez and R. L\'{o}pez,
\href{http://journals.aps.org/prl/abstract/10.1103/PhysRevLett.110.026804}{Phys. Rev. Lett. {\bf 110}, 026804 (2013)}.

\bibitem{meair2013}
J. Meair, and Ph. Jacquod,
\href{http://dx.doi.org/10.1088/0953-8984/25/8/082201}{J. Phys.: Condens. Matter {\bf 25}, 082201 (2013)}.

\bibitem{whitney2013}
R. S. Whitney,
\href{http://dx.doi.org/10.1103/PhysRevB.87.115404}{Phys. Rev. B {\bf 87}, 115404 (2013)}.

\bibitem{bib:nazarov} Y. V. Nazarov and Y. M. Banter, \textit{Quantum Transport} (Cambridge, New York, 2009). 

\bibitem{bib:ioffe2} A. F. Ioffe, \textit{Semiconductor Thermoelements and Thermoelectric Cooling} (Infosearch, London, 1957). 

\bibitem{paolo2016}
P. A. Erdman, 
\textit{Thermoelectric Efficiency of a Multilevel Interacting Quantum Dot}, Master's Thesis Univ. Pisa (2016), 
available at 
\href{https://etd.adm.unipi.it/theses/available/etd-06272016-190534/}{https://etd.adm.unipi.it/theses/available/etd-06272016-190534/}.

\bibitem{bib:linke}
T. E. Humphrey, R. Newbury, R. P. Taylor, and H. Linke,
\href{http://journals.aps.org/prl/abstract/10.1103/PhysRevLett.89.116801}{Phys. Rev. Lett. {\bf 89}, 116801 (2002)}.

\bibitem{bib:humphrey} T. E. Humphrey  and H. Linke, 
\href{http://journals.aps.org/prl/abstract/10.1103/PhysRevLett.94.096601}{Phys. Rev. Lett. {\bf 94}, 096601 (2005)}.

\bibitem{Tsaousidou2010}
M. Tsaousidou, and G. P. Triberis, 
\href{https://doi.org/10.1088/0953-8984/22/35/355304}{J. Phys.: Condens. Matter {\bf 22}, 355304 (2010)}.

\bibitem{bib:whitney} R. S. Whitney, 
\href{http://journals.aps.org/prb/abstract/10.1103/PhysRevB.88.064302}{Phys. Rev. B {\bf 88}, 064302 (2013)}.

\bibitem{Benenti2011}
G. Benenti, K. Saito, and G. Casati,
\href{http://journals.aps.org/prl/abstract/10.1103/PhysRevLett.106.230602}{Phys. Rev. Lett. {\bf 106}, 230602 (2011)}.

\bibitem{Brandner2013}
K. Brandner, K. Saito, and U. Seifert,
\href{http://journals.aps.org/prl/abstract/10.1103/PhysRevLett.110.070603}{Phys. Rev. Lett. {\bf 110}, 070603 (2013)}.

\bibitem{Balachandran2013}
V. Balachandran, G. Benenti, and G. Casati,
\href{http://journals.aps.org/prb/abstract/10.1103/PhysRevB.87.165419}{Phys. Rev. B {\bf 87}, 165419 (2013)}.

\bibitem{Brandner2013b}
K. Brandner and U. Seifert,
\href{http://iopscience.iop.org/article/10.1088/1367-2630/15/10/105003/meta}{New J. Phys. {\bf 15}, 105003 (2013)}.

\bibitem{Yamamoto2016}
K. Yamamoto, O. Entin-Wohlman, A. Aharony, and N. Hatano,
\href{https://doi.org/10.1103/PhysRevB.94.121402}{Phys. Rev. B {\bf 94}, 121402 (2016)}.

\bibitem{Vandersypen}
L. H. Willems van Beveren, R. Hanson, I. T. Vink, F. H. L. Koppens,
L. P. Kouwenhoven. and L. M. K. Vandersypen,
\href{http://iopscience.iop.org/article/10.1088/1367-2630/7/1/182}{New J. Phys. {\bf 7}, 182 (2005)}.

\bibitem{SchmiedlSeifert2008}
T. Schmiedl and U. Seifert, 
\href{http://dx.doi.org/10.1209/0295-5075/81/20003}{Europhys. Lett. {\bf 81}, 20003 (2008)}.

\bibitem{bib:esposito2} M. Esposito, R. Kawai, K. Lindenberg, and C. Van den Broeck, \href{http://journals.aps.org/prl/abstract/10.1103/PhysRevLett.105.150603}{Phys. Rev. Lett. {\bf 105}, 150603 (2010)}.
\end{thebibliography}
\end{document}